%% file: paper1.tex
\documentclass[11pt]{article}
\usepackage[margin=1in]{geometry}
\usepackage{latexsym}
\usepackage{amssymb}
\usepackage{times}

\usepackage{amsmath,amsfonts,amsthm}
\usepackage{graphicx}

\usepackage[ruled,vlined,resetcount, algosection]{algorithm2e}

\usepackage{hyperref}
\usepackage{tabu}

\hypersetup{
bookmarks=true,
colorlinks=true,
linkcolor=blue,
urlcolor=blue,
citecolor=blue,
pdftex,
linktocpage=true, 
hyperindex=true
}

\newcommand{\supp}{\mathsf{supp}}
\newcommand{\SD}{\mathsf{SD}}
\newcommand{\Ext}{\mathsf{Ext}}
\newcommand{\size}{\mathsf{size}}
\newcommand{\depth}{\mathsf{depth}}

\newcommand{\myd}{2}
\newcommand{\poly}{\mathsf{poly}}
\newcommand{\Samp}{\mathsf{Samp}}
\newcommand{\coll}{\mathsf{coll}}
\newcommand{\Cond}{\mathsf{Cond}}
\newcommand{\AC}{\mathsf{AC}}
\newcommand{\NC}{\mathsf{NC}}
\newcommand{\NOT}{\mathsf{NOT}}
\newcommand{\AND}{\mathsf{AND}}
\newcommand{\OR}{\mathsf{OR}}

\newcommand{\NW}{\mathsf{NW}}
\newcommand{\IW}{\mathsf{IW}}
\newcommand{\eps}{\epsilon}
\newcommand{\bit}{\{0, 1\}}

\newcommand{\maj}{\mathsf{maj}}
\newcommand{\Match}{\mathsf{Match}}
\newcommand{\dist}{\mathsf{dist}}
\newcommand{\negl}{\mathsf{neg}}
\newcommand{\F}{\mathsf{F}}

\newtheorem{lemma}{Lemma}[section]
\newtheorem{theorem}[lemma]{Theorem}
\newtheorem{construction}[lemma]{Construction}

\newtheorem{definition}[lemma]{Definition}

\newtheorem{remark}[lemma]{Remark}

\begin{document}

\allowdisplaybreaks

\begin{titlepage}
\def\thepage{}

\title{Randomness Extraction in $\AC^0$ and with Small Locality}
\author{Kuan Cheng \thanks{kcheng17@jhu.edu.\ Department of Computer Science, Johns Hopkins University.\ Supported in part by NSF award CCF-1617713.} \and Xin Li \thanks{lixints@cs.jhu.edu.\ Department of Computer Science, Johns Hopkins University.\ Supported in part by NSF award CCF-1617713.}}

\maketitle \thispagestyle{empty}


\input{abstract.tex}
\end{titlepage}

\input{introduction.tex}

\input{prelim.tex}

\input{LowerBoundforErr.tex}

\input{basicAC0ext.tex}

\input{ErrorReduction.tex}

\input{OutputLenOPT.tex}

\input{DExtForBitfixing.tex}

\input{ExpanderExt.tex}

\input{Application.tex}

\bibliographystyle{alpha}
\bibliography{refs}

\input{appendix.tex}

\end{document}

%% file: abstract.tex
\begin{abstract}
Randomness extractors, which extract high quality (almost-uniform) random bits from biased random sources, are important objects both in theory and in practice.\ While there have been significant progress in obtaining near optimal constructions of randomness extractors in various settings, the computational complexity of randomness extractors is still much less studied. In particular, it is not clear whether randomness extractors with good parameters can be computed in several interesting complexity classes that are much weaker than $\mathsf{P}$.

In this paper we study randomness extractors in the following two models of computation: (1) constant-depth circuits ($\AC^0$), and (2) the local computation model. Previous work in these models, such as \cite{viola2005complexity}, \cite{goldreich2015randomness} and \cite{bogdanov2013sparse}, only achieve constructions with weak parameters. In this work we give explicit constructions of randomness extractors with much better parameters. Our results on $\AC^0$ extractors refute a conjecture in \cite{goldreich2015randomness} and answer several open problems there. We also provide a lower bound on the error of extractors in $\AC^0$, which together with the entropy lower bound in \cite{viola2005complexity, goldreich2015randomness} almost completely characterizes extractors in this class. Our results on local extractors also significantly improve the seed length in \cite{bogdanov2013sparse}. As an application, we use our $\AC^0$ extractors to study pseudorandom generators in $\AC^0$, and show that we can construct both cryptographic pseudorandom generators (under reasonable computational assumptions) and unconditional pseudorandom generators for space bounded computation with very good parameters.

Our constructions combine several previous techniques in randomness extractors, as well as introduce new techniques to reduce or preserve the complexity of extractors, which may be of independent interest. These include (1) a general way to reduce the error of strong seeded extractors while preserving the $\AC^0$ property and small locality, and (2) a seeded randomness condenser with small locality.
\end{abstract}

%% file: introduction.tex
\section{Introduction}
Randomness extractors are functions that transform biased random sources into almost uniform random bits. Throughout this paper, we model biased random sources by the standard model of general weak random sources, which are probability distributions over $n$-bit strings with a certain amount of min-entropy $k$.\footnote{A probability distribution is said to have min-entropy $k$ if the probability of getting any element in the support is at most $2^{-k}$.} Such sources are referred to as $(n, k)$-sources. In this case, it is well known that no deterministic extractors can exist for one single weak random source even if $k=n-1$; therefore seeded randomness extractors were introduced in \cite{NisanZ96}, which allow the extractors to have a short uniform random seed (say length $O(\log n)$). In typical situations, we require the extractor to be \emph{strong} in the sense that the output is close to uniform even given the seed. Formally, we have the following definition.

\begin{definition}[\cite{NisanZ96}]
 A function $\Ext:\{0,1\}^{n} \times \{ 0,1\}^d \rightarrow \{ 0,1\}^m$ is a seeded $(k, \epsilon)$ extractor if for any $(n, k)$ source $X$, we have
 \[|\Ext(X,U_d) - U_m| \leq \epsilon.\] 
 $\Ext$ is strong if in addition $|(\Ext(X,U_d), U_d) - (U_m,U_d) | \le \epsilon$, where $U_m$ and $U_d$ are independent uniform strings on $m$ and $d$ bits respectively, and $| \cdot |$ stands for the statistical distance. 
 \end{definition} 

Since their introduction, seeded randomness extractors have become fundamental objects in pseudorandomness, and have found numerous applications in derandomization, complexity theory, cryptography and many other areas in theoretical computer science. In addition, through a long line of research, we now have explicit constructions of seeded randomness extractors with almost optimal parameters (e.g., \cite{GuruswamiUV09}). However, the complexity of randomness extractors is still much less studied and understood. For example, while in general explicit constructions of randomness extractors can be computed in polynomial time of the input size, some of the known constructions are actually more explicit than that. These include for example extractors based on universal hashing \cite{CarterW79}, and Trevisan's extractor \cite{Trevisan01}, which can be computed by highly uniform constant-depth circuits of polynomial size with parity gates. Thus a main question one can ask is: can we do better and construct good randomness extractors with very low complexity?

This question is interesting not just by its own right, but also because such extractors, as building blocks, can be used to potentially reduce the complexity of other important objects.\ In this paper we study this question and consider the parallel and local complexity of randomness extractors. 

{\bf The parallel-$\AC^0$ model.} The hierarchy of $\NC$ and $\AC$ circuits are standard models for parallel computation. It is easy to see that the class of $\NC^0$ or even $\ell$-local functions for small $\ell$ 
, which correspond to functions where each output bit depends on at most $\ell$ input bits (including both the weak source and the seed), cannot compute strong extractors (since one can just fix $\ell$ bits of the source). Thus, a natural relaxation is to consider the class $\AC^0$, which refers to the family of polynomial-size and constant-depth circuits with unbounded fan-in gates.\ Note that although we have strong lower bounds here for explicit functions, it is still not clear whether some important objects, such as randomness extractors and pseudorandom generators, can be computed in $\AC^0$ with good parameters. Thus the study of this question also helps us better understand the power of this class. 

Viola \cite{viola2005complexity} was the first to consider this question, and his result was generalized by Goldreich et al. \cite{goldreich2015randomness} to show that for strong seeded extractors, even extracting a single bit is impossible if $k < n/\poly(\log n)$. When $k \geq n/\poly(\log n)$, Goldreich et al. showed how to extract $\Omega(\log n)$ bits using $O(\log n)$ bits of seed, or more generally how to extract $m< k/2$ bits using $O(m)$ bits of seed. Note that the seed length is longer than the output length.\footnote{They also showed how to extract $\poly(\log n)$ bits using an $O(\log n)$ bit seed, but the error of the extractor becomes $1/\poly (\log n)$.} When the extractor does not need to be strong, they showed that extracting $r+\Omega(r)$ bits using $r$ bits of seed is impossible if $k < n/\poly(\log n)$; while if $k \geq n/\poly(\log n)$ one can extract $(1+c)r$ bits for some constant $c>0$, using $r$ bits of seed. All the positive results here have error $1/\poly(n)$. 

Therefore, a natural and main open problem left in \cite{goldreich2015randomness} is whether one can construct randomness extractors in $\AC^0$ with shorter seed and longer output. Specifically, \cite{goldreich2015randomness} asks if one can extract more than $\poly (\log n) r$ bits in $\AC^0$ using a seed length $r=\Omega(\log n)$, when $k \geq n/\poly(\log n)$. In \cite{goldreich2015randomness} the authors conjectured that the answer is negative. Another open question is to see if one can achieve better error, e.g., negligible error instead of $1/\poly(n)$. 

Goldreich et al. \cite{goldreich2015randomness} also studied deterministic extractors for bit-fixing sources, and most of their effort went into extractors for oblivious bit-fixing sources (although they also briefly studied non-oblivious bit-fixing sources). An $(n, k)$-oblivious bit-fixing source is a string of $n$ bits such that some unknown $k$ bits are uniform, while the other $n-k$ bits are fixed. Extractors for such sources are closely related to exposure-resilient cryptography \cite{CanettiDHKS00, KampZ07}. In this case, a standard application of H{\aa}stad's switching lemma \cite{Has:parity} implies that it is impossible to construct extractors in $\AC^0$ for bit-fixing sources with min-entropy $k < n/\poly(\log n)$. The main result in \cite{goldreich2015randomness} is a theorem which shows the \emph{existence} of deterministic extractors in $\AC^0$ for min-entropy $k \geq n/\poly(\log n)$ that output $k/\poly(\log n)$ bits with error $2^{-\poly(\log n)}$. We emphasize that this is an existential result, and \cite{goldreich2015randomness} did not give any explicit constructions of such extractors. 

{\bf The local model.} Another relaxation, introduced by Bogdanov and Guo \cite{bogdanov2013sparse}, is the notion of sparse extractor families. These are families of functions for which each function in the family has a small number of overall input-output dependencies (referred to as the sparsity, meaning that the input-output dependency graph is sparse), while taking a random function from the family serves as a randomness extractor. Such extractors can be used generally in situations where hashing is used and preserving small input-output dependencies is needed. As an example, the authors in \cite{bogdanov2013sparse} used such extractors to obtain a transformation of non-uniform one-way functions into non-uniform pseudorandom generators that preserves output locality. 

In this paper, we consider the condition of the family being $\ell$-local, which is a worst case notion rather than the average case notion of sparsity. Furthermore, we will focus on the case of strong extractor families. Note that a strong extractor family is equivalent to a strong seeded extractor, since the randomness used to choose a function from the family can be included in the seed. Thus, we study strong seeded extractors with small locality, i.e., for any fixing of the seed, each output bit depends on at most $\ell$ input bits.

Note that an extractor family with $m$ output bits and locality $\ell$ is automatically an $\ell m$-sparse extractor family. Conversely, if an extractor is $s$-sparse, then half of its output bits depend on at most $2s/m$ input bits, so by removing half of the output bits one could obtain locality $2s/m$; a technical point here is that one may need to drop different output bits depending on the seed, but this does not affect the error of the extractor. 

The authors of \cite{bogdanov2013sparse} gave a construction of a strong extractor family for all entropy $k$ with output length $m \leq k$, error $\eps$, and sparsity $O(n \log (m/\eps)\log (n/m))$, which corresponds to locality $O(\frac{n}{m}\log (\frac{m}{\eps})\log (\frac{n}{m}))=\Omega(n/k \log(n/\eps))$ whenever $k \leq n/2$. They also showed that such sparsity is necessary whenever $n^{0.99} \leq m \leq n/6$ and $\eps$ is a constant. However, the main drawback of the construction in \cite{bogdanov2013sparse} is that the family size is quite large. Indeed the family size is $2^{nm}$, which corresponds to a seed length of at least $nm$.\footnote{In fact, the seed length is even larger since the seed is used to sample from a non-uniform distribution.} Therefore, a main open problem in \cite{bogdanov2013sparse} is to reduce the size of the family (or, equivalently, the seed length).

De and Trevisan \cite{DT09} obtained a strong extractor for $(n, k)$ sources such that for any fixing of the seed, each bit of the extractor's output only depends on $\poly(\log n)$ bits of the source. However, their construction only works for $k=\delta n$ where $\delta$ is any constant. Their extractor has seed length $d=O(\log n)$ and outputs $k^{\Omega(1)}$ bits, but the error is only $n^{-\alpha}$ for a small constant $0<\alpha<1$. 

It is also worthwhile to compare our definition of a strong extractor family with small locality to the definition of $t$-local extractors given by Vadhan \cite{v:local}. For a $t$-local extractor, one requires that for any fixing of the seed $r$, the output of the function $\Ext(x, r)$ \emph{as a whole} depends on only $t$ bits of $x$. In contrast, our definition requires that \emph{each output bit} of the function $\Ext(x, r)$ depends on at most $\ell$ bits of $x$. It can be seen that a strong extractor with sparsity $t$ is automatically a $t$-local extractor, but the converse may not be true: a $t$-local extractor may have locality $t$ and sparsity up to $mt$. By a lower bound in \cite{v:local}, the parameter $t$ in local-extractors is at least $\Omega(nm/k)$, which is larger than $m$ and matches the sparsity in \cite{bogdanov2013sparse} and our results up to polylogarithmic factors. In this sense, our definition of extractor with small locality is stronger than $t$-local extractors. Furthermore, the construction of $t$-local extractors in \cite{v:local}, which uses the sample-then-extract approach, only works for large min-entropy (at least $k > \sqrt{n}$); while our goal here is to construct strong extractor families even for very small min-entropy, with locality $\ell \ll m$.
   
\subsection{Our Results}
As in \cite{viola2005complexity, goldreich2015randomness}, in this paper we obtain both negative results and positive results about randomness extraction in $\AC^0$. While the negative results in \cite{viola2005complexity, goldreich2015randomness} provide lower bounds on the entropy required for $\AC^0$ extractors, our negative results provide lower bounds on the error such extractors can achieve. We show that such extractors (both seeded extractors and deterministic extractors for bit-fixing sources) cannot achieve error better than $2^{-\poly(\log n)}$, even if the entropy of the sources is quite large. Specifically, we have

\begin{theorem}
(General weak source) If $\Ext:\{0,1\}^n \times \{0,1\}^{d} \rightarrow \{0,1\}^m$ is a  strong $(k= n-1, \epsilon)$-extractor that can be computed by $\AC^0$ circuits of depth $\mathtt{dth}$ and size $s$, then $\epsilon =2^{  - (O(\log s))^{\mathtt{dth}-1}   \log (n+d)}$.

\noindent (Bit-fixing source) 
There is a constant $c > 1$ such that if $\Ext:\{0,1\}^n  \rightarrow \{0,1\}^m$ is a $(k, \epsilon)$-extractor for oblivious bit-fixing sources with $k  = n - (c\log s )^{\mathtt{dth}-1}$, that can be computed by $\AC^0$ circuits of depth $\mathtt{dth}$ and size $s$, then $\epsilon =2^{  - (O(\log s))^{\mathtt{dth}-1}   \log n}$.
\footnote{This holds even if we allow $\Ext$ to have a uniform random seed, see Theorem~\ref{DExterbound}.}

\end{theorem}

Thus, our results combined with the lower bounds on the entropy requirement in \cite{viola2005complexity, goldreich2015randomness} almost completely characterize the power of randomness extractors in $\AC^0$.

We now turn to our positive results. As our first contribution, we show that the authors' conjecture about seeded $\AC^0$ extractors in \cite{goldreich2015randomness} is false. We give explicit constructions of \emph{strong} seeded extractors in $\AC^0$ with much better parameters. This in particular answers open problems 8.1 and 8.2 in \cite{goldreich2015randomness}. To start with, we have the following theorem.

\begin{theorem} \label{thm:one}
For any constant $c \in \mathbb{N}$, any $k =\Omega( n/\log^{c} n )$ and any $\epsilon  = 1/\poly(n)$, there exists
an explicit construction of a strong $(k, \epsilon)$-extractor $\Ext:\{0,1\}^{n}\times \{0,1\}^d \rightarrow \{0,1\}^m$ that can be computed by an $\AC^0$ circuit of depth $ c + 10$, where $d = O(\log n)$ , $m = k^{\Omega(1)}$ and the extractor family has locality $ O(\log^{c+5} n)$. \end{theorem}

Note that the depth of the circuit is almost optimal, within an additive $O(1)$ factor of the lower bound given in \cite{goldreich2015randomness}. In addition, our construction is also a family with locality only $\poly(\log n)$. Note that the seed length $d=O(\log n)$ is (asymptotically) optimal, while the locality beats the one obtained in \cite{bogdanov2013sparse} (which is $O(n/m\log (m/\eps)\log (n/m))=n^{\Omega(1)})$ and is within a $\log^4 n$ factor to $O(n/k \log(n/\eps))$. 

Our result also improves that of De and Trevisan \cite{DT09}, even in the high min-entropy case, as our error can be any $1/\poly(n)$ instead of just $n^{-\alpha}$ for some constant $0<\alpha<1$. Moreover, our seed length remains $O(\log n)$ even for $k =n/\poly(\log n)$, while in this case the extractor in \cite{DT09} has seed length $\poly(\log n)$.

Next, we can boost our construction to reduce the error and extract almost all the entropy. We have

\begin{theorem}
\label{AC0ext}
For any constant $\gamma \in (0,1)$, $a, c\in \mathbb{N}$, any $k = \delta n = \Omega(n/\log^c n) $, $\epsilon = 1/2^{O(\log^a n)}$,
there exists an explicit strong $(k ,\epsilon)$-extractor $\Ext: \{0,1\}^{n} \times \{0,1\}^d \rightarrow \{0,1\}^{m}$ in $\AC^0$ with depth $ O(a+c+1)$ where $d = O((\log n + \frac{\log (n/\epsilon) \log (1/\epsilon)}{\log n})/\delta)$, $m = (1-\gamma)k$.

\end{theorem}

As our second contribution, we give \emph{explicit} deterministic extractors in $\AC^0$ for oblivious bit-fixing sources with entropy $k \geq n/\poly(\log n)$, which output $(1-\gamma)k$ bits with error $2^{-\poly(\log n)}$. This is in contrast to the non-explicit existential result in \cite{goldreich2015randomness}. Further, the output length and error of our extractor are almost optimal, while the output length in \cite{goldreich2015randomness} is only $k/\poly(\log n)$. Specifically, we have

\begin{theorem}
For  any constant $a, c \in \mathbb{N}$  and any constant $\gamma \in (0,1]$, there exists an explicit deterministic $(k  = \Omega(n/\log^a n), \epsilon = 2^{-\log^c n})$-extractor $\Ext:\{0,1\}^{n} \rightarrow \{0,1\}^{(1-\gamma)k}$ that can be computed by $\AC^0$ circuits of depth $O(a + c +1)  $, for any $(n, k)$-bit-fixing source. 
\end{theorem}

For sparse extractor families, we can reduce the error of Theorem~\ref{thm:one} while keeping the locality small.

\begin{theorem}\label{thm:sext1}
There exists a constant $\alpha \in (0, 1)$ such that for any $k \geq \frac{n}{\poly(\log n)} $ and $\epsilon \geq 2^{-k^{\alpha}}$,
there exists an explicit construction of a strong $(k, \epsilon)$-extractor $\Ext: \{0,1\}^{n} \times \{0,1\}^d \rightarrow \{0,1\}^{m}$, with $d =O(\log n + \frac{ \log(n/\epsilon)  \log(1/\epsilon)}{\log n})$, $m = k^{\Omega(1)}$ and locality $  \log^2 (1/\epsilon)  \poly(\log n) $.
\end{theorem}

We also give strong extractor families with small locality for min-entropy $k$ as small as $\log^2 n$. Our approach is to first condense it into another weak source with constant entropy rate. For this purpose we introduce the following definition of a (strong) randomness condenser with small locality. 

\begin{definition}
A function $\Cond : \{0,1\}^{n} \times \{0,1\}^d \rightarrow \{0,1\}^{n_1}$ is a strong $(n, k, n_1, k_1, \epsilon)$-condenser if for every $(n,k)$-source $X$ and independent uniform seed $R \in \{0,1\}^d$,  $R\circ \Cond(X, R)$ is $\epsilon$-close to $R\circ D$,  where $D$ is a distribution on  $\{0,1\}^{n_1}$ such that for any $r \in \{0,1\}^d$, we have that $D|_{R=r}$ is an $(n_1, k_1)$-source. We say the condenser family has locality $\ell$ if for every fixing of $R=r$, the function $\Cond(., r)$ can be computed by an $\ell$-local function.
\end{definition}

We now have the following theorem.

\begin{theorem}\label{thm:cond}
For any $ k \geq \log^2 n$, there exists a strong $(n, k, t = 10k, 0.08k,  \epsilon)$-condenser $\Cond:\{0,1\}^n \times \{0,1\}^d$ $\rightarrow \{0,1\}^t$ with $d = O(k)$ ,  $\epsilon = 2^{-\Omega(k)}$ and locality $O(\frac{n}{k}\log n)$.
\end{theorem}

Combining the condenser with our previous extractors, we get strong extractor families with small locality for any min-entropy $k \geq \log^2 n$. Specifically, we have

\begin{theorem}\label{thm:sext2}
There exits a constant $\alpha \in (0,1)$ such that for any $k \geq \log^2 n$, any constant $\gamma \in (0,1)$ and any $\eps \geq 2^{-k^{\alpha}}$,
there exists a strong $(k , \epsilon)$-extractor $\Ext:\{0,1\}^n \times \{0,1\}^d \rightarrow \{ 0,1 \}^m$, where $d =  O(k), m = (1-\gamma)k$ and the extractor family has locality $\frac{n}{k}\log^2 (1/\epsilon) (\log n)\poly(\log k)$.
\end{theorem}

In the above two extractors, our seed length is still much better than that of \cite{bogdanov2013sparse}. However, our locality becomes slightly worse.

\paragraph{Related work, independent work and further results.} The work of Dziembowski and Maurer \cite{dziembowski2002tight} also gave extractors with small locality. However, the model of the weak source studied there is uniform random bits subject to a bounded leakage, which is more restrictive than the model of general weak random sources we consider here. In particular, the analysis of \cite{dziembowski2002tight}, which uses a guessing game, may not work for general weak random sources. A recent independent work by Papakonstantinou et. al \cite{papakonstantinou2016true} used similar techniques as \cite{dziembowski2002tight} and  gave constructions of seeded extractors in the multi-stream model \cite{GS05}. Their main motivation and result is an extractor that can extract $\Omega(k)$ bits from any $(n, k)$ source with $k=\Omega(n)$, using $O(\log n\log(n/\eps))$ bits of seed together with two streams, $O(\log\log (n/\eps))$ passes and $O(\log (n/\eps))$ space. However, it turns out that their construction can also be realized in $\AC^0$ and also has the property of small locality. Specifically, for an $(n, k)$ source with $k=\delta n =n/\poly \log(n)$ and error $\eps=2^{-\poly \log(n)}$, their construction gives an $\AC^0$ extractor with seed length $O(\frac{1}{\delta^{O(1)}}\log n \log (n/\eps))$ and locality $O(\frac{1}{\delta^{O(1)}} \log (n/\eps))$. Their construction, which is based on randomly re-bucketing the bits of the source into blocks and arguing this results in a block source, can be viewed as orthogonal to our construction, which is based on hardness amplification. Compared to our results (Theorem~\ref{AC0ext} and Theorem~\ref{thm:sext1}), they have a better dependence on $\eps$ but we have a better dependence on $\delta$ and $n$. For very small entropy (e.g., $k=\log^2 n$), we can first use our condenser and then apply their construction,  which will give an extractor with seed length $O(k)$, output length $\Omega(k)$ and locality $O(\frac{n}{k} \log n \log(k/\eps))$.

\subsection{Applications to pseudorandom generators in $\AC^0$}

Like extractors, pseudorandom generators are also fundamental objects in the study of pseudorandomness, and constructing ``more explicit" pseudorandom generators is another interesting question that has gained a lot of attention. A pseudorandom generator (or PRG for short) is an efficient deterministic function that maps a short random seed into a long output that looks uniform to a certain class of distinguishers.

\begin{definition}
A function $G: \bit^n \to \bit^m$ is a pseudorandom generator for a class $\cal C$ of Boolean functions with error $\eps$, if for every function $ A \in \cal C$, we have that

\[|\Pr[{ A}(U_m)=1]-\Pr[{ A}(G(U_n))=1]| \leq \eps.\]
\end{definition}

Here we mainly consider two kinds of pseudorandom generators, namely cryptographic PRGs, which are necessarily based on computational assumptions; and unconditional PRGs, most notably PRGs for space bounded computation. 

Standard cryptographic PRGs (i.e., PRGs that fool polynomial time computation or polynomial size circuits with negligible error) are usually based on one-way functions (e.g., \cite{HILL93}), and can be computed in polynomial time. However, more explicit PRGs have also been considered in the literature, for the purpose of constructing more efficient cryptographic protocols. Impagliazzo and Naor \cite{impagliazzo1996efficient} showed how to construct such a PRG in $\AC^0$, which stretches $n$ bits to $n + \log n$ bits. Their construction is based on the assumed intractability of the subset sum problem. On the other hand, Viola \cite{viola2005constructing} showed that there is no black-box PRG construction with linear stretch in $\AC^0$ from one-way functions.  Thus, to get such stretch one must use non black-box constructions. 

In \cite{applebaum2006cryptography, applebaum2008pseudorandom}, Applebaum et al.\ showed that the existence of cryptographic PRGs in $\NC^0$ with sub-linear stretch follows from a variety of standard assumptions, and they constructed a  cryptographic PRG in $\NC^0$ with linear stretch based on a specific intractability assumption related to the hardness of decoding Òsparsely generatedÓ linear codes. In \cite{applebaum2013pseudorandom}, Applebaum further constructed PRG \emph{collections} (i.e., a family of PRG functions) with linear  stretch and polynomial stretch based on the assumption of one-wayness of a variant of the random local functions proposed by Goldreich \cite{goldreich2011candidate}. 

In the case of unconditional PRGs, for $d \geq 5$ Mossel et al.\ \cite{mossel2006varepsilon} constructed $d$-local PRGs with output length $n^{\Omega(d/2)}$ that fool all linear tests with error $2^{-n^{\frac{1}{2\sqrt{d}}}}$, which were used by Applebaum et al.\ \cite{applebaum2006cryptography} to give a $3$-local PRG with linear stretch that fools all linear tests. In the same paper, Applebaum et al.\ also gave a $3$-local PRG with sub linear stretch that fools sublinear-space computation. Thus, it remains to see if we can construct better PRGs (cryptographic or unconditional) in $\NC^0$ or $\AC^0$ with better parameters.

\subsubsection{Our PRGs}

We show that under reasonable computational assumptions, we can construct very good cryptographic PRGs in $\AC^0$ (e.g. with polynomial stretch and negligible error). In addition, we show that we can construct very good unconditional PRGs for space bounded computation in $\AC^0$ (e.g., with polynomial stretch). 

We first give explicit cryptographic PRGs in $\AC^0$ based on the one-wayness of random local functions, the same assumption as used in \cite{applebaum2013pseudorandom}. To state the assumption we first need the following definitions.

\begin{definition}[Hypergraphs \cite{applebaum2013pseudorandom}]
An $(n, m, d)$ hypergraph is a graph over $n$ vertices and $m$ hyperedges each of cardinality $d$. For each hyperedge $S = (i_0, i_1,\ldots, i_{d-1})$, the indices $i_0,i_1,\ldots, i_{d-1}$ are ordered. The hyperedges of $G$ are also ordered. Let $G$ be denoted as $([n], S_0, S_1, \ldots, S_{m-1})$ where for $i = 0, 1,\ldots, m-1$, $S_i$ is a hyperedge.

\end{definition}

\begin{definition}[Goldreich's Random Local Function \cite{goldreich2011candidate}]

Given a predicate $Q: \{0,1\}^d \rightarrow \{0,1\}$ and an $(n,m,d)$ hypergraph $G = ([n], S_0, \ldots, S_{m-1} )$,
the function $f_{G,Q}: \{0,1\}^n \rightarrow \{0,1\}^m$ is defined as follows: for input $x$, the $i$th output bit of $f_{G, Q}(x)$ is $f_{G, Q}(x)_i = Q(x_{S_i})$. 

For $m = m(n)$, the function collection $F_{Q,n,m}:\{0,1\}^{s} \times \{0,1\}^{n} \rightarrow \{0,1\}^{m}$ is defined via the mapping $(G,x) \rightarrow f_{G, Q}(x)$, where $G$ is sampled randomly by the $s$ bits and $x$ is sampled randomly by the $n$ bits. 

For every $k\in \{0,1\}^s$, we also denote $F(k, \cdot)$ as $F_k(\cdot)$.

\end{definition}

\begin{definition}[One-wayness of a Collection of Functions]

For $\epsilon = \epsilon(n) \in (0,1)$, a collection of functions $F:\{0,1\}^{s}\times \{0,1\}^n \rightarrow \{0,1\}^m$ is an $\epsilon$-one-way function if for every efficient adversary $A$ which outputs a list of $\poly(n)$ candidates and for sufficiently large $n$'s, we have that
$$ \Pr_{k, x, y = F_k(x)}[\exists z \in A(k, y), z' \in F^{-1}_k(y), z = z'] < \epsilon, $$
where $k$ and $x$ are independent and uniform.
\end{definition}

We now have the following theorem.
\begin{theorem}
For any $d$-ary predicate $Q$,
if the random local function $F_{Q,n,m}$ is $\delta$-one-way for some constant $\delta \in (0,1)$, then we have the following results.

\begin{enumerate}

\item If there exists a constant $\alpha > 0$ such that $m \geq (1+\alpha)n$, then for any constant $ c >1  $, there exists an explicit cryptographic PRG  $G: \{0,1\}^r \rightarrow \{0,1\}^t$ in $\AC^0$, where $t \geq cr$ and the error is negligible\footnote{The error $\epsilon: \mathbb{N}\rightarrow [0,1]$ is negligible if $ \epsilon(n) = n^{-\omega(1)}$.}.

\item If there exists a constant $\alpha > 0$ such that $m \geq n^{1+\alpha}$, then for any constant $c>1 $ there exists an explicit cryptographic PRG $G: \{0,1\}^r \rightarrow \{0,1\}^t$ in $\AC^0$, where $t \geq r^c$ and the error is negligible.

\end{enumerate}
\end{theorem}

As noted in \cite{applebaum2013pseudorandom}, there are several evidence supporting this assumption. In particular, current evidence is consistent with the existence of a $\delta$-one-way random local function $F_{Q,n,m}$ with $m \geq n^{1+\alpha}$ for some constant $\alpha>0$.

Compared to the constructions in \cite{applebaum2013pseudorandom}, our construction is in $\AC^0$ instead of $\NC^0$. However, our construction has the following advantages.

\begin{itemize}

\item We construct a standard PRG instead of a PRG collection, where the PRG collection is a family of functions and one needs to randomly choose one function before any application. 


\item The construction of a PRG with polynomial stretch in \cite{applebaum2013pseudorandom} can only achieve polynomially small error, and for negligible error one needs to  assume that the random local function cannot be inverted by any adversary with slightly super polynomial running time. Our construction, on the other hand, achieves negligible error while only assuming that the random local function cannot be inverted by any adversary that runs in polynomial time.
\end{itemize}

Next we give an explicit  PRG in $\AC^0$ with polynomial stretch, that fools space bounded computation. It is a straight forward application of our $\AC^0$-extractor to the Nisan-Zuckerman PRG \cite{NisanZ96}. 

\begin{theorem}

For every constant $c\in \mathbb{N}$ and every $m = m(s) = \poly(s)$, there is an explicit PRG $g: \{0,1\}^{r = O(s)} \rightarrow \{0,1\}^{m}$ in $\AC^0$, such that for any randomized algorithm $A$ using space $s$,
$$ | \Pr[A(g(U_r)) = 1] - \Pr[A(U_m) = 1] | = \epsilon \leq  2^{-\Theta(\log^c s)},$$
where $U_r$ is the uniform distribution of length $r$,  $U_m$ is the uniform distribution of length $m$.

\end{theorem} 

Compared to the Nisan-Zuckerman PRG \cite{NisanZ96}, our PRG is in $\AC0$, which is more explicit. On the other hand, our error is  $2^{-\Theta(\log^c s)}$ for any constant $c>0$ instead of being exponentially small as in \cite{NisanZ96}. It is a natural open problem to see if we can reduce the error to exponentially small. We note that this cannot be achieved by simply hoping to improve the extractor, since our negative result shows that seeded extractors in $\AC^0$ cannot achieve error better than $2^{-\poly(\log n)}$. 

\subsection{Overview of the Constructions and Techniques}
Our negative results about the error of $\AC^0$ extractors follow by a simple application of Fourier analysis and the well known spectrum concentration theorem of $\AC^0$ functions \cite{LMN89}. We present it in Section~\ref{sec:errorbound}. We now briefly describe our positive results. We will extensively use the following two facts: the parity and inner product over $\poly(\log n)$ bits can be computed by $\AC^0$ circuits of size $\poly(n)$; in addition, any Boolean function on $O(\log n)$ bits can be computed by a depth-2 $\AC^0$ circuit of size $\poly(n)$.

\subsubsection{Basic construction}
All our constructions are based on a basic construction of a strong extractor in $\AC^0$ for any $k \geq \frac{n}{\poly(\log n)}$ with seed length $d=O(\log n)$ and error $\eps=n^{-\Omega(1)}$. This construction is a a modification of the Impagliazzo-Wigderson pseudorandom generator \cite{iw:bppequalsp}, interpreted as a randomness extractor in the general framework found by Trevisan \cite{Trevisan01}. The IW-generator first takes a Boolean function on $\log n$ bits, applies a series of hardness amplifications to get another Boolean function on $O(\log n)$ bits, and then uses the Nisan-Wigderson generator \cite{NisanW94} together with the new Boolean function. The hardness amplification consists of three steps: the first step, developed by Babai et al. \cite{bfnw}, is to obtain a mild average-case hard function from a worst-case hard function; the second step involves a constant number of substeps, with each substep amplifying the hardness by using Impagliazzo's hard core set theorem \cite{impagliazzo1995hard}; the third step, developed by Impagliazzo and  Wigderson \cite{iw:bppequalsp}, uses a derandomized direct-product generator to obtain a function that can only be predicted with exponentially small advantage. 

Trevisan \cite{Trevisan01} showed that given an $(n, k)$-source $X$, if one regards the $n$ bits of $X$ as the truth table of the initial Boolean function on $\log n$ bits and applies the IW-generator, then by setting parameters appropriately one obtains an extractor. The reason is that for any $x \in \supp(X)$ that makes the output of the extractor fail a certain statistical test $T$, one can ``reconstruct" $x$ by showing that it can computed by a small size circuit, when viewing $x$ as the truth table of the function with $T$ gates. Thus the number of such bad elements $x \in \supp(X)$ is upper bounded by the total number of such circuits. This extractor works for any min-entropy $k \geq n^{\alpha}$. 

However, this extractor itself is not in $\AC^0$ (which is not surprising since it can handle min-entropy $k \geq n^{\alpha}$). Thus, at least one of the steps in the construction of the IW-generator/extractor is not in $\AC^0$. By carefully examining each step one can see that the only step not in $\AC^0$ is actually the first step of hardness amplification (This was also pointed out by \cite{viola2005complexity}). Indeed, all the other steps of hardness amplification are essentially doing the same thing: obtaining a function $f'$ on $O(\log n)$ bits from another function $f$ on $O(\log n)$ bits, where the output of $f'$ is obtained by taking the inner product over two $O(\log n)$ bit strings $s$ and $r$. In addition, $s$ is obtained directly from part of the input of $f'$, while $r$ is obtained by using the other part of the input of $f'$ to generate $O(\log n)$ inputs to $f$ and concatenating the outputs. All of these steps can be done in $\AC^0$, assuming $f$ is in $\AC^0$ (note that $f$ here depends on $X$). 

We therefore modify the IW-generator by removing the first step of hardness amplification, and start with the second step of hardness amplification with the source $X$ as the truth table of the initial Boolean function. Thus the initial function $f$ can be computed by using the $\log n$ input bits to select a bit from $X$, which can be done in $\AC^0$. Therefore the final Boolean function $f'$ can be computed in $\AC^0$. The last step of the construction, which applies the NW-generator, is just computing $f'$ on several blocks of size $O(\log n)$, which certainly is in $\AC^0$. This gives our basic extractor in $\AC^0$.

The analysis is again similar to Trevisan's argument \cite{Trevisan01}. However, since we have removed the first step of hardness amplification, now for any $x \in \supp(X)$ that makes the output of the extractor  to fail a certain statistical test $T$, we cannot obtain a small circuit that \emph{exactly computes} $x$. On the other hand, we can obtain a small circuit that can \emph{approximate} $x$ well, i.e., can compute $x$ correctly on $1-\gamma$ fraction of inputs for some $\gamma=1/\poly(\log n)$. We then argue that the total number of strings within relative distance $\gamma$ to the outputs of the circuit is bounded, and therefore combining the total number of possible circuits we can again get a bound on the number of such bad elements in $\supp(X)$. A careful analysis shows that our extractor works for any min-entropy $k \geq n/\poly(\log n)$. However, to keep the circuit size small we have to set the output length to be small enough, i.e., $n^{\alpha}$ and set the error to be large enough, i.e., $n^{-\beta}$. 

\subsubsection{Error reduction}
We now describe how we reduce the error of the extractor. We will borrow some techniques from the work of Raz et al. \cite{rrv:error}, where the authors showed a general way to reduce the error of strong seeded extractors. However, the techniques in Raz et al. \cite{rrv:error} do not preserve the $\AC^0$ property, thus our techniques are significantly different from theirs. Nevertheless, our starting point is a lemma from \cite{rrv:error}, which roughly says the following: given any strong seeded $(k, \eps)$-extractor $\Ext$ with seed length $d$ and output length $m$, then for any $x \in \bit^n$ there exists a set $G_x \subset \bit^d$ of density $1-O(\eps)$, such that if $X$ is a source with entropy slightly larger than $k$, then the distribution $\Ext(X, G_X)$ is very close to having min-entropy $m-O(1)$. Here  $\Ext(X, G_X)$ is the distribution obtained by first sampling $x$ according to $X$, then sampling $r$ uniformly in $G_x$ and outputting $\Ext(x, r)$. 

Giving this lemma, we can apply our basic $\AC^0$ extractor with error $\eps=n^{-\beta}$ for some $t$ times, each time with fresh random seed, and then concatenate the outputs. By the above lemma, the concatenation is roughly $(O(\eps))^t$-close to a source such that one of the output has min-entropy $m-O(1)$ (i.e., a somewhere high min-entropy source). By choosing $t$ to be a large enough constant the $(O(\eps))^t$ can be smaller than any $1/\poly(n)$. We now describe how to extract from the somewhere high min-entropy source with error smaller than any $1/\poly(n)$, in $\AC^0$. This is where our construction differs significantly from \cite{rrv:error}, as there one can simply apply a good extractor for constant entropy rate.

Assume that we have an $\AC^0$ extractor $\Ext'$ that can extract from $(m, m-\sqrt{m})$-sources with error any $\eps'=1/\poly(n)$ and output length $m^{1/3}$. Then we can extract from the somewhere high min-entropy source as follows. We use $\Ext'$ to extract from each row of the source with fresh random seed, and then compute the XOR of the outputs. We claim the output is $(2^{-m^{\Omega(1)}}+\eps')$-close to uniform. To see this, assume without loss of generality that the $i$'th row has min-entropy $m-O(1)$. We can now fix the outputs of all the other rows, which has a total size of $t m^{1/3} \ll \sqrt{m}$ as long as $t$ is small. Thus, even after the fixing, with probability $1-2^{-m^{\Omega(1)}}$, we have that the $i$'th row has min-entropy at least $m-\sqrt{m}$. By applying $\Ext'$ we know that the XOR of the outputs is close to uniform. 

What remains is the extractor $\Ext'$. To construct it we divide the source with length $m$ sequentially into $m^{1/3}$ blocks of length $m^{2/3}$. Since the source has min-entropy $m-\sqrt{m}$, this forms a block source such that each block roughly has min-entropy at least $m^{2/3}-\sqrt{m}$ conditioned on the fixing of all previous ones. We can now take a strong extractor $\Ext''$ in $\AC^0$ with seed length $O(\log n)$ and use the same seed to extract from all the blocks, and concatenate the outputs. It suffices to have this extractor output one bit for each block. Such $\AC^0$ extractors are easy to construct since each block has high min-entropy rate (i.e., $1-o(1)$). For example, we can use the extractors given by  Goldreich et al. \cite{goldreich2015randomness}. 

It is straightforward to check that our construction is in $\AC^0$, as long as the final step of computing the XOR of $t$ outputs can be done in $\AC^0$. For error $1/\poly(n)$, it suffices to take $t$ to be a constant and the whole construction is in $\AC^0$, with seed length $O(\log n)$. We can even take $t$ to be $\poly(\log n)$, which will give us error $2^{-\poly(\log n)}$. 

\subsubsection{Increasing output length }
The error reduction step reduces the output length from $m$ to $m^{1/3}$, which is still $n^{\Omega(1)}$. We can increase the output length by using a standard boosting technique as that developed by Nisan and Zuckerman \cite{NisanZ96, Zsamp}. Specifically, we first use random bits to sample several blocks from the source, using a sampler in $\AC^0$. We then apply our $\AC^0$ extractor on the blocks backwards, and use the output of one block as the seed to extract from the previous block. When doing this we divide the seed into blocks each with the same length as the seed of the $\AC^0$ extractor, apply the $\AC^0$ extractor using each block as the seed, and then concatenate the outputs. This way each time the output will increase by a factor of $n^{\Omega(1)}$. Thus after a constant number of times it will become say $\Omega(k)$. Since each step is computable in $\AC^0$, the whole construction is still in $\AC^0$. 

\subsubsection{Explicit $\AC^0$ extractors for bit-fixing source}
Our explicit $\AC^0$ extractors for (oblivious) bit-fixing sources follow the high-level idea in \cite{goldreich2015randomness}. Specifically, we first reduce the oblivious bit-fixing source to a non-oblivious bit-fixing source, and then apply an extractor for non-oblivious bit-fixing sources. This approach is natural in the sense that the best known extractors for oblivious bit-fixing sources (e.g., parity or \cite{KampZ07}) can both work for small entropy and achieve very small error. Thus by the negative results in \cite{goldreich2015randomness} and our paper, none of these can be in $\AC^0$. However, extractors for non-oblivious bit-fixing sources are equivalent to resilient functions, and there are well known resilient functions in $\AC^0$ such as the Ajtai-Linial function \cite{ajtai93influence}. 

The construction in \cite{goldreich2015randomness} is not explicit, but only existential for two reasons. First, at that time the Ajtai-Linial function is a random function, and there was no explicit construction matching it. Second, the conversion from oblivious-bit fixing source to non-oblivious bit-fixing source in \cite{goldreich2015randomness} is to multiply the source by a random matrix, for which the authors of \cite{goldreich2015randomness} showed its existence but were not able to give an explicit construction. Now, the first obstacle is solved by recent explicit constructions of resilient functions in $\AC^0$ that essentially match the Ajtai-Linial function (\cite{CZ15, Mek:resil, li2015improved}). Here we use the extractor in \cite{li2015improved} that can output many bits. For the second obstacle, we notice that the extractors for non-oblivious bit-fixing sources in \cite{CZ15, li2015improved} do not need the uniform bits to be independent, but rather only require $\poly(\log N)$-wise independence if $N$ is the length of the source.

By exploiting this property, we can give an explicit construction of the matrix used to transfrom the original oblivious bit-fixing source. Our construction is natural and simpler than that in \cite{goldreich2015randomness}, in the sense that it is a matrix over $\F_2$ while the matrix in \cite{goldreich2015randomness} uses fields of larger size. Specifically, we will take a  seeded extractor and view it as a bipartite graph with $N=n^{O(1)}$ vertices on the left, $n$ vertices on the right and left degree $d=\poly(\log N)=\poly(\log n)$. We identify the right vertices with the $n$ bits of the bit-fixing source, and for each left vertex we obtain a bit which is the parity of its neighbors. The new non-oblivious bit-fixing source is the $N$ bit source obtained by concatenating the left bits.

Now suppose the original source has entropy $k=\delta n$ for some $\delta \geq 1/\poly(\log n)$, and let $T$ denote the unfixed bits. A standard property of the seeded extractor implies that most of the left vertices have a good fraction of neighbors in $T$ (i.e., an extractor is a good sampler), so that each left bit obtained from these vertices is uniform. Next we would like to argue that they are $\poly(\log N)$-wise independent. For this we require the seeded extractor to have a stronger property: that it is a \emph{design extractor} as defined by Li \cite{Li12a}. Besides being an extractor itself, a design extractor requires that any pair of left vertices have a small intersection of neighbors. Assuming this property, it is easy to show that if we take any small subset $S$ of the ``good" left vertices, then there is a bit in $T$ that is only connected to a single vertex in $S$ (i.e., a unique neighbor). Thus the XOR of any small enough subset of the ``good" left bits is uniform, which indicates that they are some $t$-wise independent. Several explicit constructions of design extractors were given in \cite{Li12a}, and for our applications it suffices to use a simple greedy construction. By adjusting the parameters, we can ensure that $t=\poly(\log N)$ which is enough for applying the extractor in \cite{li2015improved}. In addition, the degree $d=\poly(\log N)$ so the parity of $d$ bits can be computed in $\AC^0$. 

Once we have the basic extractor, we can use the same techniques as in \cite{goldreich2015randomness} to reduce the error, and use the techniques by Gabizon et al.\cite{GabizonRS04} to increase the output length (this is also done in \cite{goldreich2015randomness}). Note that the techniques in \cite{GabizonRS04} require a seeded extractor. In order for the whole construction to be in $\AC^0$, we use our previously constructed seeded extractor in $\AC^0$ which can output $(1-\gamma)k$ bits. Thus we obtain almost optimal explicit $\AC^0$ extractors for oblivious bit-fixing sources. In contrast, the seeded extractor used in \cite{goldreich2015randomness} only outputs $k/\poly(\log n)$ bits, and thus their (non-explicit) $\AC^0$ extractor for oblivious bit-fixing sources also only outputs $k/\poly(\log n)$ bits. 

\subsubsection{Extractors with small locality for low entropy}
Our basic extractor (Theorem~\ref{thm:one}) also enjoys the property of small locality, but it only works for large entropy. To get constructions for small min-entropy, we adapt the techniques in \cite{bogdanov2013sparse}. There the authors constructed a strong extractor family with small sparsity by randomly sampling an $m \times n$ matrix $M$ and outputting $MX$, where $X$ is the $(n, k)$-source. Each entry in $M$ is independently sampled according to a Bernoulli distribution, and thus the family size is $2^{nm}$. We derandomize this construction by sampling the second row to the last row using a random walk on an expander graph, starting from the first row. For the first row, we observe that the process of generating the entries and doing inner product with $X$ can be realized by read-once small space computation, thus we can sample the first row using the output of a pseudorandom generator for space bounded computation (e.g., Nisan's generator \cite{Nisan92}). We show that this gives us a very good condenser with small locality, i.e., Theorem~\ref{thm:cond}. Combining the condenser with our previous extractors we then obtain strong extractor families with small locality.

\subsubsection{Applications to pseudorandom generators}
For cryptographic pseudorandom generators, we mainly adapt the approach of Applebaum \cite{applebaum2013pseudorandom}, to the $\AC^0$ setting.\ The construction of cryptographic pseudorandom generator \emph{families} in \cite{applebaum2013pseudorandom} is based on random local functions. Specifically, given a random bipartite graph with $n$ left vertices, $m$ right vertices and right degree $d$ (think of $d$ as a constant), and a suitable predicate $P$ on $d$ bits, Applebaum showed that based on a conjecture on random local one-way functions, the $m$ output bits obtained by applying $P$ to the $m$ subsets of input bits corresponding to the hyper edges give a distribution with high pseudo Shannon entropy. He then showed how to boost the output to have high pseudo min-entropy by concatenating several independent copies. At this point he used an extractor in $\NC^0$ to turn the output into a pseudorandom string. 

However, an extractor in $\NC^0$ needs to have a large seed length (i.e., $\Omega(n)$), thus the $\NC^0$ PRG constructed using this approach only achieves linear stretch. Another issue is that the $\NC^0$ PRG is actually a collection of functions rather than a single function, because the random bits used to sample the bipartite graph is larger than the output length, and is treated as a public index to the collection of functions.

Here, by replacing the extractor with our $\AC^0$ extractor we can achieve a polynomial stretch PRG (based on appropriate assumptions as in \cite{applebaum2013pseudorandom}), although now the PRG is in $\AC^0$ instead of $\NC^0$. In addition, we can get a single PRG instead of a collection of PRG functions, by including the random bits used to sample the bipartite graph as part of the seed. Since in the graph each right vertex only has a constant number $d$ of neighbors, the sampling uses $md \log n$ bits and can be done in $\AC^0$. To ensure that the PRG has a stretch, we take the sampled graph $G$ and apply the \emph{same} graph to several independent copies of $n$ bit input strings. We show that we can still use the method in \cite{applebaum2013pseudorandom} to argue that this gives a a distribution with high pseudo Shannon entropy. We then use the same method as in \cite{applebaum2013pseudorandom} to turn it into a distribution with high pseudo min-entropy, and finally we apply our $\AC^0$ extractor. This way we ensure that the  $md \log n$ bits used to sample the graph $G$ are ``absorbed" by the stretch of the PRG, and thus we get a standard PRG instead of a collection of PRG functions.

For PRGs for space bounded computation, we simply adapt the PRG by Nisan and Zuckerman \cite{NisanZ96}, which stretches $O(S)$ random bits to any $\poly(S)$ bits that fool space $S$ computation. We now replace the seeded extractor used there by our $\AC^0$ extractor. Notice that the Nisan-Zuckerman PRG simply applies the seeded extractor iteratively for a constant number of times, so the whole construction is still in $\AC^0$.

\input{open.tex}

\subsection{Organization of this Paper}
The rest of the paper is organized as follows. In Section~\ref{sec:prelim} we review some basic definitions and the relevant background. In Section~\ref{sec:errorbound}, we give the lower bounds on errors of $\AC^0$ extractors for general weak sources and bit-fixing sources. In Section~\ref{sec:basicext} we describe our construction of a basic extractor in $\AC^0$, and with small locality. In section~\ref{sec:error} we describe the error reduction techniques for $\AC^0$ extractors. In Section~\ref{sec:output} we show how to increase the output length for our $\AC^0$ extractors. In section~\ref{sec:bitfixing} we construct $\AC^0$ extractors for bit-fixing sources.
In Section~\ref{sec:localext} we construct extractors with small locality for small entropy sources. In Section~\ref{sec:app} we present several applications, i.e., constructing pseudorandom generators in $\AC^0$. We put some omitted details in Appendix~\ref{appen}.

%% file: open.tex
\subsection{Open Problems}\label{sec:conc}
Our work leaves many natural open problems. First, in terms of the seed length and output length, our $\AC^0$ extractor is only optimal when $k=\Omega(n)$. Is it possible to simultaneously achieve optimal seed length and output length when $k =n/\poly(\log n)$? Second, can we construct good $\AC^0$ extractors for other classes of sources, such as independent sources and affine sources?

Turning to strong extractor families with small locality, again the parameters of our constructions do not match the parameters of optimal seeded extractors. In particular, our seed length is still $O(k)$ when the min-entropy $k$ is small. Can we reduce the seed length further? We note that using our analysis together with the IW-generator/extractor, one can get something meaningful (i.e., a strong extractor family with a relatively short seed and small locality) even when $k =n^{\alpha}$ for some $\alpha>1/2$. But it's unclear how to get below this entropy. 

For pseudorandom generators in $\AC^0$, there are also many interesting open problems left. For example, can we construct better cryptographic PRGs, or use weaker computational assumptions? In particular, it would be nice to construct a cryptographic PRG with polynomial stretch based on the one-wayness of a random local function with $m=(1+\alpha)n$ instead of $m =n^{1+\alpha}$ as in our current construction. For space bounded computation, is it possible to match the exponentially small error of the Nisan-Zukerman PRG? Taking one step further, is it possible to construct PRGs in $\AC^0$ for space bounded computation, with stretch matching the PRGs of Nisan \cite{Nisan92} and Impagliazzo-Nisan-Wigderson \cite{inw}?

%% file: prelim.tex
\section{Preliminaries}\label{sec:prelim}

For any $i\in \mathbb{N}$, we use $\langle i\rangle$ to denote the string which is the binary representation of $i$. Let $\langle \cdot, \cdot\rangle$ denote the inner product of two binary strings having the same length. Let $|\cdot|$ denote the length of the input string. Let $w(\cdot)$ denote the weight of the input binary string. For any strings $x_1$ and $x_2$, let $x_1 \circ x_2$ denote the concatenation of $x_1$ and $x_2$. For any strings $x_1, x_2, \ldots, x_t$, let $\bigcirc_{i=1}^t x_i$ denote $x_1\circ x_2 \circ \cdots \circ x_t$.

Let $\supp(\cdot)$ denote the support of the input random variable.

\begin{definition}[Weak Random Source, Block Source]
The min-entropy of a random variable $X$ is 
$$H_{\infty}(X) = \min_{x\in \supp(X)}\{-\log \Pr(X = x)\}.$$
We say a random variable $X$ is an $(n,k)$-source if the length of $X$ is $n$ and $H_{\infty}(X) \geq k$. We say $X = \bigcirc_{i=1}^m X_i$ is an $((n_1, k_1), (n_2,k_2),\ldots, (n_m, k_m))$-block source if $\forall i\in [m]$, $\forall x \in \supp(\bigcirc_{j=1}^{i-1}X_j)$, $X_i|_{\bigcirc_{j=1}^{i-1}X_j = x}$ is an $(n_i, k_i)$-source.

\end{definition}

For simplicity, if $n_1,n_2, \cdots, n_m $ are clear from the context, then we simply say that the block source $X$ is a $(k_1, k_2,\ldots, k_m)$-block source.

We say an $(n,k)$-source $X$ is a flat $(n,k)$-source if $\forall a\in \supp(X)$, $\Pr[X=a] = 2^{-k}$. In this paper, $X$ is usually a random binary string with finite length. So $\supp(X)$ includes all the binary strings of that length such that $\forall x\in \supp(X), \Pr[X=x] > 0$.

Bit-fixing source is a special kind of weak source. In this paper we also consider deterministic extractors for bit-fixing source.
\begin{definition}[Non-oblivious Bit-Fixing Sources]

A source $X$ on $\{0,1\}^n$ is a $(q, t, \gamma)$-non-oblivious bit-fixing source (in short, NOBF source ) if there exists a subset $Q\subseteq [n]$ of size at most $q$ and a sequence of functions $f_1, f_2,\ldots, f_n:\{0,1\}^{|I|} \rightarrow \{0,1\}$ such that the joint distribution of the bits indexed by $\bar{Q} = [n]\backslash Q$ (denoted by $X_{\bar{Q}}$) is $(t, \gamma)$-wise independent ($\gamma$-close to a $t$-wise independent source) and $X_i = f_i(X_{\bar{Q}})$ for every $i\in Q$.

\end{definition}

Bit-fixing sources are special non-oblivious bit-fixing sources.
An $(n, t)$-bit-fixing source is defined to be an $(n-t, t, 0)$-non-oblivious bit-fixing source.

We use $U$ to denote the uniform distribution. In the following, we do not always claim the length of $U$, but its length can be figured out from the context.

\begin{definition}[Statistical Distance]

The statistical distance between two random variables $X$ and $Y$, where $|X| = |Y|  $ , is $\SD(X, Y)$ which is defined as follows.
$$\SD(X, Y) = 1/2 \sum_{a\in \{0,1\}^{|X|}} |\Pr[X=a] - \Pr[Y=a]| $$

\end{definition}

\begin{lemma}[Properties of Statistical Distance \cite{arora2009computational}]
\label{SDproperty}
Statistical distance has the following properties.
\begin{enumerate}
\item (Triangle Inequality) For any random variables $X$, $Y$, $Z$, such that $|X| = |Y| = |Z|$, we have
$$\SD(X, Y) \leq \SD(X, Z) + \SD(Y, Z).$$
\item For any $n, m \in \mathbb{N}^+$, any deterministic function $f:\{0,1\}^n \rightarrow \{0,1\}^m$ and any random variables $X$, $Y$ over $\{0,1\}^n$, $\SD(f(X),f(Y))\leq \SD(X,Y)$.
\end{enumerate}

\end{lemma}

\begin{definition}[Extractor]
A $(k,\epsilon)$-extractor is a function $\Ext: \{0,1\}^n \times \{0,1\}^d \rightarrow \{0,1\}^m$ with the following property. For every $(n,k)$-source $X$, the distribution $\Ext(X, U)$ is within statistical distance $\epsilon$ from uniform distributions over $\{0,1\}^{m}$. 

A strong $(k,\epsilon)$-extractor is a function $\Ext: \{0,1\}^n \times \{0,1\}^d \rightarrow \{0,1\}^m$ with the following property. For every $(n,k)$-source $X$, the distribution $U\circ \Ext(X, U)$ is within statistical distance $\epsilon$ from uniform distributions over $\{0,1\}^{d+m}$. The entropy loss of the extractor is $ k-m$.
\end{definition}

The existence of extractors can be proved using the probabilistic method. The result is stated as follows.
\begin{theorem}[\cite{vadhan2012pseudorandomness}]
\label{extinPne}
For any $n,k \in \mathbb{N}$ and $ \epsilon > 0$,
there exists a strong $(k,\epsilon)$-extractor $\Ext:\{0,1\}^n \times \{0,1\}^d \rightarrow \{0,1\}^m$ such that $d = \log (n-k)+ 2\log(1/\epsilon) + O(1), m = k-2\log (1/\epsilon) + O(1)$.
\end{theorem}
In addition, researchers have found explicit extractors with almost optimal parameters, for example we have the following theorem.
\begin{theorem}[\cite{GuruswamiUV09}]
\label{extinP}
For every constant $\alpha > 0$, every $n,k \in \mathbb{N}$ and $\epsilon >0$, 
there exist an explicit construction of strong $(k,\epsilon)$-extractor $\Ext:\{0,1\}^n \times \{0,1\}^d \rightarrow \{0,1\}^m$ with $d = O(\log \frac{n}{\epsilon}), m \geq (1-\alpha)k$.
\end{theorem}

We also use the following version of Trevisan's extractor \cite{Trevisan01}.

\begin{theorem}[Trevisan's Extractor \cite{Trevisan01}]
\label{Trevext}
For  any constant $\gamma \in (0,1]$, let $k = n^{\gamma}$.
For any $\epsilon \in (0, 2^{-k/12})$,
there exists an explicit construction of $(k, \epsilon)$-extractor $\Ext:\{0,1\}^{n} \times \{0,1\}^d \rightarrow \{0,1\}^m$ such that $d = O((\log n/\epsilon)^{2}/\log n)$, $ m \in [36, k/2)$. 

\end{theorem}

For block sources,  randomness extraction can be done in parallel, using the same seed for each block.
\begin{lemma}[Block Source Extraction]
\label{blocksourceextraction}
For any $t\in \mathbb{N}^+$,  let $X = \bigcirc_{i=1}^t X_i$ be any $( k_1,  k_2, \ldots,  k_t )$-block source where for each $i\in [t]$, $|X_i| = n_i$.
For every $i\in [t] $, let $\Ext_i:\{0,1\}^{n_i} \times \{0,1\}^{d} \rightarrow \{0,1\}^{m_i}$ be a strong $(k_i, \epsilon_i)$-extractor. 
Then the distribution $R \circ \Ext_1(X_1, R)\circ \Ext_2(X_2, R) \circ \ldots \circ  \Ext_t(X_t,R)$ is $\sum_{i\in [t]} \epsilon_i$-close to uniform, where $R$ is uniformly sampled from $\{0,1\}^{d}$, and independent of $X$.
\end{lemma}

\begin{proof}
We use induction.
If the source has only $1$ block, then the statement is true by the definition of strong extractors.

Assume for $(t-1)$ blocks, the statement is true. We view $\Ext_1(X_1, R)\circ \Ext_2(X_2, R) \circ \ldots \circ  \Ext_t(X_{t}, R)$ as $Y\circ \Ext_t(X_{t}, R)$. Here $Y = \Ext_1(X_1, R)\circ \Ext_2(X_2, R) \circ \ldots \circ  \Ext_{t-1}(X_{t-1}, R)$. Let $U_1, U_2$ be two independent uniform distributions, where $|U_1|= |Y|= m$ and $|U_2| = m_t$. Then
\begin{equation}
\begin{split}
&\SD( R\circ Y\circ \Ext_{t}(X_{t}, R), R\circ U_1 \circ U_2  ) \\
\leq &\SD(R\circ Y \circ \Ext_{t}(X_{t}, R), R\circ U_1 \circ Z) + \SD(R\circ U_1\circ Z, R \circ U_1\circ U_2).\\
\end{split}
\end{equation}
Here $Z$ is the random variable such that $\forall r\in \{0,1\}^d, \forall y\in \{0,1\}^{m}$, $Z|_{R=r,U_1=y}$ has the same distribution as $\Ext_{t}(X_{t}, R)|_{R=r, Y=y}$.

First we give the upper bound of $  \SD(R\circ Y \circ \Ext_t(X_{t}, R), R\circ U_1 \circ Z) $.

\begin{align*}
&\SD(R\circ Y \circ \Ext_t(X_{t}, R), R\circ U_1 \circ Z) \tag{\stepcounter{equation}\theequation} \\
=  & \frac{1}{2}\sum_{r\in \{0,1\}^d} \sum_{y \in \{0,1\}^{m}} \sum_{z\in \{0,1\}^{m_{t}}}|\Pr[R=r]\Pr[Y=y|_{R=r}]\Pr[\Ext_t(X_t, R) = z|_{R=r, Y=y}]\\
 & - \Pr[R=r]\Pr[U_1=y]\Pr[Z=z|_{R=r, U_1 = y}] |   \\
= & \frac{1}{2}\sum_{r\in \{0,1\}^d} \sum_{y \in \{0,1\}^{m}} \sum_{z\in \{0,1\}^{m_{t}}} \Pr[R=r] \Pr[Z=z|_{R=r, U_1 = y}] |\Pr[Y=y|_{R=r}] - \Pr[U_1=y]|   \\
= & \frac{1}{2}\sum_{r\in \{0,1\}^d} \sum_{y \in \{0,1\}^{m}}  \Pr[R=r] |\Pr[Y=y|_{R=r}] - \Pr[U_1=y]|  \sum_{z\in \{0,1\}^{m_{t}}} \Pr[Z=z|_{R=r, U_1 = y}]  \\
=  & \frac{1}{2}\sum_{r\in \{0,1\}^d} \sum_{y \in \{0,1\}^{m}}  \Pr[R=r] |\Pr[Y=y|_{R=r}] - \Pr[U_1=y]| \\
= &\SD(R\circ Y, R\circ U)\\
\leq  &\sum_{i=1}^{t-1}\epsilon_i.\\
\end{align*}

Next we give the upper bound of $ \SD( R\circ U_1\circ Z, R\circ U_1\circ U_2 ) $.

\begin{align*}
& \SD( R\circ U_1\circ Z, R\circ U_1\circ U_2 ) \tag{\stepcounter{equation}\theequation} \\
= &\frac{1}{2} \sum_{r\in \{0,1\}^r} \sum_{u\in \{0,1\}^{m}}\sum_{z\in \{0,1\}^{m_{t}}} |\Pr[R = r]\Pr[U_1 = u]\Pr[Z = z|_{R = r, U_1 = u}]\\
&- \Pr[R = r]\Pr[U_1 = u]\Pr[U_2 = z]| \\
= & \frac{1}{2} \sum_{r\in \{0,1\}^r} \sum_{u\in \{0,1\}^{m}}\sum_{z\in \{0,1\}^{m_{t}}} \Pr[R = r]\Pr[U_1 = u] |\Pr[Z = z|_{R = r, U_1 = u}]- \Pr[U_2 = z]| \\
= & \frac{1}{2}\sum_{u\in \{0,1\}^{m}} \sum_{r\in \{0,1\}^r} \sum_{z\in \{0,1\}^{m_{t}}} \Pr[R = r]\Pr[U_1 = u] |\Pr[Z = z|_{R = r, U_1 = u}]- \Pr[U_2 = z]| \\
= & \frac{1}{2} \sum_{u\in \{0,1\}^{m}} \Pr[U_1 = u]\sum_{r\in \{0,1\}^r}\sum_{z\in \{0,1\}^{m_{t}}}\Pr[R=r] | \Pr[Z=z|_{R = r, U_1 = u}] - \Pr[U_2 = z]|\\
= & \frac{1}{2} \sum_{u\in \{0,1\}^{m}} \Pr[U_1 = u]\sum_{r\in \{0,1\}^r}\sum_{z\in \{0,1\}^{m_{t}}}\Pr[R=r]| \Pr[\Ext_t(X_t, R)=z|_{R = r, Y = u}] - \Pr[U_2 = z]|\\
= & \sum_{u\in \{0,1\}^{m}} \Pr[U_1 = u] \SD(R\circ \Ext_t(X_t,R)|_{Y = u}, R\circ U_2)\\
\leq & \sum_{u\in \{0,1\}^{m}} \Pr[U_1 = u] \epsilon_t \\
= &\epsilon_t. \\
\end{align*}

So $\SD( R\circ Y\circ \Ext_{t}(X_{t}, R), R\circ U_1 \circ U_2  )  \leq \sum_{i=1}^t \epsilon_t$.
This proves the lemma.
\end{proof}

For any circuit $C$, the size of $C$ is denoted as $\size(C)$. The depth of $C$ is denoted as $\depth(C)$. 
\begin{definition}[$\AC^0$]

$\AC^0$ is the complexity class which consists of all families of circuits having constant depth and polynomial size. The gates in those circuits are $\NOT$ gates, $\AND$ gates and $\OR$ gates where $\AND$  gates and $\OR$ gates have unbounded fan-in. 

\end{definition}

\begin{lemma}
\label{xorpolylog}
The following are some well known properties of $\AC^0$ circuits. For any $n \in \mathbb{N}$,
\begin{enumerate}
\item  (\cite{arora2009computational} forklore) any boolean function $f: \{0,1\}^{l = \Theta(\log n)} \rightarrow \{0,1\}$ can be computed by an $\AC^0$ circuit of size $\poly(n)$ and depth $\myd$. In fact, it can be represented by either a CNF or a DNF.


\item (\cite{goldwasser2007verifying}) for every $c\in \mathbb{N}$, every integer $l = \Theta(\log^{c} n)$,  if the function $ f_l:\{0,1\}^l \rightarrow \{0,1\}$ can be computed by circuits of depth  $O(\log l)$ and size $\poly(l)$, then it can be computed by $\AC^0$ (in $n$) circuits of depth $c+1$.
\end{enumerate}
\end{lemma}
%
\begin{proof}

For the first assertion, for an input string $u\in \{0,1\}^l$,
$$ f(u) = \bigvee_{j=0}^{2^l-1} (I_{u = \langle j\rangle} \wedge f(\langle j\rangle)) = \bigwedge_{j=0}^{2^l-1} (I_{u \neq \langle j\rangle}  \vee f(\langle j\rangle) ) .$$
Here $I_e$ is the indicator function such that $I_e = 1$ if $e$ is true and $I_e = 0$ otherwise.
We know that $I_{u = \langle j\rangle} $ can be represented as a boolean formula with only $\AND$ and $\NOT$ gates, checking whether $u = \langle j\rangle$ bit by bit. Similarly $I_{u \neq \langle j\rangle}$ can be represented as a boolean formula with only $\OR$ and $\NOT$ gates by taking the negation of $I_{u = \langle j\rangle}$. So the computation of obtaining $f(u)$ can be represented by a CNF/DNF. Thus it can be realized by a circuit of depth 2 by merging the gates of adjacent levels.

Next we prove the second assertion.

As there exists an $\NC^1$-complete problem which is downward self-reducible \cite{goldwasser2007verifying}, $f_l$ can be reduced to ($\AC^0$ reduction) to $f_{l'}$ where $l' = l^{\alpha}$, for any $\alpha\in (0,1)$. Once we let $ l' = l^{\alpha} = O(\log n)$, $f$ can be computed in $\AC^0$ (in $n$). The circuit depth is $c+1$, as $\AC^0$ reduction has depth $c$ and $f_{l'}$ can be realized by CNF/DNFs.

\end{proof}

\begin{definition}
A boolean function $f: \{0,1\}^l \rightarrow \{0,1\}$ is $\delta$-hard on uniform distributions for circuit size $g$, if for any circuit $C$ with at most $g$ gates ($\size(C) \leq g$), we have $\Pr_{x\leftarrow U}[C(x) = f(x)] < 1-\delta$.

\end{definition}

\begin{definition}[Graphs]
Let $G = (V, E)$ be a graph. Let $A$ be the adjacency matrix of $G$. Let $\lambda(G)$ be the second largest eigenvalue of $A$. We say $G$ is $d$-regular, if the degree of $G$ is $d$. When $G$ is clear in the context, we simply denote $\lambda(G)$ as $\lambda$.
\end{definition}

%% file: LowerBoundforErr.tex
\section{Lower Bound for Error Parameters of $\AC^0$ Extractors}\label{sec:errorbound}

Here we show a lower bound on the error of strong $\AC^0$ seeded extractors. Our conclusion is mainly based on the well known LMN theorem deduced by Fourier analysis, given by  Linial, Mansour, and Nisan \cite{LMN89}. 

Let the Fourier expansion of a function $f:\{-1,1\}^n \rightarrow \{-1,1\}$ be $f(x) = \sum_{S\subseteq [n]} \hat{f}_{S} \chi_S(x)$, where $\chi_S(x) = \prod_{i=1}^n x_i$. For any $f,g:\{-1,1\}^n\rightarrow \{-1,1\}$, $\langle f, g\rangle = \frac{1}{2^n} \sum_{x\in\{-1,1\}^n} f(x)g(x)$. 

\begin{theorem}[LMN Theorem \cite{LMN89} \cite{o2014analysis}] 
\label{LMNthm}
Let $f: \{-1, 1\}^n \rightarrow \{-1 , 1\}$ be computable by $\AC^0$ circuits of size $s > 1$ and depth $\mathtt{dth}$. Let $ \epsilon \in (0 , 1/2]$. There exists $t = O(\log(s/\epsilon ))^{\mathtt{dth}-1} \cdot \log(1/\epsilon) $ s.t.
$$\sum_{S\subseteq [n], |S| > t}\hat{f}_{S}^2 \leq \epsilon.$$ 

\end{theorem}

Our first lower bound is as the follows.
%

\begin{theorem}
\label{negerr}

 If $\Ext:\{0,1\}^n \times \{0,1\}^{d} \rightarrow \{0,1\}^m$ is a  strong $(k= n-1, \epsilon)$-extractor that can be computed by $\AC^0$ circuits of depth $\mathtt{dth}$ and size $s$, then $\epsilon =2^{  - (O(\log s))^{\mathtt{dth}-1}   \log (n+d)}$.
\end{theorem}

\begin{proof}

Without loss of generality, let $m = 1$. 
Let's transform the function space of $\Ext$ to $ \{-1,1\}^{n+d} \rightarrow \{-1,1\}$, achieving function $f$. 
Let $\epsilon_0 = 1/2$.
By Theorem \ref{LMNthm}, there exists $ t = O(\log(s/\epsilon_0 ))^{\mathtt{dth}-1} \cdot \log(1/\epsilon_0) = O(\log s)^{\mathtt{dth}-1}$ s.t.
$$\sum_{S\subseteq [n+d], |S| \leq t}\hat{f}_{S}^2 > 
1-\epsilon_0 = 1/2. $$

Fix an $S = S_1 \cup \{i+ n \mid i \in S_2 \}$ with $|S| \leq t$, where $S_1 \subseteq [n], S_2 \subseteq \{1, 2, \ldots, d\}$. 
We know that 
$$\hat{f}_S = \langle f, \chi_S\rangle  = 1 - 2\Pr_{u }[f(u) \neq \chi_S(u)]$$
where $u$ is uniformly drawn from $\{-1,1\}^{n+d} $.

For $a \in \{-1,1\}$, let $X_a$ be the uniform distribution over $\{-1,1\}^n$ conditioned on $\prod_{i\in S_1} X_i = a$. Also for $b\in \{-1,1\}$, let $R_b$ be the uniform distribution over $\{-1,1\}^d$ conditioned on $\prod_{i\in S_2}R_i = b$. So $\chi_S(x\circ r) = ab$ for $x\in \supp(X_a), r\in \supp(R_b)$. For special situations, saying $S_1 = \emptyset$ (or $S_2 = \emptyset$ ), let $X_a$ (or $R_b$) be uniform.

As $\Ext$ is a strong $(k, \epsilon)$-extractor, $R_b$ only blows up the error by 2.
Also note that by definition,  $ X_{a}$ has entropy $n-1$. So
$$ \dist(f(X_a \circ R_b), U)\leq 2\epsilon,$$
where $U$ is uniform over $\{-1, 1\}$.

So $$\forall a,b \in \{-1,1\}, |\Pr[f(X_a \circ R_b) \neq ab] - 1/2| \leq 2\epsilon.$$


Thus 
\begin{equation}
\begin{split}
|\Pr_{u  }[f(u) \neq \chi_S(u)] - 1/2| &=   |\sum_{a\in\{-1,1\}}\sum_{b\in \{-1,1\}}\frac{1}{4}  (\Pr[f(X_a \circ R_b) \neq ab]-1/2)| \\
&\leq  \sum_{a\in\{-1,1\}}\sum_{b\in \{-1,1\}}\frac{1}{4}  |\Pr[f(X_a \circ R_b) \neq ab]-1/2|\\
& \leq 2\epsilon.\\
\end{split}
\end{equation}

Hence 
$$ |\hat{f}_S| = |1-2\Pr_{u   }[f(u) \neq \chi_S(u)] | = 2|\Pr_{u }[f(u) \neq \chi_S(u)] - 1/2| \leq 4\epsilon.$$ 
As a result,
$$1/2 \leq  \sum_{S\subseteq [n+d], |S| \leq t}\hat{f}_{S}^2 \leq \sum_{i=0}^t {n+d \choose i } (4\epsilon)^2.$$

So
$$\epsilon \geq \sqrt{\frac{1}{32\sum_{i=0}^t {n+d \choose i}}} .$$

As $\sum_{i=0}^t {n+d \choose i} \leq (\frac{e(n+d)}{t})^t = 2^{O(t\log (n+d))} =2^{   O(\log s)^{\mathtt{dth}-1}   \log (n+d)}$, $ \epsilon =2^{  - O(\log s)^{\mathtt{dth}-1}   \log (n+d)}$.

\end{proof}

We also consider extractors for bit-fixing sources and give the following negative result on the error. 

\begin{theorem}
\label{DExterbound}

There is a constant $c > 1$ such that if $\Ext:\{0,1\}^n \times \{0,1\}^{d } \rightarrow \{0,1\}^m$ is a strong $(k, \epsilon)$-extractor for oblivious bit-fixing sources with $k  = n - (c\log s )^{\mathtt{dth}-1}$, that can be computed by $\AC^0$ circuits of depth $\mathtt{dth}$ and size $s$, then $\epsilon =2^{  - (O(\log s))^{\mathtt{dth}-1}   \log (n+d)}$.

\end{theorem}

The proof is slightly different from that of theorem \ref{negerr}. 
\begin{proof}
Let $m = 1$ and also transform the function space of $\Ext$ to $ \{-1,1\}^{n+d} \rightarrow \{-1,1\}$, achieving function $f$. 
Let $\epsilon_0 = 1/2$.
By Theorem \ref{LMNthm}, there exists $ t = O(\log(s/\epsilon_0 ))^{\mathtt{dth}-1} \cdot \log(1/\epsilon_0) = O(\log s)^{\mathtt{dth}-1}$ s.t.
$$\sum_{S\subseteq [n+d], |S| \leq t}\hat{f}_{S}^2 > 
1-\epsilon_0 = 1/2. $$

Fix  an $S = S_1 \cup \{i + n \mid i \in S_2 \}$,  with $|S| \leq t$, where $S_1 \subseteq [n], S_2 \subseteq \{1, 2, \ldots, d\}$. 
We know that 
$$\hat{f}_S = \langle f, \chi_S\rangle = 1 - 2\Pr_{u  }[f(u) \neq \chi_S(u)]$$
where $u$ is uniformly drawn from $\{-1,1\}^{n + d } $.

For $a \in \{-1,1\}^{|S_1|}$, let $X_a$ be the uniform distribution over $\{-1,1\}^n$ conditioned on   $X_{S_1} = a$.  
For $b \in \{-1,1\}^{|S_2|}$, let $R_b$ be the uniform distribution over $\{-1,1\}^d$ conditioned on  $R_{S_2} = b$. 
So $\chi_S(x\circ r) = \prod_{i\in [|S_1|]} a_i \prod_{j\in [|S_2|]}  b_j$ for $x\in \supp(X_a), r\in \supp(R_b)$. For special situations, saying $S_1 = \emptyset$ (or $S_2 = \emptyset$ ), let $X_a$ (or $R_b$) be uniform.

As $\Ext$ is a strong $(k, \epsilon)$-extractor, $R_b$ only blows up the error by at most $2^{|S|}$.
Also note that $ X_{a}$ has entropy $n-|S_1| \geq n-t = n-O(\log s)^{\mathtt{dth}-1} \geq k = n - (c\log  s )^{\mathtt{dth}-1}$ by choosing $c$ large enough. So
$$ \dist(f(X_a \circ R_b), U)\leq 2^{|S|}\epsilon,$$
where $U$ is uniform over $\{-1, 1\}$.

So $$\forall a \in \{-1,1\}^{|S_1|}, \forall b\in \{-1, 1\}^{|S_2|}, |\Pr[f(X_a \circ R_b) \neq  \prod_{i\in [|S_1|]} a_i \prod_{j\in [|S_2|]}  b_j] - 1/2| \leq 2^{|S|}\epsilon.$$

Thus 
\begin{equation}
\begin{split}
|\Pr_{u  }[f(u) \neq \chi_S(u)] - 1/2| &=   |\sum_{a\in\{-1,1\}^{|S_1|}}\sum_{b\in \{-1,1\}^{|S_2|}}\frac{1}{2^{|S|}}  (\Pr[f(X_a \circ R_b) \neq \prod_{i\in [|S_1|]} a_i \prod_{j\in [|S_2|]}  b_j]-1/2)| \\
&\leq  \sum_{a\in\{-1,1\}^{|S_1|}}\sum_{b\in \{-1,1\}^{|S_2|}}\frac{1}{2^{|S|}}  |\Pr[f(X_a \circ R_b) \neq \prod_{i\in [|S_1|]} a_i \prod_{j\in [|S_2|]}  b_j]-1/2|\\
& \leq 2^{|S|}\epsilon.\\
\end{split}
\end{equation}

Hence 
$$ |\hat{f}_S| = |1-2\Pr_{u  }[f(u) \neq \chi_S(u)] | = 2|\Pr_{u }[f(u) \neq \chi_S(u)] - 1/2| \leq 2^{|S|+1}\epsilon.$$ 
As a result,
$$1/2 \leq \sum_{S\subseteq [n +d ], |S| \leq t}\hat{f}_{S}^2 \leq \sum_{i=0}^t {n + d \choose i } (2^{|S|+1}\epsilon)^2 \leq \sum_{i=0}^t {n + d \choose i } (2^{t+1}\epsilon)^2.$$

So
$$\epsilon \geq 2^{-(t+1)}\sqrt{\frac{1}{2\sum_{i=0}^t {n + d \choose i}}} .$$

As $\sum_{i=0}^t {n+d \choose i} \leq (\frac{e(n+d)}{t})^t = 2^{O(t\log (n+d))} =2^{   O(\log s)^{\mathtt{dth}-1}   \log (n+d)}$, $ \epsilon =2^{  - O(\log s)^{\mathtt{dth}-1}   \log (n+d)}$.

\end{proof}

%% file: basicAC0ext.tex
\section{The Basic Construction of Extractors in $\AC^0$}\label{sec:basicext}

Our basic construction is based on the general idea of I-W generator \cite{iw:bppequalsp}. In \cite{Trevisan01}, Trevisan showed that I-W generator is an extractor if we regard the string $x$ drawn from the input $(n,k)$-source $X$ as the truth table of a function $f_x$ s.t. $f_x(\langle i\rangle), i\in [n]$ outputs the $i$th bit of $x$.

The construction of I-W generator involves a process of hardness amplifications from a worst-case hard function to an average-case hard function. There are mainly 3 amplification steps. Viola \cite{viola2005complexity} summarizes these results in details, and we review them again. The first step is established by Babai et al. \cite{bfnw}, which is an amplification from worst-case hardness to mildly average-case hardness.

\begin{lemma}[\cite{bfnw}]
\label{amp1}

If there is a boolean function $f:\{0,1\}^l\rightarrow \{0,1\}$ which is $0$-hard for circuit size $g= 2^{\Omega(l)}$ then there is a boolean function $f':\{0,1\}^{\Theta(l)} \rightarrow \{0,1\} $ that is $1/\poly(l)$-hard for circuit size $g' = 2^{\Omega(l)}$.

\end{lemma}

The second step is an amplification from mildly average-case hardness to constant average-case hardness, established by Impagliazzo \cite{impagliazzo1995hard}.

\begin{lemma}[\cite{impagliazzo1995hard}]
\label{amp2}

\begin{enumerate}
\item If there is a boolean function $f:\{0,1\}^l\rightarrow \{0,1\}$ that is $\delta$-hard for circuit size $g$ where $\delta < 1/(16l)$, then there is a boolean function $f':\{0,1\}^{3l} \rightarrow \{0,1\}$ that is $0.05\delta l$-hard for circuit size $g' = \delta^{O(1)}l^{-O(1)}g$.
$$f'(s,r) = \langle s, f(a_1)\circ f(a_2)\circ \cdots \circ f(a_l)\rangle$$

Here $|s| = l$, $|r| = 2l$ and $|a_i| = l, \forall i\in [l]$. Regarding $r$ as a uniform random string, $a_1,\ldots, a_l$ are generated as pairwise independent random strings from the seed $r$. 

\item If there is a boolean function $f:\{0,1\}^l\rightarrow \{0,1\}$ that is $\delta$-hard for circuit size $g$ where $\delta < 1$ is a constant, then there is a boolean function $f':\{0,1\}^{3l} \rightarrow \{0,1\}$  that is $1/2 - O(l^{-2/3})$-hard for circuit size $g' = l^{-O(1)}g$, where
$$f'(s,r) = \langle s, f(a_1)\circ f(a_2)\circ \cdots \circ f(a_l)\rangle.$$
Here $|s| = l$, $|r| = 2l$ and $|a_i| = l, \forall i\in [l]$. Regarding $r$ as a uniform random string, $a_1,\ldots, a_l$ are generated as pairwise independent random strings from the seed $r$. 
\end{enumerate}

\end{lemma}

The first part of this lemma can be applied for a constant number of times to get a function having constant average-case hardness. After that the second part is usually
applied for only once to get a function with constant average-case hardness  such that the constant is large enough (at least $1/3$).

The third step is an amplification from constant average-case hardness to even stronger average-case hardness, developed by Impagliazzo and Widgerson \cite{iw:bppequalsp}.
Their construction uses the following Nisan-Widgerson Generator \cite{NisanW94} which is widely used in hardness amplification.
\begin{definition}[$(n, m, k, l)$-design and Nisan-Widgerson Generator \cite{NisanW94}]

A system of sets $S_1, S_2, \ldots, S_m \subseteq [n]$ is an $(n, m, k, l)$-design, if $\forall i \in [m], |S_i| = l$ and $\forall i,j\in [m], i\neq j,  |S_i \cap S_j| \leq k$.

Let $\mathcal{S} = \{S_1, S_2, \ldots, S_m\}$ be an $(n,m,k,l)$ design and $f:\{0,1\}^l \rightarrow \{0,1\}$ be a boolean function. The Nisan-Widgerson Generator is defined as $\NW_{f,\mathcal{S}}(u) = f(u|_{S_1})\circ f(u|_{S_2})\circ \cdots \circ f(u|_{S_m})$. Here $u|_{S_i} = u_{i_1}\circ u_{i_2}\circ \cdots \circ u_{i_m}$ assuming $ S_i = \{i_1, \ldots, i_m\}$.

\end{definition}

Nisan and Widgeson \cite{NisanW94} showed that the $(n, m, k,l)$-design can be constructed efficiently.

\begin{lemma}[Implicit in \cite{NisanW94}]
For any $\alpha \in (0,1)$, for any large enough $l\in\mathbb{N}$, for any $m < \exp\{\frac{\alpha l}{4}\}$,
there exists an $(n,m, \alpha l,l)$-design where $n = \lfloor \frac{10l}{\alpha}\rfloor $. This design can be computed in time polynomial of $2^{n}$.
\end{lemma}

As we need the parameters to be concrete (while in \cite{NisanW94} they use big-$O$ notations), we prove it again. 

\begin{proof}

Our algorithm will construct these $S_i$s one by one. For $S_1$, we can choose an arbitrary subset of $[n]$ of size $\alpha l$.

First of all, $S_1$ can constructed by choosing $l$ elements from $[n]$.

Assume we have constructed $ S_1,\ldots, S_{i-1} $, now we construct $S_{i}$. We first prove that $S_i$ exists. Consider a random subset of size $l$ from $[n]$. Let $H_{i,j} = |S_{i}\cap S_{j}|$. We know that $\mathbf{E}H_{i,j} = l^2/n$.
As $n = \lfloor  \frac{10l}{\alpha} \rfloor \in [\frac{10l}{\alpha}-1, \frac{10l}{\alpha}]$, $\mathbf{E}H_{i,j} \in [  \frac{\alpha l}{10}, \frac{\alpha l}{10}+1]$. 

So $\Pr[H_{i,j} \geq \alpha l ] \leq  \Pr[H_{i,j}\geq (1+9)(\mathbf{E}H_{i,j}-1) ]$

By the Chernoff bound, 
$$    \Pr[H_{i,j}\geq 10(\mathbf{E}H_{i,j}-1) ] \leq \Pr[H_{i,j}\geq 9\mathbf{E}H_{i,j}] \leq \exp\{-\frac{8\mathbf{E}H_{i,j}}{3} \} \leq  \exp\{-\frac{4}{15}\alpha l \}\leq  \exp\{-\frac{\alpha l}{4} \} $$

By the union bound,
$$\Pr[\forall j=1,\ldots, i-1, H_{i,j} \leq \alpha l] \geq 1 - m\exp\{-\frac{\alpha l}{4}\} > 0 .$$

This proves that there exists a proper $S_i$.
As there are $n$ bits totally, we can find it in time polynomial of $2^{n}$.
\end{proof}

The following is the third step of hardness amplification.
\begin{lemma}[Implicit in \cite{iw:bppequalsp}]
\label{amp3}

%
%
%

For any  $\gamma \in (0, 1/30)$,
if there is a boolean function $f:\{0,1\}^l\rightarrow \{0,1\}$ that is $1/3$-hard for circuit size $g = 2^{\gamma l}$, then there is a boolean function $f':\{0,1\}^{l' = \Theta(l)} \rightarrow \{0,1\}$ that is $(1/2-\epsilon)$-hard for circuit size $g' =  \Theta( g^{1/4} \epsilon^2 l^{-2} )$  
where $\epsilon \geq (500l)^{1/3} g^{-1/12} $.
$$f'(a,s,v_1,w) = \langle s, f(a|_{S_1} \oplus v_1) \circ f(a|_{S_2} \oplus v_2) \circ \cdots f(a|_{S_l} \oplus v_l) \rangle$$

Here $(S_1,\ldots, S_l)$ is an $(|a|,l, \gamma l/4,l)$-design where $|a| = \lfloor \frac{40l}{\gamma}\rfloor$. The vectors $v_1,\ldots,v_l$ are obtained by a random walk on an expander graph, starting at $v_1$ and walking according to $w$ where $|v_1| = l, |w| = \Theta(l)$. The length of $s$ is $l$. So $ l' = |a| + |s| + |v_1|+|w| = \Theta(l)$.

\end{lemma}

The proof of Lemma \ref{amp3} is in the Appendix.

The construction of the Impagliazzo Widgerson Generator \cite{iw:bppequalsp} is as follows. Given the input $x \leftarrow X$, let $f:\{0,1\}^{\log n} \rightarrow \{0,1\}$ be such that $f(\langle a\rangle ) = x_{a}, \forall a\in [n]$. Then we run the 3 amplification steps, Lemma \ref{amp1}, Lemma \ref{amp2} (part1 for a constant number of times, part 2 for once) and Lemma \ref{amp3} sequentially to get function $f'$ from $f$. The generator $\IW(x,u) = \NW_{f', \mathcal{S}}(u)$. As pointed out by Trevisan \cite{Trevisan01}, the function $\IW$ is a $(k, \epsilon)$-extractor. Let's call it the $\IW$-Extractor. It is implicit in \cite{Trevisan01} that the output length of the $\IW$-Extractor is $k^\alpha$ and the statistical distance of $\IW(X,U)$ from uniform distributions is $\epsilon = 1/k^{\beta}$ for some $0<\alpha, \beta<1$. This can be verified by a detailed analysis of the $\IW$-Extractor.

However, this construction is not in $\AC^0$ because the first amplification step is not in $\AC^0$.

Our basic construction is an adjustment of the $\IW$-Extractor. 

\begin{construction}
\label{constr1}
For any $c_2 \in \mathbb{N}^+$ such that $c_2 \geq 2$ and any
$k  = \Theta(n/\log^{c_2-2} n) $,
let $X$ be an $(n, k)$-source .
We construct a strong $(k,2\epsilon)$ extractor $\Ext_0: \{0,1\}^n \times \{0,1\}^d \rightarrow \{ 0,1 \}^m$ where $\epsilon = 1/n^{\beta}$, $\beta=1/600$, $d = O(\log n)$, $m = k^{\Theta(1)}$.
Let $U$ be the uniform distribution of length $ d  $.
\begin{enumerate}

\item Draw $x$ from $X$ and $u$ from $U$. Let $f_1:\{0,1\}^{l_1} \rightarrow \{0,1\}$ be a boolean function such that $\forall i \in [2^{l_1}],$ $f_1(\langle i\rangle) = x_i$ where $l_1 = \log n$.

\item Run amplification step of Lemma \ref{amp2} part 1 for $c_2$ times and run amplification step of Lemma \ref{amp2} part 2 once to get function $f_2:\{0,1\}^{l_2} \rightarrow \{0,1\}$ from $f_1$ where $l_2 = 3^{c_2+1}l_1 = \Theta(\log n)$.

\item Run amplification step Lemma \ref{amp3} to get function $f_3:\{0,1\}^{l_3} \rightarrow \{0,1\}$ from $f_2$ where $l_3 = \Theta(\log n)$. 

\item  Construct function $\Ext_0$ such that $\Ext_0(x, u) = \NW_{f_3, \mathcal{S}}(u)$. 

\end{enumerate}
Here $\mathcal{S} = \{S_1, S_2, \ldots, S_m\}$ is a $(d,m ,\theta l_3, l_3)$-design with $\theta =  l_1/(900 l_3)$, $d = \lfloor 10l_3/\theta \rfloor$, $m = \lfloor 2^{\frac{\theta l_3}{4}} \rfloor = \lfloor n^{\frac{1}{3600}} \rfloor$.
\end{construction}

\begin{lemma}
\label{badset}
In Construction \ref{constr1}, $\Ext_0$ is a strong $(k,2\epsilon)$ extractor.

\end{lemma}

The proof follows from the ``Bad Set'' argument given by Trevisan 
\cite{Trevisan01}. In Trevisan 
\cite{Trevisan01} the argument is not explicit for strong extractors. Here our argument is explicit for proving that our construction gives a strong extractor.

\begin{proof}

We will prove that for every $(n,k)$-source $X$ and for every $A:\{0,1\}^{d+m} \rightarrow \{0,1\}$ the following holds.
$$ |\Pr[A(U_s\circ \Ext_0(X, U_s)) = 1] - \Pr[A(U) = 1]| \leq 2\epsilon $$
Here $U_s$ is the uniform distribution over $\{0,1\}^d$ and $U$ is the uniform distribution over $\{0,1\}^{d+m}$.

For every flat $(n,k)$-source $X$, and for every (fixed) function $A$, 
let's focus on a set $B \subseteq \{0,1\}^n$ such that $\forall x \in \supp(X)$, if $x\in B$, then
$$|\Pr[A(U_s\circ \Ext_0(x,U_s)) = 1] - \Pr[A(U) = 1]|> \epsilon.$$

According to Nisan and Widgerson \cite{NisanW94}, we have the following lemma.

\begin{lemma}[Implicit in \cite{NisanW94} \cite{Trevisan01}]
\label{designCsize}
If there exists an $A$-gate such that
$$|\Pr[A(U_s\circ \Ext_0(x,U_s)) = 1] - \Pr[A(U) = 1]|> \epsilon,$$
then there is a circuit $C_3$ of size $O(2^{\theta l_3}m)$, using $A$-gates, that can compute $f_3$ correctly for $1/2+\epsilon/m$ fraction of inputs.

Here $A$-gate is a special gate that can compute the function $A$.
\end{lemma}

%
%
%
%

By Lemma \ref{designCsize}, there is a circuit $C_3$ of size $O(m2^{\theta l_3}  ) = O(2^{\frac{5\theta l_3}{4}}) = O(n^{1/720})$, using $A$-gates, that can compute $f_3$ correctly for $1/2+\epsilon/m \geq 1/2 + 1/n^{1/400}$ fraction of inputs.

By  Lemma \ref{amp3}, there is a circuit $C_2$, with $A$-gates, of 
size at most $\Theta(n^{\frac{1}{30}})$ 
which can compute $f_2$ correctly for at least $2/3$ fraction of inputs. 

According to Lemma \ref{amp2} and our settings, there is a circuit $C_1$, with $A$-gates, of size 
$n^{\frac{1}{30}}\poly\log n$
which can compute $f_1$ correctly for at least $1-1/(c_1\log^{c_2} n)$ fraction of inputs for some constant $c_1>0$. 

Next we give an upper bound on the size of $B$. $\forall x\in B$, assume we have a circuit of size $S = n^{1/30}\poly (\log n)$, using $A$-gates, that can compute at least $1-1/(c_1\log^{c_2} n)$ fraction of bits of $x$. The total number of circuits, with $A$-gates, of size  $S$ is at most $2^{\Theta(mS\log S)} = 2^{n^{1/15}\poly(\log n)}$, as $A$ is fixed and has fan-in $m+d=O(m)$.
Each one of them corresponds to at most $\sum_{i=0}^{n/(c_1\log^{c_2} n)} {n \choose i} \leq (e\cdot c_1\log^{c_2} n)^{n/(c_1\log^{c_2} n)} = 2^{O(n/\log^{c_2-1} n)}$ number of $x$.
So 
$$|B| \leq 2^{n^{1/15}\poly(\log n)} 2^{O(n/\log^{c_2-1} n)} =  2^{O(n/(\log^{c_2-1} n)}.$$ 
As $X$ is an $(n,k)$-source with $ k =\Theta( n/\log^{c_2-2} n )$,
$$\Pr[X\in B] \leq |B|\cdot 2^{-k}\leq \epsilon.$$
Then we know,
\begin{equation}
\begin{split}
&|\Pr[A(U_s\circ \Ext(X, U_s)) = 1] - \Pr[A(U) = 1]| \\
= &\sum_{x\in B} \Pr[X=x] |\Pr[A(U_s\circ \Ext(x, U_s)) = 1] - \Pr[A(U) = 1]| \\
&+ \sum_{x\notin B} \Pr[X=x] |\Pr[A(U_s \circ \Ext(x, U_s)) = 1] - \Pr[A(U) = 1]| \\
\leq & 2\epsilon.
\end{split}
\end{equation}

\end{proof}

\begin{lemma}
\label{seedlen}
The seed length of construction \ref{constr1} is $O(\log n)$. 
\end{lemma}
\begin{proof}

We know that $l_1 = \log n, l_2 = 3^{c_2 + 1 }l_1 = \Theta(\log n), l_3 = \Theta(\log n)$. Also $\mathcal{S}$ is a $(\lceil 10l_3/c \rceil = \Theta(l_3),m,cl_3, l_3)$-design. So $d = \lfloor 10l_3/c \rfloor = \Theta(l_3) = \Theta(\log n)$.

\end{proof}

\begin{lemma}
\label{locality}
The function $\Ext_0$ in Construction \ref{constr1} is in $\AC^0$. The circuit depth is $c_2+5$. The locality is $\Theta(\log^{c_2+2} n) = \poly (\log n)$.
\end{lemma}

\begin{proof}
First we prove that the locality is  $\Theta(\log^{c_2+2} n)$.

By the construction of $f_1$, we know $f_1(\langle i\rangle)$ is equal to the $i$th bit of $x$.

Fix the seed $u$.
According to Lemma \ref{amp2} part 1, if we apply the amplification once to get $f'$ from $f$, then $f'(s, r)$ depends on  $f(w_1), f(w_2), \cdots, f(w_l)$, as 
$$f'(s,r) = \langle s, f(w_1)\circ f(w_2)\circ \cdots \circ f(w_l)\rangle.$$
Here $l = O(\log n)$ is equal to the input length of $f$.

The construction in Lemma \ref{amp2} part 2 is the same as that of Lemma \ref{amp2} part 1. As a result, if apply Lemma \ref{amp2} part 1 for $c_2$ times and Lemma \ref{amp2} part 2 for 1 time to get $f_2$ from $f_1$, the output of $f_2$ depends on $\Theta(\log^{c_2+1} n)$ bits of the input $x$.

According to Lemma \ref{amp3}, the output of $f_3$ depends on $f_2(a|_{S_1} \oplus v_1), f_2(a|_{S_2} \oplus v_2), \cdots, f_2(a|_{S_l} \oplus v_{l_2})$, as
$$f_3(a,s,v_1,w) = \langle s, f_2(a|_{S_1} \oplus v_1) \circ f_2(a|_{S_2} \oplus v_2) \circ \cdots f_2(a|_{S_l} \oplus v_{l_2}) \rangle$$ 
So the output of $f_3$ depends on $O(\log^{c_2 + 2} n)$ bits of the $x$. 

So the overall locality is $O(\log^{c_2 + 2} n) = \poly \log n$.

Next we prove that the construction is in $\AC^0$.

The input of $\Ext_0$ has two parts, $x$ and $u$. Combining all the hardness amplification 
steps and the NW generator, we can see that essentially $u$ is used for two 
purposes: to select some $t =  \Theta(\log^{c_2 + 2}(n))$ bits (denote it as $x'$) from $x$ (i.e., provide $t$ indices $u'_1, \ldots,u'_t$
in $[n]$), and to provide a vector $s'$ of length $t$,   finally taking the inner product of $x'$ and the vector $s'$. Here although for each amplification step we do an inner product operation, the overall procedure can be realized by doing only one inner product operation. 

Since $u$ has $O(\log n)$ bits, $s'$ can be 
computed from $u$ by using a circuit of depth 2, according to Lemma \ref{xorpolylog} part 1.

Next we show that selecting 
$x'$ from $x$ using the indices can be computed by CNF/DNFs, of polynomial size, with 
inputs being $x$ and the indices. The indices, $u'_i, i\in [t]$, are decided by $u$. Let's assume $\forall i\in [t], u'_i = h_i(u)$ for some deterministic functions $h_i, i\in [t]$. As $|u| = O(\log n)$, the indices can be computed by CNF/DNFs of polynomial size. Also $\forall i\in [t], f(u'_i)$ can be represented by a CNF/DNF when $u'_i$ is given.
This is because 
$$ f(u'_i) = \bigvee_{j=0}^{|x|} (I_{u'_i = j} \wedge x_j) = \bigwedge_{j=0}^{|x|} (I_{u'_i \neq j} \vee x_j ) .$$
Here $I_e$ is the indicator function such that $I_e = 1$ if $e$ is true and $I_e = 0$ otherwise. We know that $I_{u'_i = j} $ can be represented by a boolean formula with only $\AND$ and $\NOT$ gates, checking whether $u'_i = j$ bit by bit. Similarly $I_{u'_i \neq j} $ can be represented by a boolean formula with only $\OR$ and $\NOT$ gates, taking the negation of $I_{u'_i = j}$. 
As a result, this step can be computed by a circuit of depth 2.

So the computation of obtaining $x'$ can be realized by a circuit of depth 3 by merging the gates between adjacent depths.

Finally we can take the inner product of two vectors $x'$ and $s'$ of length $t =  \Theta(\log^{c_2 + 2}(n))$. By Lemma \ref{xorpolylog} part 2, we know that this computation can be represented by a poly-size circuit of depth $c_2+3$.

The two parts of computation can be merged together to be a circuit of depth $c_2 + 5$, as we can merge the last depth of the circuit obtaining  $x'$ and the first depth of the circuit computing the inner product. The size of the circuit is polynomial in $n$ as both obtaining $x'$ and the inner product operation can be realized by poly-size circuits.

\end{proof}

By Construction \ref{constr1}, Lemma \ref{badset}, Lemma \ref{seedlen}, Lemma \ref{locality}, we have the following theorem.

\begin{theorem}
\label{basicthm}
%

For any $c\in \mathbb{N}$, any $k = \Theta( n/\log^{c} n )$, there exists
an explicit strong $(k, \epsilon)$-extractor $\Ext:\{0,1\}^{n}\times \{0,1\}^d \rightarrow \{0,1\}^m$ in $\AC^0$ of depth $c+7$, where  $\epsilon  = n^{-1/600}$, $d = O(\log n)$,  $m = \lfloor n^{\frac{1}{3600}} \rfloor$ and the locality is $\Theta(\log^{c+4} n) = \poly\log n$. 
\end{theorem}

We call this extractor the Basic-$\AC^0$-Extractor.

%% file: ErrorReduction.tex
\section{Error Reduction}\label{sec:error}
 
By Theorem \ref{basicthm}, for any $k = \frac{n}{\poly(\log n)}$, we have a $(k, \epsilon)$-extractor in $\AC^0$, with $ \epsilon = 1/n^\beta$ where $\beta$ is a constant. In this section, we  do error reduction to give an explicit $(k, \epsilon)$-extractor in $\AC^0$ such that $\epsilon$ can be quasi-polynomially small.

We use two major techniques. First is the sample-then-extract method.

\subsection{Sample-Then-Extract}

We first analyze the sampling method which is well studied by Zuckerman \cite{Zsamp}, Vadhan \cite{Vadhan04}, Goldreich et al. \cite{goldreich2015randomness} and Healy \cite{healy2008randomness}.

\begin{definition}[\cite{Vadhan04}]
A $(\mu_1, \mu_2, \gamma)$-averaging sampler is
a function $\Samp:\{0,1\}^r \rightarrow [n]^t$ such that $ \forall f:[n] \rightarrow [0,1]$, if  $\mathbf{E}_{i\in [n]}[f(i)] \geq \mu_1$, then
$$ \Pr_{I \leftarrow \Samp(U_r)}[\frac{1}{t}\sum_{i\in I} f(i) < \mu_2]\leq \gamma.$$

The $t$ samples generated by the sampler must be distinct.

\end{definition}

Vadhan \cite{Vadhan04} gives the following lemma on how to use samplers on weak sources.

\begin{lemma}[Sample a Source \cite{Vadhan04}]
\label{sampresult}
Let $0<3\tau \leq \delta \leq 1 $. If $\Samp:\{0,1\}^r \rightarrow [n]^t$ is a $(\mu_1, \mu_2, \gamma)$-averaging sampler for $\mu_1 = (\delta - 2\tau)/\log(1/\tau)$ and $\mu_2 = (\delta - 3\tau)/\log(1/\tau)$, then for every $(n, \delta n)$-source $X$, we have $\SD(U\circ X_{\Samp(U_r)} , U\circ W) \leq \gamma + 2^{-\Omega(\tau n)}$. Here $U$ is the uniform distribution over $\{0,1\}^r$. For every $ a $ in $\{0,1\}^r$, the random variable $W|_{U=a}$ is a $(t, (\delta-3\tau)t)$-source.
\end{lemma}

We mainly use the following samplers given by Healy \cite{healy2008randomness}.

\begin{theorem}[\cite{healy2008randomness} Theorem 3]
\label{NC1sampler}
For any $n \in \mathbb{N} $, any $ \mu \in (0,1], \eps > \mu$,
there exists an $(\mu, \mu - \epsilon, \gamma)$-averaging sampler $\Samp: \{0, 1\}^r \rightarrow [n]^t$ with seed length
$r =  \log n + O(\log(1/\gamma)/\eps^2) $ and $t = O(\log(1/\gamma)/\eps^2)$ which can be computed by $\NC^1$ circuits of size $\poly(n, 1/\eps, \log(1/\gamma))$.

\end{theorem}

\begin{remark}

If $\gamma = \Theta(2^{- \log^c n}), \eps = \Theta(1/\log^a n)$,  then the sampler can be computed by $\AC^0$ circuitsof depth $a+c+ 1$  by Lemma \ref{xorpolylog}.

For sample complexity (parameter $t$), by Lemma 8.3 of \cite{Vadhan04}, we can modify the sampler and get a new sampler with the number of samples to be at least $t$ while having the same seed length.

\end{remark}

After sampling, we use leftover hash lemma to do extraction.

\begin{lemma}[Leftover Hash Lemma \cite{iz}]
\label{leftoverh}
Let $X$ be an $(n',k=\delta n')$-source.
For any $\Delta > 0$, let $H$ be a universal family of hash functions mapping $n'$ bits to $m = k - 2\Delta$ bits. The distribution $U\circ \Ext(X, U)$ is at distance at most $1/2^{\Delta}$ to uniform distribution where the function $\Ext: \{0,1\}^{n'} \times \{0,1\}^d \rightarrow \{0,1\}^m $ chooses the $U$'th hash function $h_U$ in $H$ and outputs $h_U(X)$.

We use the following universal hash function family $H = \{ h_u, u\in \{0,1\}^{n'} \}$. For every $u$, the hash function
$h_u(x)$  equals to the last $m$ bits of $u \cdot x$ where $u \cdot x$ is computed in $\mathbb{F}_{2^{n'}}$. 

Specifically, for any constant $a \in \mathbb{N}^+$, for any $n' =\Theta(\log^a n)$ then $\Ext $ can be computed by an  $\AC^0$ circuit of depth $a+1$.

\end{lemma}

\begin{proof}

The proof in \cite{iz} has already shown that the universal hash function is a strong extractor.
We only need to show that the hash functions can be computed in $\AC^0$.

Given a seed $u$, we need to compute $ u \cdot x $ which is a multiplication in $\mathbb{F}_{2^{n'}}$. We claim that this can be done in $\AC^0$. 
Note that since the multiplication is in $ \mathbb{F}_{2^{n'}}$, it is also a bi-linear function when regarding the two inputs as two $n'$-bit strings. Thus, each output bit is essentially the inner product over some input bits.

This shows that each output bit of $p\cdot q$ is an inner product of two vectors of $n'$ dimension. As $n' =\Theta(\log^a n)$, by Lemma \ref{xorpolylog}, this can be done in $\AC^0$ of depth $a+1$ and size $\poly(n)$. All the output bits can be computed in parallel. So $u \cdot x$ can be computed in $\AC^0$ of depth $a+1$ and size $\poly(n)$.

\end{proof}

\begin{theorem}
\label{improveGAC0ext}
For  any constant $\delta \in (0,1], a\in \mathbb{N^+}$ and any $\epsilon = 1/2^{- \Theta(\log^a n)}$,
there exists an explicit construction of a $(k=\delta n,   \epsilon  )$-extractor $\Ext:\{0,1\}^{n} \times \{0,1\}^{d} \rightarrow \{0,1\}^{m}$ in $\AC^0$ of depth $2a+1$, where $d = O(\log (n/\eps))$, $ m = \Theta(\log (n/\eps)) $ and the locality is $ O(\log (n/\epsilon)) $.

\end{theorem}

\begin{proof}

We follow the sample-then-extract procedure. 

Let $\Samp:\{0,1\}^{r_s}\rightarrow \{0,1\}^t$ be a $(\mu_1, \mu_2, \gamma)$-averaging sampler following from Theorem \ref{NC1sampler}. Let $\tau $ be a small enough constant, $\mu_1 = (\delta - 2 \tau)/\log(1/\tau), \mu_2 = (\delta - 3 \tau )/\log (1/\tau)$,  $ \gamma = 0.8 \epsilon $. As a result, $\mu_1 $ is a constant and $\mu_2 = \alpha \mu_1$ for some constant $\alpha \in (0,1)$.
For an $(n, k)$-source $X$, by Lemma \ref{sampresult}, we have $\SD(R\circ X_{\Samp(R)} , R\circ W) \leq \gamma + 2^{-\Omega(\tau n)}$. Here $R $ is a uniform random variable. For every $ r $ in $\{0,1\}^{r_s}$, the random variable $W|_{R=r}$ is a $(t, (\delta-3\tau)t)$-source.

By Lemma \ref{NC1sampler}, $r_s = \log n + O(\log (1/\gamma))$ and we can set $t = \Theta(\log (n/\eps))$ to be large enough.

Let $ \epsilon_1 = 0.1\epsilon$. We pick $m = O(\log (n/\eps))$ to be such that $(\delta - 3\tau)t \geq  m + 2\log(1/\epsilon_1)$. 
Let $\Ext_1:\{0,1\}^t \times \{0,1\}^{d_1} \rightarrow \{0,1\}^m$ be a $((\delta-3\tau)t, \epsilon_1)$-extractor following from Lemma \ref{leftoverh}. As a result, 
$$\SD(U \circ \Ext_1(W, U), U') \leq \epsilon_1,$$
where $U, U'$ are uniform distributions.

As a result, the sample-then-extract procedure gives an extractor of error
$$\gamma + 2^{-\Omega(\tau n)} + \epsilon_1 \leq 0.8 \epsilon+  2^{-\Omega(\tau n)} +   0.1\epsilon .$$

As $\tau $ is a constant, $2^{-\Omega(\tau n)} \leq 0.1 \epsilon$.

Thus the error of the extractor is at most $ \epsilon $.

The seed length is $r_s + d_1 = O(\log (n/\eps))$.

The locality  is $t = O(\log (n/\epsilon))$ because when the seed is fixed, we select $t$ bits from $X$ by sampling.

The sampler $\Samp$ is in $\AC^0$ of depth $a+1$. The extractor $\Ext_1$ is in $\AC^0$ of depth $a+1$. So $\Ext$ is in $\AC^0$ of depth $2a+1$.

\end{proof}

In this way, we have an $\AC^0$ extractor which can have quasi-polynomial small errors.

\subsection{ Previous Error Reduction Techniques }

Another tool we will be relying on is the error reduction method for extractors, given by Raz et al. \cite{rrv:error}.
They give an error reduction method for poly-time extractors and we will adapt it to the $\AC^0$ settings.

\begin{lemma}[$G_x$ Property \cite{rrv:error}]
\label{Gx}
Let $\Ext:\{0,1\}^n \times \{0,1\}^d \rightarrow \{0,1\}^m$ be a $(k, \epsilon)$-extractor with $\epsilon <1/4$. Let $X$ be any $(n, k+t)$-source. For every $x\in \{0,1\}^n$, there exists a set $G_x$ such that the following holds.
\begin{itemize}
\item For every $x\in \{0,1\}^n $, $G_x\subset \{0,1\}^d$ and $|G_x|/2^d = 1 - 2\epsilon$.
\item $\Ext(X, G_X)$ is within distance at most $2^{-t}$ from an $(m, m-O(1))$-source. Here $\Ext(X, G_X)$ is obtained by first sampling $x$ according to $X$, then choosing $r$ uniformly from $G_x$, and outputting $\Ext(x,r)$. We also denote $\Ext(X, G_X)$ as $\Ext(X, U)|_{U\in G_X}$.
\end{itemize}

\end{lemma}

Raz et al. \cite{rrv:error} showed the following result.
\begin{lemma}[\cite{rrv:error}]
\label{twoblock}
Let $\Ext:\{0,1\}^{n} \times \{0,1\}^d \rightarrow \{0,1\}^{m} $ be a $(k,\epsilon)$-extractor. Consider $\Ext': \{0,1\}^{n} \times \{0,1\}^{2d} \rightarrow \{0,1\}^{2m}$ which is constructed in the following way.
$$ \Ext'(x, u) = \Ext(x, u_1)\circ \Ext(x, u_2) $$
Here $u = u_1\circ u_2$.

For any $t\leq n-k$, let $X$ be an $(n,k+t)$-source . Let $U$ be the uniform distribution of length $2d$.

With probability at least $1-O(\epsilon^2)$, $\Ext'(X, U)$ is $2^{-t}$-close to having entropy $m-O(1)$.
\end{lemma}
\begin{remark}
Here we briefly explain the result in lemma \ref{twoblock}.
The distribution of $Y = \Ext'(X, U_1\circ U_2)$ is the convex combination of  $Y|_{U_1\in G_X,U_2\in G_X}$, $Y|_{U_1\notin G_X,U_2\in G_X}$, $Y|_{U_1\in G_X,U_2\notin G_X}$ and $Y|_{U_1\notin G_X,U_2\notin G_X}$.
That is
\begin{equation}
\begin{split}
Y = & I_{U_1\in G_X,U_2\in G_X}Y|_{U_1\in G_X,U_2\in G_X} + I_{U_1\notin G_X,U_2\in G_X}Y|_{U_1\notin G_X,U_2\in G_X}\\
& +I_{U_1\in G_X,U_2\notin G_X}Y|_{U_1\in G_X,U_2\notin G_X}+ I_{U_1\notin G_X,U_2\notin G_X}Y|_{U_1\notin G_X,U_2\notin G_X}.
\end{split}
\end{equation}
Also we know that $\Pr[I_{U_1\notin G_X,U_2\notin G_X} = 1] = O(\epsilon^2)$. As a result, according to Lemma \ref{Gx}, this lemma follows.
\end{remark}

Informally speaking,
this means that if view $Y = \Ext'(X, U) = Y_1 \circ Y_2$, then with high probability either $Y_1$ or $Y_2$ is $2^t$-close to having entropy $m-O(1)$. 

We adapt this lemma by doing the extraction for any $t\in \mathbb{N}^+$ times instead of 2 times. We have the following result.
\begin{lemma}
\label{multipleext}

Let $\Ext:\{0,1\}^{n} \times \{0,1\}^d \rightarrow \{0,1\}^{m} $ be a $(k,\epsilon)$-extractor. For any $t\in \mathbb{N}^+$, consider $\Ext': \{0,1\}^{n} \times \{0,1\}^{td} \rightarrow \{0,1\}^{tm}$ which is constructed in the following way.
$$ \Ext'(x, u) = \Ext(x, u_1)\circ \Ext(x, u_2) \circ \cdots \circ \Ext(x, u_t)$$
Here $u = u_1\circ u_2 \circ \cdots \circ u_t$.

For any $a\leq n-k$, let $X$ be an $(n,k+a)$-source. Let $U=\bigcirc_{i=1}^t U_i$ be the uniform distribution such that $\forall i\in [t], |U_i| = d$.

\begin{enumerate}
\item 
For $S\subseteq [t]$, let $I_{S,X}$ be the indicator such that $I_{S,X} = 1$ if $\forall i\in S, U_i\in G_X, \forall j\notin S, U_j\notin G_X$ and  $I_{S,X} = 0$ otherwise. Here $G_X$ is defined according to \ref{Gx}. The distribution of
$\Ext'(X, U)$ is a convex combination of the distributions of $ \Ext'(X, U)|_{I_{S,X} = 1}, S\subseteq [t]$. That is
$$ U\circ \Ext'(X, U) = \sum_{S\subseteq [t]} I_{S, X} U\circ\Ext'(X, U)|_{I_{S,X} = 1}$$

\item For every $ S\subseteq [t_1], S \neq \emptyset $, there exists an $i^* \in [t_1]$ such that $  \Ext(X, U_{i^*})|_{ I_{S, X} = 1 }$ is  $2^{-a}$-close to having entropy $m-O(1)$.
\end{enumerate}


\end{lemma}

\begin{proof}
The first assertion is proved as the follows.
By the definition of $G_x$ of Lemma \ref{Gx}, for each fixed $x\in \supp(X)$, $\sum_{S\subseteq [t]}I_{S, x} = 1$ as for each $i$, $U_i \in G_x$ either  happens or not. Also $I_{S, X}$ is a convex combination of $I_{S, x},  \forall x\in \supp(X)$. So $ \sum_{S\subseteq [t]}I_{S, X} = \sum_{S\subseteq [t]} \sum_{x\in \supp(X)}I_{S, x} I_{X=x}=  1 $. As a result, the assertion follows. 

The second assertion is proved as the follows.
For every $ S\subseteq [t_1], S \neq \emptyset $, by the definition of $I_{S,X}$, there exists an $i^* \in [t_1]$, $U_{i^*} \in G_X $.
By Lemma \ref{Gx}, $ \Ext(X, U_{i^*})|_{U_{i^*} \in G_X} = \Ext(X, U_{i^*})|_{ I_{S, X} = 1 } $  is $2^{-a}$-close to having entropy $m-O(1)$.

\end{proof}

\subsection{The Construction}
Finally we give the construction for error reduction of super-polynomially small errors.
\begin{construction}[Error Reduction for Super-Polynomially Small Error]
\label{constr2}
For any constant $a \in \mathbb{N}^+$, any constant $c\in \mathbb{N}$, any $k =  \Theta(n/\log^c n)$ and any $\epsilon = 1/2^{\Theta(\log^a n)}$,
we construct a strong $(k, \epsilon)$-extractor $\Ext:\{0,1\}^{n} \times \{0,1\}^{d} \rightarrow \{0,1\}^m$ with $m = k^{\Omega(1)}$.

\begin{itemize}
\item Let $\Ext_0:\{0,1\}^{n_0 = n} \times \{0,1\}^{d_0} \rightarrow \{0,1\}^{m_0}$ be a $(k_0, \epsilon_0 )$-extractor following from Theorem \ref{basicthm} with $k_0 \leq k - \Delta_1, \Delta_1 = \log(n/\epsilon)$, $\epsilon_0 = k^{-\Theta(1)}$, $d_0 = \Theta(\log n)$, $m_0 = k^{O(1)}$.
\item  Let $\Ext_1:\{0,1\}^{n_1 = m_0/t_2} \times \{0,1\}^{d_1} \rightarrow \{0,1\}^{m_1}$ be a $(k_1, \epsilon_1)$-extractor following from Theorem \ref{improveGAC0ext} where $ k_1 = 0.9 n_1$, $\epsilon_1 = \epsilon/n$, $d_1 = O(\log (n/\epsilon))$, $ m_1 = \Theta(\log (n/\epsilon))$.
\item Let $t_1$ be such that $(2\epsilon_0)^{t_1} \leq 0.1\epsilon$.  (We focus on the case that $\epsilon < \epsilon_0$. If $\epsilon \geq \epsilon_0 $, we set $\Ext$ to be $\Ext_0$.)
\item Let $t_2 = m_0^{1/3}$. 
\end{itemize}

Let $X$ be the input $(n,k)$ source.
Our construction is as follows.
\begin{enumerate}\addtocounter{enumi}{0}
\item Let $R_1, R_2, \ldots, R_{t_1}$ be independent uniform distributions such that for every $i\in [t_1]$ the length of $R_i$ is $d_0$.
Get $Y_1 = \Ext_0(X, R_1), \ldots, Y_{t_1} = \Ext_0(X, R_{t_1})$.
\item Get $Y = Y_1 \circ Y_2 \circ Y_3\circ \cdots \circ Y_{t_1}$.
\item For each $i\in [t_1]$, let $Y_i = Y_{i,1} \circ Y_{i,2}\circ \cdots \circ Y_{i,t_2}$ such that for every $j\in [t_2]$, $Y_{i,j}$ has length $n_1 = m_0/t_2$. Let $S_1, S_2, \ldots, S_{t_1}$ be independent uniform distributions, each having length $d_1$.
Get $Z_{i,j} = \Ext_1(Y_{i,j}, S_i) , \forall i\in[t_1], j\in [t_2]$. Let $Z_i = Z_{i,1}\circ Z_{i,2}\circ \cdots Z_{i,t_2}$.
\item Let $R = \bigcirc_i R_i, S = \bigcirc_i S_i$. We get $ \Ext(X,U) = Z = \bigoplus_i^{t_1} Z_i$ where $U = R \circ S$.

\end{enumerate}

\end{construction}

\begin{lemma}
\label{erlemma}

Construction \ref{constr2} gives a strong $(k, \epsilon)$-extractor.

\end{lemma}

In order to prove this Lemma, we need the following facts.

\begin{lemma}[Chain Rule of Min-Entropy \cite{vadhan2012pseudorandomness}]
\label{chainrule}

Let $(X,Y)$ be a jointly distributed random variable with entropy $k$. The length of $X$ is $l$. For every $\epsilon >0$, with probability at least $1-\epsilon$ over $x\leftarrow X$, $Y|_{X = x}$ has entropy $k-l-\log(1/\epsilon)$.

Also there exists another source $(X, Y')$ such that $\forall x\in \{0,1\}^{l}$, $Y'|_{X=x}$ has entropy $k-l-\log(1/\epsilon)$ and $\SD((X,Y), (X, Y'))\leq \epsilon$.

\end{lemma}

\begin{lemma}
\label{goodsource}
Let $X = X_1\circ \cdots \circ X_t $ be an $(n, n-\Delta)$-source where for each $i\in [t]$, $|X_i| = n_1 = \omega(\Delta)$. 

Let $k_1 = n_1-\Delta-\log(1/\epsilon_0) $ where $\epsilon_0$ can be as small as $1/2^{0.9n_1}$.

Let $\Ext_1:\{0,1\}^{n_1}\times \{0,1\}^{d_1} \rightarrow \{0,1\}^{m_1}$ be a strong $(k_1, \epsilon_1)$-extractor.

Let $\Ext: \{0,1\}^{n} \times \{0,1\}^d \rightarrow \{0,1\}^m$ be constructed as the following,
$$\Ext(X, U_s) = \Ext_1(X_1, U_s)\circ \cdots \circ \Ext_1(X_t, U_s).$$
Then $\Ext$ is a strong $(n-\Delta, \epsilon)$-extractor where $\epsilon = \SD(U_s \circ \Ext(X, U_s),U) \leq t(\epsilon_0 + \epsilon_1)$.
\end{lemma}

\begin{proof}
We prove by induction over the block index $i$.

For simplicity, let $\tilde{X}_i = X_1\circ \cdots \circ X_{i}$ for every $i$. We slightly abuse the notation $\Ext$ here so that $\Ext(\tilde{X}_i, U_s) = \Ext_1(X_1, U_s)\circ \cdots \circ \Ext_1(X_i, U_s)$ denotes the extraction for the first $i$ blocks.

For the first block, we know $H_{\infty}(X_1) = n_1 - \Delta$. According to the definition of $\Ext_1$,
$$\SD(U_s\circ \Ext_1(X_1, U_s), U) \leq \epsilon_1 \leq  (\epsilon_0 + \epsilon_1).$$

Assume for the first $i-1$ blocks, $\SD(U_s \circ \Ext_1(\tilde{X}_{i-1}, U_s) ,U )\leq (i-1)(\epsilon_0 + \epsilon_1)$. Consider $\tilde{X}_{i}$. By Lemma \ref{chainrule}, we know that there exists $X'_i$ such that $\SD(\tilde{X}_{i}, \tilde{X}_{i-1} \circ X'_{i}) \leq \epsilon_0$, where $X'_{i}$ is such that $\forall \tilde{x}_{i-1} \in \supp(\tilde{X}_{i-1}), H_{\infty}(X'_i|\tilde{X}_{i-1} = \tilde{x}_{i-1} ) \geq n_1-\Delta-\log(1/\epsilon_0)$. So according to Lemma \ref{SDproperty} part 2, as $U_s \circ \Ext(\tilde{X}_{i-1}, U_s) \circ \Ext_1(X_i, U_s)$ is a convex combination of $ u \circ \Ext(\tilde{X}_{i-1}, u) \circ \Ext_1(X_i, u) , \forall u\in \supp(U_s)$ and $ U_s\circ\Ext(\tilde{X}_{i-1}, U_s) \circ \Ext_1(X'_i, U_s)$ is a convex combination of $ u\circ\Ext(\tilde{X}_{i-1}, u) \circ \Ext_1(X'_i, u), \forall u\in \supp(U_s)$, we have
$$\SD(U_s \circ \Ext(\tilde{X}_{i-1}, U_s) \circ \Ext_1(X_i, U_s), U_s\circ\Ext(\tilde{X}_{i-1}, U_s) \circ \Ext_1(X'_i, U_s) ) \leq \SD(\tilde{X}_{i}, \tilde{X}_{i-1} \circ X'_{i}) \leq \epsilon_0.$$

According to the assumption, Lemma \ref{blocksourceextraction} and the triangle inequality of Lemma \ref{SDproperty}, we have the following.

\begin{equation}
\begin{split}
&\SD( U_s \circ \Ext(\tilde{X}_{i-1}, U_s) \circ \Ext_1(X_i, U_s), U) \\
\leq &   \SD(U_s \circ \Ext(\tilde{X}_{i-1}, U_s) \circ \Ext_1(X_i, U_s), U_s \circ \Ext(\tilde{X}_{i-1}, U_s) \circ \Ext_1(X'_i, U_s)) \\
& + \SD(U_s \circ \Ext(\tilde{X}_{i-1}, U_s) \circ \Ext_1(X'_i, U_s), U)   \\
\leq &  \epsilon_0 +  (i-1)(\epsilon_0 + \epsilon_1) + \epsilon_1\\
=  &i(\epsilon_0 + \epsilon_1)
\end{split}
\end{equation}
The first inequality is due to the triangle property of Lemma \ref{SDproperty}.
For the second inequality, first we have already shown that $  \SD(U_s \circ \Ext(\tilde{X}_{i-1}, U_s) \circ \Ext_1(X_i, U_s), U_s \circ \Ext(\tilde{X}_{i-1}, U_s) \circ \Ext_1(X'_i, U_s)) \leq   \epsilon_0$. Second, as $\tilde{X}_{i-1}\circ X'_i $ is a $((i-1)n_1-\Delta, n_1 - \Delta - \log(1/\epsilon_0))$-block source,  by our assumption and Lemma \ref{blocksourceextraction}, $\SD(U_s \circ \Ext(\tilde{X}_{i-1}, U_s) \circ \Ext_1(X'_i, U_s), U) \leq  (i-1)(\epsilon_0 + \epsilon_1) + \epsilon_1$. This proves the induction step.

As a result, $\SD(U_s\circ\Ext(X, U_s),U) \leq (\epsilon_0 + \epsilon_1)t$.
\end{proof}

Next we prove Lemma \ref{erlemma}.

\begin{proof}[Proof of Lemma \ref{erlemma}]

Let $G_X$ be defined by Lemma \ref{Gx} on $X$ and $\Ext_0$. 

For any $T \subseteq [t_1]$,
let
$ I_{T, X}$ be the indicator such that $I_{T, X} = 1,  $ if $\forall i\in T, R_i \in G_X,\forall i \notin T, R_i \notin G_X$ and $I_{T, X} = 0$, otherwise. 
By Lemma \ref{multipleext}, 
\begin{equation}
\begin{split} 
& R \circ Y \\  
= &\sum_{T\subseteq [t_1]} I_{T, X} (R \circ Y|_{I_{T, X} = 1})\\
= &  I_{\emptyset, X} (R  \circ Y|_{I_{\emptyset, X} = 1}) +  (1-I_{\emptyset, X}) ( R\circ Y|_{I_{\emptyset, X} = 0} )\\
= &  I_{\emptyset, X} (R  \circ Y|_{I_{\emptyset, X} = 1}) +  \sum_{T \subseteq [t_1], T\neq \emptyset} I_{T,X} ( R\circ Y|_{I_{T, X} = 1} )\\
\end{split}
\end{equation}

Fixing a set $T \subseteq [t_1], T\neq \emptyset$, by Lemma \ref{multipleext}, there exists an $i^*\in [t_1]$ such that $R_{i^*} \in G_X$ and $R \circ Y|_{ I_{T, X} = 1 }$ is $2^{-\Delta_1}$-close to 
$$ R' \circ A\circ W \circ B =  \bigcirc_i R'_i \circ A\circ W \circ B $$
Here $W = Y_{i^*}|_{ R_i \in G_X }$ has entropy at least $ m_0 - O(1)$. Also, $ A, B$ and $R'_i, i = 1,2, \ldots, t$ are some random variables where $A = (i^*-1)m_1, |B| = (t-i^*)m_1$ and  $\forall i \in [t], |R'_i| = d_1$. In fact, $R' = R|_{I_{T,X} = 1}$.

According to our construction, next step we view $A\circ W \circ B $ as having $t_1 t_2 $ blocks of block size $n_1$. We apply the extractor $\Ext_1$ on each block. Although for all blocks the extractions are conducted simultaneously, we can still view the procedure as first extracting $A$ and $B$, then extracting $W$. Assume for $A$, after extraction by using seed $S_A$, it outputs $A'$. Also for $B$, after extraction by using seed $S_B$, it outputs $B'$. So after extracting $A$ and $B$, we get $A' \circ W \circ B'$. The length of $A'\circ B'$ is at most $t_1 m_0 m_1/n_1 = t_1t_2 m_1 = \Theta(t_1t_2 \log n)$, as $n_1 = m_0/t_2$.

We know that $n_1 = m_0/t_2 = m_0^{2/3} \geq 10 t_1 t_2 m_1 = m_0^{1/3}\poly(\log n)$. Also according to Lemma \ref{chainrule}, $R' \circ S_A \circ S_B \circ A'\circ W \circ B'$ is $\epsilon'$-close to $R'\circ S_A \circ S_B \circ A'\circ W' \circ B'$ such that for every $r'\in \supp(R'), a\in \supp(A'), b\in \supp(B'), s_A\in\supp(S_A), s_B\in \supp(S_B)$, conditioned on $R' = r', S_A = s_A, S_B = s_B, A'=a, B'=b$, $W'$ has entropy at least $n_1 - O(\log n) - t_1 t_2 m_1 - \log(1/\epsilon') = n_1 - \Delta_2$ where $\Delta_2 = O(\log n) + t_1t_2 m_1 + \log(1/\epsilon')=O(m_0^{1/3} \log n)$. Here $\epsilon'$ can be as small as $2^{-k^{\Omega(1)}}$. That is
\begin{equation}
\begin{split}
&\forall r'\in \supp(R'), a\in \supp(A'), b\in \supp(B'), s_A\in\supp(S_A), s_B\in \supp(S_B),\\
&H_{\infty}(W'|_{ R' = r', S_A = s_A, S_B = s_B,  A'=a, B'=b}) \geq n_1 -\Delta_2.
\end{split}
\end{equation}


Let $\Ext'_1(W', S_{i^*}) = \bigcirc_{i \in [t_2]}\Ext_1(W'_i, S_{i^*})$ where $W' = \bigcirc_{i \in [t_2]} W'_i$ and $\forall i\in [t_2], |W'_i| = n_1$.
By Lemma \ref{goodsource}, as $k_1 = 0.9 n_1 \leq n_1 - \Delta_2$, $S_{i^*} \circ \Ext'_1(W', S_{i^*})|_{R' = r', S_A = s_A, S_B = s_B,  A'=a, B'=b}$ is $(\epsilon'_0 + \epsilon_1)t_2$-close to uniform distributions where $\epsilon'_0  $ can be as small as $ 2^{-k^{\Omega(1)}}$.

As a result, we have the following.
\begin{equation}
\begin{split}
&\SD(U\circ \Ext(X, U) ,U')\\
= & \SD( R \circ S \circ \Ext(X, U)   ,    R \circ S \circ \tilde{U} )   \\
= & \SD( I_{\emptyset, X} (R \circ S  \circ \Ext(X, U)|_{I_{\emptyset, X} = 1}), I_{\emptyset, X} (R \circ S  \circ \tilde{U}|_{I_{\emptyset, X} = 1}))\\
 & + \SD((1-I_{\emptyset, X})(R \circ S  \circ \Ext(X, U)|_{I_{\emptyset, X} = 0})  , (1-I_{\emptyset, X}) (R \circ S \circ \tilde{U} |_{I_{\emptyset, X} = 0} )) \\
= & \Pr[I_{\emptyset, X}  = 1]\SD(R \circ S  \circ \Ext(X, U)|_{I_{\emptyset, X} = 1},R \circ S  \circ \tilde{U}|_{I_{\emptyset, X} = 1})\\
= & (2\epsilon_0)^{t_1}\SD(R \circ S  \circ \Ext(X, U)|_{I_{\emptyset, X} = 1},R \circ S  \circ \tilde{U}|_{I_{\emptyset, X} = 1})\\
 & + \SD((1-I_{\emptyset, X})(R \circ S  \circ \Ext(X, U)|_{I_{\emptyset, X} = 0})  , (1-I_{\emptyset, X}) (R \circ S \circ \tilde{U} |_{I_{\emptyset, X} = 0} )) \\
\end{split}
\end{equation}
As 
$$ (2\epsilon_0)^{t_1} \SD( R \circ S  \circ \Ext(X, U)|_{I_{\emptyset, X} = 1}, R \circ S  \circ \tilde{U}|_{I_{\emptyset, X} = 1}) \leq  (2\epsilon_0)^{t_1}  $$
let's focus on $\SD((1-I_{\emptyset, X})(R \circ S  \circ \Ext(X, U)|_{I_{\emptyset, X} = 0})  , (1-I_{\emptyset, X}) (R \circ S \circ \tilde{U} |_{I_{\emptyset, X} = 0} ))$.
 
\begin{equation}
\begin{split}
 & \SD((1-I_{\emptyset, X})(R \circ S  \circ \Ext(X, U)|_{I_{\emptyset, X} = 0})  , (1-I_{\emptyset, X}) (R \circ S \circ \tilde{U} |_{I_{\emptyset, X} = 0} )) \\
= & \SD(\sum_{T\subseteq [t_1], T\neq \emptyset}I_{T,X}(R \circ S  \circ \Ext(X, U)|_{I_{T, X} = 1})  , \sum_{T\subseteq [t_1], T \neq \emptyset} I_{T,X}(R \circ S \circ \tilde{U} |_{I_{T, X} = 1} )) \\
= &  \sum_{T \subseteq [t_1], T\neq \emptyset} \Pr[I_{T,X} = 1]\SD( R\circ S \circ \Ext(X,U)|_{I_{T,X} = 1} ,   R \circ S \circ \tilde{U} |_{I_{T,X} = 1} )\\
\leq  &  \sum_{T \subseteq [t_1], T\neq \emptyset} \Pr[I_{T,X} = 1]( 2^{-\Delta_1} +\SD( R'\circ S\circ (A' \oplus \Ext'_1(W, S_{i^*}) \oplus B') ,  R' \circ S \circ \tilde{U}  )) \\
=  & (1-(2\epsilon_0)^{t_1}  )(2^{-\Delta_1} + \SD( R'\circ S\circ (A' \oplus \Ext'_1(W, S_{i^*}) \oplus B') ,  R' \circ S \circ \tilde{U}  )) \\
\leq & 2^{-\Delta_1} + \SD(R'\circ S\circ (A' \oplus \Ext'_1(W, S_{i^*}) \oplus B') ,  R' \circ S \circ \tilde{U}  )\\
\leq & 2^{-\Delta_1} + \SD(R'\circ S\circ (A' \oplus \Ext'_1(W, S_{i^*}) \oplus B') , R'\circ S\circ (A'\oplus \Ext'_1(W', S_{i^*})\oplus B'))\\
& + \SD(R'\circ S\circ (A'\oplus \Ext'_1(W', S_{i^*})\oplus B'),  R' \circ S \circ \tilde{U}  )\\
\leq & 2^{-\Delta_1} + \epsilon' + \SD(R'\circ S\circ (A'\oplus \Ext'_1(W', S_{i^*})\oplus B') , R' \circ S \circ \tilde{U})\\
\leq &  2^{-\Delta_1} + \epsilon' + (\epsilon'_0 + \epsilon_1)t_2\\
\end{split}
\end{equation}
Here $U, U', \tilde{U}$ are uniform distributions. In the second equation, $ I_{\emptyset, X}$ is the indicator such that $I_{\emptyset, X} = 1  $ if $\forall i\in [t_1], R_i \notin G_X$ where $G_X$ is defined by Lemma \ref{Gx} on $X$ and $\Ext_0$. 
For the first inequality, 
we need to show that
$$\SD(R\circ S \circ \Ext(X,U)|_{I_{T,X} = 1} ,  R'\circ S\circ (A' \oplus \Ext'_1(W, S_{i^*}) \oplus B')) \leq 2^{-\Delta_1} .$$
We know that for every $s\in \supp(S)$, by Lemma \ref{SDproperty} part 2,
\begin{equation}
\begin{split}
&\SD(R\circ S \circ \Ext(X, U)|_{I_{T,X} = 1, S = s}, R'\circ S \circ (A' \oplus \Ext'_1(W, S_{i^*}) \oplus B')|_{S = s} ) \\
\leq &\SD( R\circ Y|_{I_{T, X} = 1}, R'\circ A\circ W\circ B )\\
\leq &2^{-\Delta_1}.
\end{split}
\end{equation}
Here $R\circ S \circ \Ext(X, U)|_{I_{T,X} = 1, S = s} = h(R\circ Y|_{I_{T, X} = 1})$ for some deterministic function $h$ as $S=s$ is fixed. Also  $R'\circ S \circ (A' \oplus \Ext'_1(W, S_{i^*}) \oplus B')|_{S = s}  = h(R'\circ A\circ W\circ B) $ for the same reason.
As a result, 
\begin{equation}
\begin{split}
&\SD(R\circ S \circ \Ext(X,U)|_{I_{T,X} = 1} ,  R'\circ S\circ (A' \oplus \Ext'_1(W, S_{i^*}) \oplus B'))  \\
= & \sum_{s\in\supp(S)}\Pr[S = s] \SD(R\circ S \circ \Ext(X, U)|_{I_{T,X} = 1, S = s}, R'\circ S \circ (A' \oplus \Ext'_1(W, S_{i^*}) \oplus B')|_{S = s} )\\
\leq & 2^{-\Delta_1}.
\end{split}
\end{equation}
The third inequality holds by the triangle property of Lemma \ref{SDproperty} part 1. The $4$th inequality holds because by Lemma \ref{SDproperty} part 2,
\begin{equation}
\begin{split}
&\SD(R'\circ S\circ (A' \oplus \Ext'_1(W, S_{i^*}) \oplus B')|_{S = s} , R'\circ S\circ (A'\oplus \Ext'_1(W', S_{i^*})\oplus B')|_{S = s}) \\
\leq &\SD( R'\circ S\circ A \circ W\circ B, R'\circ S \circ A\circ W'\circ B )\\
\leq &\epsilon'.
\end{split}
\end{equation}.

As a result, the total error is at most 
$$ (2\epsilon_0)^{t_1} + (2^{-\Delta_1} + \epsilon' + (\epsilon'_0 + \epsilon_1)t_2)  $$

We can set $\epsilon' = 0.1 \epsilon,\epsilon'_0 =\epsilon/n$ so that $ (2^{-\Delta_1} + \epsilon' + (\epsilon'_0 + \epsilon_1)t_2) \leq 0.1\epsilon$.
As $(2\epsilon_0)^{t_1} <0.1\epsilon$, we know $\SD(U\circ \Ext(X, U) ,U') \leq \epsilon$. 

\end{proof}

\begin{lemma}
\label{outputlenerthm}
In Construction \ref{constr2}, the output length of $\Ext$ is $m = \Omega(n^{10800}\log (n/\epsilon))$.

\end{lemma}

\begin{proof}

The output length is equal to $t_2 \times m_1 = m_0^{1/3}\Omega(\log (n/\epsilon)) = \Omega(n^{1/10800}\log (n/\epsilon))$

\end{proof}

\begin{lemma}
\label{locality2}

In Construction \ref{constr2}, the function $\Ext$ can be realized by a circuit of depth $ 3a+c + 7$.
Its locality is $O(\log^{2a+c+4}n)  $.

\end{lemma}

\begin{proof}

By Theorem \ref{basicthm} and Lemma \ref{improveGAC0ext}, both $\Ext_0$ and $\Ext_1$ in our construction are in $\AC^0$. For $\Ext_0$, it can be realized by circuits of depth $c+7$. For $\Ext_1$, it can be realized by circuits of depth $2a+1$.

In the first and second steps of Construction \ref{constr2}, we only run $\Ext_0$ for $t_1$ times in parallel. So the computation can be realized by circuits of depth $c+7$

For the third step, we run $\Ext_1$ for $t_1 t_2$ times in parallel, which can be realized by circuits of depth $2a+1$.

The last step, according to Lemma \ref{xorpolylog}, taking the XOR of  $O(\log (1/\epsilon))$ bits can be realized by circuits of depth $a+1$. Each bit of $Z$ is the XOR of $t_1$ bits and all the bits of $Z$ can be computed in parallel. So the computations in this step can be realized by circuits of depth $a+1$.

Now we merge the three parts of circuits together.
As the circuits between each parts can be merged by deleting one depth, our construction can be realized by circuits of depth 
$$ (c+7) +  2a+1 + a+1 - 2  =  3a+c+7.$$

For the locality, by Theorem \ref{basicthm}, the locality of $\Ext_0$ is $O(\log^{c+4} n)$. By Lemma \ref{improveGAC0ext}, the locality of $\Ext_1$ is $O(\log (n/\epsilon))$. So each bit of $Z$ is related with at most $t_1 \times O(\log (n/\epsilon)) \times O(\log^{c+4} n) = O(\log^{2a+c+4}n) $ bits of $X$.

\end{proof}

\begin{lemma}
\label{erthmsdlen}
In Construction \ref{constr2}, $d = O(\log n +\frac{\log (n/\epsilon) \log(1/\epsilon)}{\log n})$.

\end{lemma}

\begin{proof}

In Construction \ref{constr2}, as 
$$U = R \circ S = \bigcirc_i R_i \circ \bigcirc_i S_i ,$$ 
$|U| = O(t_1d_0 + t_1 d_1)$. By the definitions of $\Ext_0$ and $\Ext_1$, we know that $d_0 = O(\log n)$ and $d_1 = O(\log (n/\epsilon))$. Also we know that $t_1 = O( \log(1/\epsilon)/\log n)$ because $ \epsilon_0 = n^{-\Theta(1)} $, $(2\epsilon_0)^{t_1} \leq 0.1\epsilon$. Note that we have to compute $\Ext_0$ for at least once. So $d = O( \log n + \frac{\log (n/\epsilon) \log(1/\epsilon)}{\log n})$.
\end{proof}

\begin{theorem}
\label{erthm}
For any constant   $a,c \in \mathbb{N}$, any $k =  \Theta(n/\log^c n)$ and any $\epsilon = 1/2^{\Theta(\log^a n)}$, 
there exists an explicit construction of a strong $(k, \epsilon)$-extractor $\Ext: \{0,1\}^{n} \times \{0,1\}^d \rightarrow \{0,1\}^{m}$ in $\AC^0$ of depth $3a+c+ 7$, where $d = O( \log n + \frac{\log (n/\epsilon) \log(1/\epsilon)}{\log n})$, $m =  \Omega(n^{1/10800} \log (n/\epsilon))= k^{\Omega(1)}$ and the locality is $ (\log n)^{2a+c+4} $.
\end{theorem}

\begin{proof}

It follows from Construction \ref{constr2}, Lemma \ref{erlemma}, Lemma \ref{outputlenerthm} , Lemma \ref{locality2} and Lemma \ref{erthmsdlen}.
\end{proof}

If we do not require the extractor to be in $\AC^0$, we can get an extractor with low locality while having very small error.
The construction is similar to Construction \ref{constr2}. The proof is also in the same way as that of Theorem \ref{erthm}. So we directly give the result.

\begin{theorem}
\label{erthmg2}
For any constant $c \in \mathbb{N}$, any $k = \Theta(\frac{n}{\log^c n }) $,
there exists an explicit construction of a strong $(k, \epsilon)$-extractor $\Ext: \{0,1\}^{n} \times \{0,1\}^d \rightarrow \{0,1\}^{m}$, where $\epsilon$ can be as small as $2^{-k^{\Omega(1)}}$,  $d =\Theta(\log n + \frac{\log (n/\epsilon)\log (1/\epsilon)}{\log n})$, $m = \Omega(  n^{\frac{1}{10800}} \log (\frac{n}{\epsilon}))$ $= k^{\Omega(1)}$ and the locality is $  \log^2 (1/\epsilon) \log^{c+ O(1)} n  $.
\end{theorem}

%% file: OutputLenOPT.tex
\section{Output Length Stretching}\label{sec:output}

In this section, we show how to extract $(1-\gamma)k$ bits for any constant $\gamma >0 $.

\subsection{Pre-sampling}

By Theorem \ref{erthm}, we have a $(k, \epsilon)$-extractor in $\AC^0$ for any $k = n/\poly(\log n)$ and any $\epsilon  = 1/\poly(n)$. We use pre-sampling to increase the output length.

Zuckerman  \cite{Zsamp} gives sampler (oblivious sampler) constructions from extractors.

\begin{definition}[ \cite{Zsamp} ]
\label{oblivioussamp}
An $(n,m, t, \gamma, \epsilon)$-oblivious sampler is
a deterministic function $\Samp:\{0,1\}^n \rightarrow (\{0,1\}^m)^t$ such that  $ \forall f: \{0,1\}^m \rightarrow [0,1]$,
$$ \Pr_{I \leftarrow \Samp(U_r)}[|\frac{1}{t}\sum_{i\in I} f(i) - \mathbf{E}f| > \epsilon]\leq \gamma.$$

\end{definition}

The following lemma explicitly gives a construction of oblivious samplers using extractors.
\begin{lemma}[ \cite{Zsamp}]
\label{exttosamp}
If there is an explicit $(k=\delta n, \epsilon)$-extractor $\Ext:\{0,1\}^{n } \times \{0,1\}^{d} \rightarrow \{0,1\}^m$, then  there is an explicit $(n,m, t = 2^{d}, \gamma = 2^{1-(1-\delta)n}, \epsilon)$-oblivious sampler.

The sampler is constructed as follows.
Given a seed $x$ of length $n$, the $t = 2^d$ samples are $\Ext(x, u),$ $\forall u\in \{0,1\}^d$.
\end{lemma}

As a result, we can construct the following samplers.

\begin{lemma}
\label{smallerersamp}

For any $a \in \mathbb{N}^+$ , let $\gamma$ be any $1/2^{\Theta(\log^a n)}$.
\begin{itemize}

\item For any $c \in \mathbb{N}$, let $ \epsilon$ be any $ \Theta(1/\log^c n)$.
There exists an explicit   $( O(\log (1/\gamma))$, $\log n$, $t$, $\gamma$, $\epsilon )$-oblivious sampler for any integer $t \in [t_0, n]$ with $t_0 = \poly(\log n)$.

\item For any constant $\alpha$ in $ (0,1)$, any $c \in \mathbb{N}$, any $\mu = \Theta(1/\log^c n)$, there exists an explicit $(\mu, \alpha \mu, \gamma)$-averaging sampler $\Samp:\{0,1\}^{\Theta(\log^a n)} \rightarrow [n]^t$ in $\AC^0$ of circuit depth $a+2$,  for any integer $t \in [t_0, n]$ with $t_0 = \poly(\log n)$.

Specifically, if $c = 0$, $t$ can be any integer in $[t_0, n ]$ with $t_0 = (\log n)^{\Theta(a)}$.

\end{itemize}

\end{lemma}

\begin{proof}

Let $k =  \log^a n$.
For any $\epsilon = \Theta(1/\log^c n)$,  let's consider a $(k , \epsilon)$-extractor $\Ext:\{0,1\}^{n' = c_0\log^a n } \times \{0,1\}^{d} \rightarrow \{0,1\}^{\log n}$ for some constant $c_0$, following Lemma \ref{Trevext}.  Here we make one modification. We replace the last $d$ bits of the output with the seed. We can see in this way, $\Ext$ is still an extractor.

Here the entropy rate is $\delta = 1/c_0$ which is a constant.
According to  Lemma \ref{Trevext}, we know that, $d$ can be $ \Theta(\frac{\log^2 (n'/\epsilon )}{\log n'}) = \Theta(\log \log n)$. 

For the first assertion, according to Lemma \ref{exttosamp}, there exists an explicit construction of a $(c\log^a n,$ $\log n,$ $t,$ $\gamma,$ $\epsilon)$-oblivious sampler where  $\gamma = 2^{1-(1-2/c)(c\log^a n)}$. As we can increase the seed length to $\log n$ by padding uniform random bits, $t$ can be any integer in $[ t_0, n]$ with $t_0 = 2^{\Theta(\frac{\log^2 (n'/\epsilon )}{\log n'})} = \poly(\log n)$. As $c_0$ can be any large enough constant, $\gamma$ can be $ 1/2^{\Theta(\log^a n)}$. 

Next we prove the second assertion.

According to the definition of oblivious sampler, we know that
$ \forall f:[n] \rightarrow [0,1]$,
$$ \Pr_{I \leftarrow \Samp(U)}[|\frac{1}{t}\sum_{i\in I} f(i) - \mathbf{E}f| > \epsilon]\leq \gamma.$$
Next we consider the definition of averaging sampler.

Let $(1- \alpha)\mu = \epsilon$. As $ \mu = \Theta(1/\log^c n)$, $\epsilon  = \Theta(1/\log^c n)$.
For any  $f:[n] \rightarrow [0,1]$ such that $\mu \leq \mathbf{E}f$, we have the following inequalities, where $\Samp$ is a  $(c\log^a n,$ $\log n,$ $t,$ $\gamma,$ $\epsilon)$-oblivious sampler.

\begin{equation}
\begin{split}
&\Pr_{I \leftarrow \Samp(U)}[\frac{1}{t}\sum_{i\in I} f(i) < \alpha\mu]\\
= &\Pr_{I \leftarrow \Samp(U)}[\frac{1}{t}\sum_{i\in I} f(i) < \mu - \epsilon]\\
= &\Pr_{I \leftarrow \Samp(U)}[\mu - \frac{1}{t}\sum_{i\in I} f(i)> \epsilon]\\
\leq &\Pr_{I \leftarrow \Samp(U)}[\mathbf{E}f- \frac{1}{t}\sum_{i\in I} f(i)> \epsilon]\\
\leq &\Pr_{I \leftarrow \Samp(U)}[|\frac{1}{t}\sum_{i\in I} f(i) - \mathbf{E}f| > \epsilon]\\
\leq &\gamma
\end{split}
\end{equation}
The first inequality holds because if the event that $\mu - \frac{1}{t}\sum_{i\in I} f(i)> \epsilon$ happens, then the event that $\mathbf{E}f- \frac{1}{t}\sum_{i\in I} f(i)> \epsilon$ will happen, as $\mu \leq \mathbf{E}f$. The second inequality is because $\mathbf{E}f- \frac{1}{t}\sum_{i\in I} f(i) \leq |\mathbf{E}f- \frac{1}{t}\sum_{i\in I} f(i)|$. So if $ \mathbf{E}f- \frac{1}{t}\sum_{i\in I} f(i)> \epsilon$ happens, then $|\frac{1}{t}\sum_{i\in I} f(i) - \mathbf{E}f| > \epsilon$ happens.

Also as we replace the last $d$ bits of the output of our extractor with the seed, the samples are distinct according to the construction of Lemma \ref{exttosamp}.

According to the definition of averaging sampler, we know that this gives an explicit $(\mu,$ $\alpha \mu,$ $\gamma)$-averaging sampler.

According to the construction described in the proof of Lemma \ref{exttosamp}, the output of the sampler is computed by running the extractor following Lemma \ref{Trevext} for $t$ times in parallel. So the circuit depth is equal to the circuit depth of the extractor $\Ext$.

Let's recall the construction of the Trevisan's extractor $\Ext$. 

The encoding procedure is doing the multiplication of the encoding matrix and the input $x$ of length $n' = c\log^a n$. By Lemma \ref{xorpolylog}, this can be done by a circuit of depth $a+1$.

The last step is the procedure of N-W generator. The selection procedure can be represented as a CNF/DNF, as the seed length for $\Ext$  is at most $\Theta(\log n)$. (Detailed proof is the same as the proof of Lemma \ref{locality}.)

As a result, we need a circuit of depth $a + 2$ to realize $\Samp$.

For the special situation that $c = 0$, the seed length $d$ for $\Ext$ can be $ \Theta(a\log \log n)$. So $t_0 = 2^d = (\log n)^{\Theta(a)}$.
\end{proof}

By Lemma \ref{sampresult}, we can sample several times to get a block source.

\begin{lemma}[Sample a Block Source]
\label{sampblock}
Let $t$ be any constant in $\mathbb{N}^+$.
For any $ \delta>0 $, let $X$ be an $(n,k=\delta n)$-source . 
Let $\Samp: \{0,1\}^r \rightarrow [n]^m$ be a $(\mu_1, \mu_2, \gamma)$-averaging sampler where $\mu_1 = (\frac{1}{t}\delta - 2\tau)/\log(1/\tau)$ and $\mu_2 = (\frac{1}{t}\delta - 3\tau)/\log(1/\tau)$, $m = (\frac{t-1}{t}k-\log(1/\epsilon_0))/t$. Let $\epsilon_s = \gamma + 2^{-\Omega(\tau n)}$. 
For any $i\in [t]$, let $U_i$s be uniform distributions over $\{0,1\}^r$. Let $X_i = X_{\Samp( U_i)}$, for $i\in [t]$.


It concludes that $\bigcirc_{i=1}^t U_i \circ \bigcirc_{i=1}^t X_i$ is $\epsilon = t(\epsilon_s + \epsilon_0)$-close to  $\bigcirc_{i=1}^t U_i \circ \bigcirc_{i=1}^t W_i$ where for every $u \in \supp(\bigcirc_{i=1}^t U_i)$, conditioned on  $\bigcirc_{i=1}^t U_i = u $, $  \bigcirc_{i=1}^t W_i $ is a $( k_1, k_2, \ldots,  k_t)$-block source with block size $m$ and $k_1 = k_2 = \cdots = k_t = (\delta/t - 3 \tau)m $. Here $\epsilon_0$ can be as small as $1/2^{\Omega(k)}$.

\end{lemma}

\begin{proof}

We prove by induction on $i \in [t]$.

If $i = 1$, according to Lemma \ref{sampresult}, we know $U_1 \circ X_1$ is $\epsilon_s = (\gamma + 2^{-\Omega(\tau n)})$-close to $U_1\circ W$ such that $\forall u \in \supp(U_1), H_{\infty}(W|_{U_1 = u}) = (\delta/t-3\tau)m$.

Next we prove the induction step.

Suppose $\bigcirc_{j=1}^{i}U_j \circ \bigcirc_{j=1}^{i} X_j$ is $(\epsilon_s + \epsilon_0)i$-close to $\bigcirc_{j=1}^{i}U_j \circ \bigcirc_{j=1}^{i} W_j$, where for every $u \in \{0,1\}^{ir}$, conditioned on  $\bigcirc_{j=1}^i U_j = u $, $  \bigcirc_{j=1}^i W_j $ is a $( k_1, k_2, \ldots,  k_i)$-block source with block size $m$ and $k_1 = k_2 = \cdots = k_i = (\delta/t - 3 \tau)m $. 

Consider $i+1$.
Recall the Chain Rule Lemma \ref{chainrule}. First notice that $\bigcirc_{j=1}^i U_j \circ \bigcirc_{j=1}^i X_j \circ X$ has entropy $ir+k$. Then we know that $\bigcirc_{j=1}^i U_j \circ \bigcirc_{j=1}^i X_j \circ X$ is $\epsilon_0$-close to $\bigcirc_{j=1}^i U_j \circ \bigcirc_{j=1}^i X_j \circ X'$ such that for every $ u\in \{0,1\}^{ir}$ and every $x\in \{0,1\}^{im}$, conditioned on $\bigcirc_{j=1}^i U_j  = u, \bigcirc_{j=1}^i X_j = x$, $X'$ has entropy $k - im - \log(1/\epsilon_0) \geq k/t $ which means the entropy rate is at least $\delta/t$. 

By our assumption for $i$,  $\bigcirc_{j=1}^i U_j \circ \bigcirc_{j=1}^i X_j \circ X'$ is $(\epsilon_s + \epsilon_0)i$-close to $\bigcirc_{j=1}^i U_j \circ \bigcirc_{j=1}^i W_j \circ \tilde{X}$, where $\tilde{X}$ is a random variable such that $\forall  u\in \{0,1\}^{ir}, \forall x\in \{0,1\}^{im}$, $\tilde{X}|_{ \bigcirc_{j=1}^i U_j  = u, \bigcirc_{j=1}^i X_j = x}$ has the same distribution as $X'|_{\bigcirc_{j=1}^i U_j  = u, \bigcirc_{j=1}^i W_j = x}$. As a result, for every $ u\in \{0,1\}^{ir}$ and $x\in \{0,1\}^{im}$, conditioned on $\bigcirc_{j=1}^i U_j  = u, \bigcirc_{j=1}^i W_j = x$, $\tilde{X}$ has entropy $k - im - \log(1/\epsilon_0) \geq k/t $.

Denote the event $(\bigcirc_{j=1}^i U_j  = u, \bigcirc_{j=1}^i W_j = x)$ as $e$,
by Lemma \ref{sampresult}, by sampling on source $\tilde{X}|_e$, we get $U_{i+1} \circ (\tilde{X}|_e)_{\Samp(U_{i+1})} = U_{i+1} \circ \tilde{X}_{\Samp(U_{i+1})}|_e$. It is $\epsilon_s$-close to $U_{i+1} \circ W|_e$ where $\forall a\in \{0,1\}^r$, $(W|_e)|_{U_{i+1} = a}$ is a $(m, (\delta/t - 3\tau)m)$-source. Thus $ \bigcirc_{j=1}^{i+1} U_j \circ \bigcirc_{j = 1}^i W_j \circ \tilde{X}_{\Samp(U_{i+1})} $ is $\epsilon_s $-close to $ \bigcirc_{j=1}^{i+1} U_j \circ \bigcirc_{j = 1}^i W_j \circ W $.

Let $W_{i+1} = W$. As a result, $\bigcirc_{j=1}^{i+1}U_j \circ \bigcirc_{j=1}^{i}X_j $ is $(\epsilon_s + \epsilon_0)(i+1)$-close to  $\bigcirc_{j=1}^{i+1}U_j \circ \bigcirc_{j=1}^{i+1} W_j$ such that for every $u \in \{0,1\}^{ir}$, conditioned on  $\bigcirc_{j=1}^{i+1} U_j = u $, $  \bigcirc_{j=1}^{i+1} W_j $ is a $( k_1, k_2, \ldots,  k_i)$-block source with block size $m$ and $k_1 = k_2 = \cdots = k_{i+1} = (\delta/t - 3 \tau)m $. 

This proves that induction step.

\end{proof}

This lemma reveals a way to get a block source by sampling. Block sources are easier to extract.

\subsection{Repeating Extraction}

Another important technique is the parallel extraction.
According to Raz at al. \cite{RazRV02}, we have the following lemma.

\begin{lemma}[\cite{RazRV02}]
\label{doubleext}
Let $\Ext_1:\{0,1\}^n \times \{0,1\}^{d_1} \rightarrow \{0,1\}^{m_1}$ be a strong $(k, \epsilon)$-extractor with entropy loss $\Delta_1$ and $\Ext_2:\{0,1\}^n \times \{0,1\}^{d_2} \rightarrow \{0,1\}^{m_2}$ be a strong $(\Delta_1 -s, \epsilon_2)$-extractor with entropy loss $\Delta_2$ for any $s<\Delta_1$. Suppose the function $\Ext: \{0,1\}^n \times \{0,1\}^{d_1+d_2} \rightarrow \{0,1\}^{m_1+m_2}$ is as  follows.

$$\Ext(x,u_1\circ u_2) = \Ext_1(x,u_1)\circ \Ext_2(x, u_2)$$

Then $\Ext$ is a strong $(k, (\frac{1}{1-2^{-s}})\epsilon_1 + \epsilon_2 \leq \epsilon_1 + \epsilon_2 + 2^{-s})$-extractor with entropy loss $\Delta_2 + s$.
\end{lemma}
This can be generalized to the parallel extraction for multiple times.
\begin{lemma}
\label{multiext}

Let $X$ be an $(n,k)$-source.
Let $\Ext:\{0,1\}^n \times \{0,1\}^{d} \rightarrow \{0,1\}^{m}$ be a strong $(k_0, \epsilon)$-extractor with $k_0 = k - tm - s$ for any $t, s$ such that $tm+s < k$. Let $\Ext': \{0,1\}^{n} \times \{0,1\}^{td} \rightarrow \{0,1\}^{tm}$ be constructed as follows.

$$\Ext'(x,\bigcirc_{i=1}^t u_i) = \Ext(x,u_1)\circ \Ext(x, u_2)\circ \cdots \circ \Ext(x, u_t)$$

Then $\Ext'$ is a strong $(k, t(\epsilon + 2^{-s}) )$-extractor.

\end{lemma}

\begin{proof}
Consider the mathematical induction on $j$.

For $j=1$, it is true. As $\Ext$ is a strong $(k_0, \epsilon)$-extractor, it is also a strong $(k, j(\epsilon+2^{-s}))$-extractor.

Next we prove the induction step.

Assume it is true for $j$. Consider $j+1$.
$$ \Ext'(x,\bigcirc_{i=1}^{j+1} u_i) =   \Ext'(x, \bigcirc_{i=1}^j u_i) \circ \Ext(x, u_{j+1})   $$
Here $\Ext'(x, \bigcirc_{i=1}^j u_i) $ is a strong $(k,  j(\epsilon+2^{-s}))$-source. Its 
entropy loss is $k - jm$. Also we know that $\Ext$ is a strong $(k-tm - s, \epsilon)$-extractor, thus a strong $(k-jm - s, \epsilon)$-extractor. According to Lemma \ref{doubleext}, $\Ext'(x, \bigcirc_{i=1}^{j+1} u_i)$ is a strong $(k, (j+1)(\epsilon + 2^{-s}))$-extractor. Its entropy loss is $ k-(j+1)m $.

This completes the proof.

\end{proof}

Lemma \ref{multiext} shows a way to extract more bits. Assume we have an $(n, k)$-source and an extractor, if the output length of the extractor is $k^\beta, \beta<1$, then we can extract several times to get a longer output. However, if we merely do it in this way, we need a longer seed. In fact, if we extract enough times to make the output length to be $\Theta(k)$, we need a seed with length $\Theta(k^{1-\beta} \log n)$. This immediately gives us the following theorem.

\begin{theorem}
\label{poorlenththm}
For any constant $a, c \in \mathbb{N}$, $\gamma \in (0,1)$, any $k  = \Theta( n/ \log^{c} n) $, $\epsilon  = 1/2^{\Theta(\log^a n)}$, there exists
an explicit strong $(k, \epsilon)$-extractor $\Ext:\{0,1\}^{n}\times \{0,1\}^d \rightarrow \{0,1\}^m$ in $\AC^0$ of depth $3a+c+7$. The locality is $ O(\log^{2a+c+4} n) $. The seed length $d =O(\frac{k \log (1/\epsilon)}{n^{1/10800} \log n }) $. The output length $m = (1-\gamma)k$.
\end{theorem}

\begin{proof}
Let $\Ext_0:\{0,1\}^{n} \times \{0,1\}^{d_0} \rightarrow \{0,1\}^{m_0}$ be a $(k_0, \epsilon_0 =  \epsilon/n)$ extractor following from Theorem \ref{erthm}. Here $k_0 = k - tm_0 - s$ where $s = \log(n/\epsilon)$, $t = (1-\gamma)k/m_0$. By lemma \ref{multiext}, we know that there exists a $(k,\epsilon')$ extractor $\Ext$ with $\epsilon' = t(\epsilon_0 + 2^{-s}) \leq \epsilon$. The output length is $(1-\gamma)k$.

According to the construction in Lemma \ref{multiext},  $\Ext$ has the same circuit depth and locality as $\Ext_0$.  The seed length is $t\times d_0 = O(\frac{k \log (1/\epsilon)}{n^{1/10800} \log n })$.

\end{proof}

If we only consider local extractors then similarly we have the following.

\begin{theorem}
\label{poorlenththm1}
For any constant $c \in \mathbb{N}$, $\gamma \in (0,1)$,  any $k = \Theta(\frac{n}{\log^c n }) $,
there exists an explicit construction of a strong $(k, \epsilon)$-extractor $\Ext: \{0,1\}^{n} \times \{0,1\}^d \rightarrow \{0,1\}^{m}$, where $\epsilon$ can be as small as $2^{-k^{\Omega(1)}}$, $d =O(\frac{k \log (1/\epsilon)}{n^{1/10800} \log n }) $, $m =(1-\gamma)k$ and the locality is $  \log^2 (1/\epsilon) \log^{c+ O(1)} n  $.
\end{theorem}

\subsection{The Construction}

In order to extract more bits while keeping seed length small,   we use classic bootstrapping techniques. Our construction is still in $\AC^0$ but it does not have small locality .

\begin{construction}
\label{constr3}
For any constant $a, c\in \mathbb{N}$, any $k =\delta n = \Theta(n/\log^c n), \epsilon = 1/2^{\Theta(\log^a n)}$,
we construct a $(k, \epsilon)$-extractor  $\Ext:\{0,1\}^{n} \times \{0,1\}^{d} \rightarrow \{0,1\}^{m}$ where  $d = O(\log n + \frac{\log (n/\epsilon) \log (1/\epsilon)}{\log n})$, $m = O(\delta k)$.
\begin{itemize}
\item Let $X$ be an $(n, k = \delta n)$-source 
\item Let $t \geq 10800 $ be a large enough constant.
\item Let $\Samp: \{0,1\}^{r} \rightarrow [n]^{m_s}$ be a $(\mu_1, \mu_2, \gamma)$-averaging sampler following from Lemma \ref{smallerersamp}, where $\mu_1 = (\frac{1}{t}\delta - 2\tau)/\log(1/\tau)$ and $\mu_2 = (\frac{1}{t}\delta - 3\tau)/\log(1/\tau)$, $m_s = (\frac{t-1}{t}k-\log(1/\epsilon_0))/t$, $\tau = \frac{1}{4}\delta$, $\gamma = \epsilon/n$. Let $\epsilon_s = \gamma + 2^{-\Omega(\tau n)}$. 
\item 
Let $\Ext_0:\{0,1\}^{n_0 = m_s} \times \{0,1\}^{d_0} \rightarrow \{0,1\}^{m_0}$ be a $(k_0, \epsilon_0)$-extractor following from Theorem \ref{erthm} where $k_0 =  0.1( \frac{1}{t}\delta - 3\tau)m_s - s$,  $\epsilon_0 = \epsilon/(10tn)$, $d_0 = O(\log n_0 + \frac{\log (n_0/\epsilon_0) \log (1/\epsilon_0)}{\log n_0})$, $m_0 = n_0^{1/10800} \Theta(\log (n_0/\epsilon_0))$.  Let $s $ be such that $2^{-s} \leq \epsilon/(10tn)$. 
\end{itemize}

Next we construct the function $\Ext$ as follows.
\begin{enumerate}\addtocounter{enumi}{0}

\item Get Let $X_i = X_{\Samp(X, S_i)}$ for $i\in [t]$, where $S_i, i\in [t]$ are independent uniform distributions. 

\item Get $Y_{t} = \Ext_0(X_t, U_0)$ where $U_0$ is the uniform distribution with length $ d_0 $.

\item For $i = t-1$ to $1$, get $Y_i = \Ext'(X_{i}, Y_{i+1} )$ sequentially.  The function $\Ext'$ is defined as follows. 
$$\Ext'(x, r) = \bigcirc_{i=1}^{\min\{\lfloor|r|/d_0\rfloor, \lfloor 0.9(\frac{1}{t}\delta - 3\tau)m_s/m_0\rfloor\}} \Ext_0(x, r_i)$$ where $r = \bigcirc_{i=1}^{\lfloor|r|/d_0\rfloor} r_i \circ r'$ for some extra bits $r'$ and $\forall i, |r_i| = d_0$. 

\item Output $\Ext(X, U_d) = Y_1 = \Ext'(X_1, Y_2)$, where $U_d = U_0 \circ \bigcirc_{i=1}^{t} S_i$.

\end{enumerate}
\end{construction}

\begin{lemma}
\label{blocklemma}
For $\epsilon_1 = 1/2^{\Omega(k)}$,
$\bigcirc_{i=1}^t S_i \circ \bigcirc_{i=1}^t X_i$ is $t(\epsilon_s+ \epsilon_1)$-close to $\bigcirc_{i=1}^t S_i \circ \bigcirc_{i=1}^t W_i $.

Here  $S_i$s are independent uniform distributions and $\forall r\in \supp(\bigcirc_{i=1}^t S_i )$, conditioned on $\bigcirc_{i=1}^t S_i  = r$, $ \bigcirc_{i=1}^t W_i$ is a $(k_1, k_2, \ldots,  k_t)$-block source with $k_1 = k_2 = \cdots = k_t = k' = ( \frac{1}{t}\delta - 3\tau)m_s $ . 
\end{lemma}

\begin{proof}
It follows from Lemma \ref{sampblock}.

\end{proof}

\begin{lemma}
\label{constr3lemma}

In Construction \ref{constr3}, the function $\Ext$  is a strong $(k, \epsilon)$-extractor.

%

\end{lemma}

\begin{proof}[Proof of Lemma \ref{constr3lemma}]
By Lemma \ref{blocklemma}, $\bigcirc_{i=1}^t S_i \circ \bigcirc_{i=1}^t X_i$ is $t(\epsilon_s + \epsilon_1) = 1/\poly(n)$-close to $\bigcirc_{i=1}^t S_i \circ B$ where $B = B_1 \circ B_2 \circ \ldots \circ B_t$. The $S_i$s are independent uniform distributions. Also $ \forall s \in \supp(\bigcirc_{i=1}^t S_i ) $, conditioned on $\bigcirc_{i=1}^t S_i  = s$, $B$ is a $(k_1, k_2,\ldots, k_t)$-block source with $k_1 = k_2= \cdots = k_t = k' = (\frac{1}{t}\delta - 3 \tau)m_s $. We denote the first $i$ blocks to be $\tilde{B}_i = \bigcirc_{j=1}^i B_i$.

Let $Y'_i = \Ext'(B_i, Y'_{i+1})$ for $i = 1,2,\ldots, t$ where $Y'_{t+1} = U_0$ is the uniform distribution with length $d_0$.


Next we use induction over $i$ (from $t$ to 1) to show that 
$$\SD(U_0 \circ Y'_i , U) \leq (t+1 - i) k (\epsilon_0+2^{-s}).$$

The basic step is to prove that $\forall b_1,b_2,\ldots,b_{t-1} \in \{0,1\}^{m_s}$, conditioned on $B_1 = b_1,\ldots, B_{t-1}=b_{t-1}$, $\SD(U_0\circ Y'_t, U) \leq k (\epsilon_0 + 2^{-s})$.
According to the definition of $\Ext'$, 
$$\SD(U_0  \circ \Ext'(B_t, U_0) , U ) \leq \epsilon_0 .$$
This proves the basic step.

For the induction step, assume that $\forall b_1,b_2,\ldots,b_{i-1} \in \{0,1\}^{m_s}$, conditioned on $B_1 = b_1,\ldots, B_{i-1}=b_{i-1}$, 
$$\SD(U_0\circ Y'_{i} , U)\leq (t+1 - i)k(\epsilon_0+2^{-s}).$$ 
Consider $U_0\circ Y'_{i-1} = U_0 \circ \Ext'(B_{i-1}, Y'_{i})$.

We know that $\forall b_1,b_2,\ldots,b_{t-2} \in \{0,1\}^{m_s}$, conditioned on $B_1 = b_1,\ldots, B_{i-2}=b_{i-2}$, $\tilde{B}_{i-1}\circ U_0 \circ Y'_i $ is a convex combination of $b_{i-1}\circ U_0 \circ Y'_i, \forall b_{i-1}\in \supp(\tilde{B}_{i-1}) $. As a result,
$$\SD(\tilde{B}_{i-1}\circ U_0 \circ Y'_i , \tilde{B}_{i-1}\circ U ) \leq (t+1 - i)k(\epsilon_0+2^{-s}).$$
Thus,  $\forall b_1,b_2,\ldots,b_{t-2} \in \{0,1\}^{m_s}$, conditioned on $B_1 = b_1,\ldots, B_{i-2}=b_{i-2}$, as $\tilde{B}_{i-1}\circ U_0 \circ Y'_i$ is a convex combination of $b_{i-1}\circ U_0 \circ Y'_i, \forall b_{i-1}\in \supp(\tilde{B}_{i-1})$ and $\tilde{B}_{i-1}\circ U$ is a convex combination of   $b_{i-1}\circ U, \forall b\in \supp(\tilde{B}_{i-1})$, by Lemma \ref{SDproperty} part 2,
\begin{equation}
\begin{split}
&\SD(U_0\circ \Ext'(B_{i-1}, Y'_i), U_1 \circ \Ext'(B_{i-1}, U_2) ) \\
\leq &\SD(\tilde{B}_{i-1}\circ U_0 \circ Y'_i , \tilde{B}_{i-1}\circ U )\\
\leq &(t+1 - i)k(\epsilon_0+2^{-s}).
\end{split}
\end{equation}

Here $U = U_1 \circ U_2$. $U_1$ is the uniform distribution having $|U_1| = |U_0|$. $U_2$ is the uniform distribution having $|U_2| = |Y'_i|$.

According to the definition of $\Ext'$ and Lemma \ref{multiext}, we know that  $\forall b_1,b_2,\ldots,b_{i-2} \in \{0,1\}^{m_s}$, conditioned on $B_1 = b_1,\ldots, B_{i-2}=b_{i-2}$,
$$ \SD(U_1\circ \Ext'(B_{i-1}, U_2), U ) \leq k(\epsilon_0+2^{-s}).$$

So according to triangle inequality of Lemma \ref{SDproperty},  $\forall b_1,b_2,\ldots,b_{t-2} \in \{0,1\}^{m_s}$, conditioned on $B_1 = b_1,\ldots, B_{i-2}=b_{i-2}$,
\begin{equation}
\begin{split}
&\SD(U_0\circ Y'_{i-1}, U )  \\
= & \SD(U_0\circ \Ext'(B_{i-1}, Y'_{i} ), U)\\
\leq & \SD(U_0\circ \Ext'(B_{i-1}, Y'_{i} ), U_1\circ \Ext'(B_{i-1}, U_2)) + \SD (U_1\circ \Ext'(B_{i-1}, U_2), U)\\
\leq &(t+1 - i) k(\epsilon_0 + 2^{-s}) +   k(\epsilon_0+2^{-s})\\
= &(t+1-(i-1)) k(\epsilon_0 + 2^{-s}).
\end{split}
\end{equation}

This proves the induction step.

So we have $\SD(U_0 \circ Y'_1 , U) \leq t k (\epsilon_0+2^{-s})$.

As a result,
\begin{equation}
\begin{split}
&\SD(U_d \circ \Ext(X, U_d) , U)\\
= & \SD( U_0 \circ \bigcirc_{i=1}^t S_i \circ Y_1, U )\\
\leq & \SD( U_0 \circ \bigcirc_{i=1}^t S_i \circ Y_1, U_0 \circ \bigcirc_{i=1}^t S_i \circ Y'_1 ) + \SD( U_0 \circ \bigcirc_{i=1}^t S_i \circ Y'_1, U)\\ 
\leq & \SD( U_0 \circ \bigcirc_{i=1}^t S_i \circ \bigcirc_{i=1}^t X_i, U_0 \circ \bigcirc_{i=1}^t S_i \circ \bigcirc_{i=1}^t B_i) + \SD( U_0 \circ \bigcirc_{i=1}^t S_i \circ Y'_1, U)\\ 
\leq & t(\epsilon_s + \epsilon_1) + tk(\epsilon_0 +2^{-s}). 
\end{split}
\end{equation}

According to the settings of $\epsilon_0, \epsilon_s$, $t$ and by setting $\epsilon_1$ to be small enough, we know the error is at most $\epsilon$.

\end{proof}

\begin{lemma}

\label{blocklength}
In Construction \ref{constr3},
the length of $Y_i$ is 
$$|Y_i| = \Theta( \min\{ m_0(\frac{m_0}{d_0})^{t-i} ,  0.9(\frac{1}{t}\delta -3\tau)m_s \}).$$
Specifically, $m = |Y_1| = \Theta((\frac{1}{t}\delta -3\tau)m_s )=\Theta(\delta k)$.
\end{lemma}

\begin{proof}
For each time we compute $Y_i = \Ext'(X_i, Y_{i+1})$, we know $|Y_i| \leq |Y_{i+1}|(\frac{m_0}{d_0})$. Also according to the definition of $\Ext'$, $|Y_i| \leq 0.9(\frac{1}{t}\delta -3\tau)m_s$. So $|Y_i| = \Theta( \min\{m_0(\frac{m_0}{d_0})^{t-i} ,  0.9(\frac{1}{t}\delta -3\tau)m_s \})$ for $i\in [t]$.

By Theorem \ref{erthm}, $m_0 = \Theta(n_0^{1/10800}\log n)$. Also we know that $n_0 = m_s = O(tk) $. As a result, 
when $t \geq 10800$, $ m_0(\frac{m_0}{d_0})^{t-1} = \omega( m_s ) $. As a result, $m = |Y_1| = \Theta((\frac{1}{t}\delta -3\tau)m_s )=\Theta(\delta k)$.

\end{proof}

\begin{lemma}
\label{lenseedlen}
In Construction \ref{constr3}, the seed length $d = O(\log n + \frac{\log (n/\epsilon) \log (1/\epsilon)}{\log n})$.

\end{lemma}

\begin{proof}

The seed for this extractor is $U_d = U_0 \circ \bigcirc_{i=1}^{t} S_i$. So $|U_d| = |U_0| + \Sigma_{i}^t |S_i|  = O(\log n + \frac{\log (n/\epsilon) \log (1/\epsilon)}{\log n}) + O(\log (1/\epsilon)) = O(\log n + \frac{\log (n/\epsilon) \log (1/\epsilon)}{\log n})$.

\end{proof}

\begin{lemma}
\label{lendepth}
In Construction \ref{constr3}, the function $\Ext$ is in $\AC^0$. The depth of the circuit is $O(a+c+1)$.
\end{lemma}

\begin{proof}

We in fact run $\Samp$ and $\Ext_0$ for constant number of times in sequential. So the total depth is $O(a+c)$ as the depth of $\Samp$ and $\Ext_0$ are both $O(a+c+1)$.

\end{proof}
%
%
%

\begin{theorem}
\label{basiclenthm}
For any constant $a, c\in \mathbb{N}$, any $k = \delta n = \Theta(n/\log^c n) $, $\epsilon = 1/2^{\Theta(\log^a n)}$,
there exists an explicit  strong $(k,\epsilon)$-extractor $\Ext: \{0,1\}^{n} \times \{0,1\}^d \rightarrow \{0,1\}^{m}$ in $\AC^0$ with depth $O(a+c+1)$, where $d = O(\log n + \frac{\log (n/\epsilon) \log (1/\epsilon)}{\log n})$, $m = \Omega(\delta k)$ . 

\end{theorem}
\begin{proof}
By Construction \ref{constr3}, Lemma \ref{constr3lemma}, Lemma \ref{blocklength},  Lemma \ref{lenseedlen} ,  and Lemma \ref{lendepth} , the conclusion immediately follows.

\end{proof}

\begin{theorem}
\label{lenthm}
For any constant $\gamma \in (0,1)$, $a, c\in \mathbb{N}$, any $k = \delta n = \Theta(n/\log^c n) $, $\epsilon = 1/2^{\Theta(\log^a n)}$,
there exists an explicit strong $(k ,\epsilon)$-extractor $\Ext: \{0,1\}^{n} \times \{0,1\}^d \rightarrow \{0,1\}^{m}$ in $\AC^0$ with depth $ O(a+c+1)$ where $d = O((\log n + \frac{\log (n/\epsilon) \log (1/\epsilon)}{\log n})/\delta)$, $m = (1-\gamma)k$.

\end{theorem}

\begin{proof}

Let the extractor following Theorem \ref{basiclenthm} be $\Ext_0: \{0,1\}^{n_0} \times \{0,1\}^{d_0} \rightarrow \{0,1\}^{m_0}$ which is a $(k_0, \epsilon_0)$-extractor with $n_0 = n, k_0 = \gamma k - s$ for some $s < \gamma k$. The construction of $\Ext$ is
$$\Ext(x, u) = \bigcirc_{i = 1}^{t} \Ext_0(x, u_i).$$
Here $ t $ is such that  $tm_0 = (1-\gamma)k $.

By Lemma \ref{basiclenthm}, we know that $m_0 = \Theta(\delta k)$ where $\delta = \Theta( \frac{1}{\log^c n})$. So $t = \Theta(1/\delta)$. By Lemma \ref{multiext}, if $tm_0 = (1-\gamma)k $, then $\Ext$ is a $(k, \epsilon)$-extractor with output length $(1-\gamma)k$ and error $ t(\epsilon_0 + 2^{-s}) $. 

We choose $s$ to be large enough and $\epsilon_0$ to be small enough such that the error is at most $\epsilon$. The seed length $d = td_0$. By Theorem \ref{basiclenthm}, $d_0 =  O(\log n + \frac{\log (n/\epsilon) \log (1/\epsilon)}{\log n}), m_0 = \Omega(\delta k)$, so $d =  O((\log n + \frac{\log (n/\epsilon) \log (1/\epsilon)}{\log n})/\delta)$, $m = (1-\gamma)k$. 
The circuit depth maintains the same as that in Theorem  \ref{basiclenthm} because the extraction is conducted in parallel.
\end{proof}

%% file: DExtForBitfixing.tex
\section{Deterministic Extractor for Bit-fixing Source}
\label{sec:bitfixing}
We use two crucial tools. One is the extractor for non-oblivious bit-fixing sources, proposed by Chattopadhyay and Zuckerman \cite{CZ15} and improved by Li \cite{li2015improved}.
%
%
%
%
%
%

\begin{theorem}[\cite{li2015improved} Theorem 1.11]
\label{nobfExt}
Let $c$ be a constant. For any $\beta > 0$ and all $n \in \mathbb{N}$, there exists an explicit extractor $\Ext: \{0,1\}^n \rightarrow \{0,1\}^m$ such that for any $(q , t, \gamma)$-non-oblivious bit-fixing source $X$ on $n$ bits with $q\leq n^{1-\beta}$, $t \geq c\log^{21} n$ and $\gamma \leq 1/n^{t+1}$, 
$$ \SD(\Ext(X), U) \leq \epsilon   $$
where $ m = t^{\Omega(1)}$, $\epsilon = n^{-\Omega(1)}$.

The extractor can be computed by standard circuits of depth $  \lceil \frac{ \log m}{\log \log n} \rceil +O(1)$.
\end{theorem}

To see the depth is $ \lceil \frac{ \log m}{\log \log n} \rceil +O(1)$, let's briefly recall the construction of the extractor. It first divides the input in to $ n^{O(1)}$ blocks. Then for each block, it applys the extractor from \cite{li2015improved} Theorem 4.1 which has depth 4. At last, it conducts a multiplication between a  matrix  of size $m\times O(m)$ and a vector of dimension $O(m)$, both  over $\mathbb{F}_2$. The last step can be computed by a circuit of depth $  \lceil \frac{\log m}{\log \log n} \rceil +1 $ by Lemma \ref{xorpolylog} part 2.

The other tool is the design extractor introduced by Li \cite{Li12a}.

\begin{definition}[\cite{Li12a}]

An $(N, M, K, D, \alpha, \epsilon)$-design extractor is a bipartite graph with left vertex set $[N]$,
right vertex set $[M]$, left degree $D$ such that the following properties hold.
\begin{itemize}

\item (extractor property) For any subset $S \subseteq [M]$, let $\rho_S = |S|/M$. For any vertex $v \in [N]$, let
$\rho_v = |\Gamma(v) \cap S|/D$. Let $Bad_S = \{v \in [N] : |\rho_v  - \rho_S| > \epsilon\}$, then $|Bad_S| \leq K$. ($\Gamma(\cdot)$ outputs the set of all neighbors of the input.)
\item (design property) For any two different vertices $u, v \in [N]$, $|\Gamma(u) \cap \Gamma(v)| \leq \alpha D$.

\end{itemize}
\end{definition}

\begin{construction}
\label{DExtConstr}
For any constant $a\in \mathbb{N}$, any $t = \poly(\log n)$, the deterministic extractor $\Ext: \{0,1\}^{n} \rightarrow \{0,1\}^{m = t^{\Omega(1)}}$ for any $(n, \delta n = \Theta(n/\log^a n))$-bit-fixing source  is constructed as the follows.


\begin{itemize}

\item Construct an $(N, M ,K, D, \alpha, \epsilon)$-design extractor, where $M = n, K = n^{1/0.9}, N = n^{1/0.3}$, $\epsilon = 1/\log^c N$,  $D = \log^b N$, $\alpha = D/M + \epsilon$,  for $c = \lceil \log t/\log \log N\rceil + a + 1$ and large enough constant $b = \Theta(c)$.

\item Let $Y = (Y_1, Y_2, \ldots, Y_N)$. Compute $Y_i =  \bigoplus_{j \in \Gamma(i)} X_j$, for $i = 1,\ldots, N$, by taking $i$ as the $i$th vertex in the left set of the design extractor.

\item Let $\Ext(X) = \Ext'(Y)$ where $\Ext':\{0,1\}^{N} \rightarrow \{0,1\}^{m}$ is the extractor from Theorem \ref{nobfExt} with error $\epsilon = n^{-\Omega(1)}$.

\end{itemize}

\end{construction}

\begin{lemma}
\label{designExtConstr}
An $(N, M ,K, D, \alpha, \epsilon)$-design extractor, where  $K = N^{1/3}, M = K^{0.9}$, $\epsilon = 1/\log^c N$,  $D =  \log^b N $,$\alpha = D/M + \epsilon$, for any constant $c$ and large enough constant $b = \Theta(c)$, can be constructed in polynomial time.

\end{lemma}

\begin{proof}
In \cite{Li12a} it is showed that design extractors can be constructed in deterministic polynomial time by a greedy algorithm.

The construction is based on a $(k_0,\epsilon)$-extractor $\Ext_0:\{0,1\}^{n_0} \times \{0,1\}^{d_0} \rightarrow \{0,1\}^{m_0}$, from Theorem \ref{extinP} (almost optimal parameters), for any $(n_0, k_0)$-source, where $ n_0 = 4k_0,  m_0= 0.9k_0, d_0 = O(\log (n_0/\epsilon)) $. Also we substitute the first $d_0$ bits of the output by the seed s.t. every left vertex has exactly $2^{d_0}$ neighbors. Recall the greedy algorithm proposed by Li  \cite{Li12a}, which picks   vertices one by one, deleting the vertices which does not meet the design property before each picking. At last we can get $2^{n_0-k_0}  \geq 2^{3k_0}$ left vertices. Let $N = 2^{3k_0}, K = 2^{k_0}, M = 2^{m_0}$. We know that $\epsilon =   1/\log^c N  $ and $d_0 = O(\log (n_0/\epsilon)) $. Thus if $b=\Theta(c)$ is large enough, $D$ can be $ \log^b N $ by adding extra random bits to adjust the length of the seed.   

\end{proof}

\begin{lemma}
\label{bftonobf}
For any constant $a \in \mathbb{N}$, any $t = \poly(\log n)$, if $X$ is an $(M, \delta M = \Theta(M/\log^a M))$-bit-fixing source, then $Y = g(X)$ is a $(q, t, 0)$-non-oblivious bit-fixing source, where $q = K $.

\end{lemma}

\begin{proof}
Assume the coordinates of random bits of $X$ form the set $S$. By the extractor property of design extractors, the number of left vertex $x$, such that $|\rho_x - \rho_S| > \epsilon $, is at most $K$. These vertices form the set $Bad_S$. 

We prove that for any subset $V \subseteq [N] \backslash Bad_S$ with size $|V| \leq t$, $\bigoplus_{j\in V} Y_j$ is uniformly distributed.  

Let $V = \{v_1, v_2,\ldots, v_{t'}\}$ be a subset of $[N] \backslash Bad_S$, where $t' \leq t$. So $|\Gamma(v_{t'}) \cap S| \geq (\delta- \epsilon )D$. By the design property of design extractors, for any $i = 1,2,\ldots, t'-1, |\Gamma(v_{t'})\cap \Gamma(v_i)| \leq \alpha D $. So  $|(\Gamma(v_{t'})\cap S) \backslash \bigcup_{i=1}^{t'-1} \Gamma(v_i)| \geq (\delta -\epsilon)D -  t \cdot \alpha D \geq 1$  for $c = \lceil \log t/\log \log N\rceil + a + 1$ and  large enough  constant $b $. Thus $\bigoplus_{j\in V} Y_j$ is uniformly distributed because some uniform random bits in $\Gamma(v_{t'})\cap S$ cannot be canceled out by bits in $\bigcup_{i=1}^{t'-1} \Gamma(v_i)$. 

By the Information Theoretic XOR-Lemma in \cite{Goldreich95}, $Y_{[N] \backslash Bad_S}$ is $t$-wise independent. Thus   $Y = g(X)$ is a $(q = K, t, 0)$-non-oblivious bit-fixing source.

\end{proof}

\begin{theorem}
\label{basicDExt}
For any constant $a \in \mathbb{N}$, there exists an explicit deterministic $(k=\delta n = \Theta(n/\log^a n), \epsilon = n^{-\Omega(1)})$-extractor $\Ext:\{0,1\}^{n} \rightarrow \{0,1\}^m$ that can be computed by $\AC^0$ circuits of depth $ \Theta( \frac{\log m}{\log \log n} +a ) $, for any $(n, k)$-bit-fixing source, where $m$ can be any $ \poly(\log n)$. 

\end{theorem}

\begin{proof}
We claim that Construction \ref{DExtConstr} gives the desired extractor.
Let $X$ be an $(n, k)$-bit-fixing source. By lemma \ref{bftonobf},  we get $Y$ which is a $(q, t, 0)$-non-oblivious bit-fixing source with length $\Theta(n^{1/0.3})$, where $q = \Theta(n^{1/0.9})$ and $t$ can be any large enough $\poly(\log n)$. 

By Theorem \ref{nobfExt}, $\Ext(X) = \Ext'(Y)$ is $\epsilon$-close to uniform. 
The output length $m = t^{\Omega(1)}$ can be any $  \poly(\log n)$ as we can set $t$ to be any $\poly(\log n)$. 

We show that the circuit for computing the extractor is in uniform $\AC^0$. In Construction \ref{DExtConstr}, each $Y_i$ is the XOR of poly-logarithmic bits of $X$. Also the extractor of Theorem \ref{nobfExt} is in $\AC^0$. So the overall construction is in $\AC^0$. 
It is in uniform $\AC^0$  because the design extractor can be constructed in polynomial time by Lemma \ref{designExtConstr} while all other operations are explicit and can be computed by uniform $\AC^0$ circuits.

The depth of the circuit is  $\Theta(\frac{\log m}{\log \log n}+ a )$. Because in Construction \ref{DExtConstr},  $c = \lceil \log t/\log \log N\rceil + a + 1$ and   $b = \Theta(c)$. By lemma \ref{xorpolylog} part 2, the XOR of $D = 1/\log^b N $ bits can be computed by circuits of depth $b$. Also the depth of $\Ext'$ is $  \lceil \frac{ \log m}{\log \log n} \rceil +O(1) $. Thus the overall depth is $\Theta(\frac{\log m}{\log \log n}+ a )$.

\end{proof}


Next we do error reduction. Our method is  based on the XOR lemma given by Barak, Impagliazzo and Wigderson \cite{barak2006extracting}. 

%
%
%
%
%

\begin{lemma}[\cite{barak2006extracting} Lemma 3.15]
\label{xorreduceerror}
Let $Y_1, Y_2, \ldots, Y_t$ be independent distributions over $\mathbb{F}$ such that $\forall  i\in [t], \SD(Y_i, U ) \leq \epsilon$.
Then 
$$ \SD(\sum_{i=1}^t Y_i,  U) \leq (2\epsilon)^{t},$$
where $U$ is uniform over $\mathbb{F}$.

\end{lemma}

\begin{proof}

For simplicity, let $ \mathbb{F} =\{ 0,1,\ldots, M-1\}$.

We use induction to show that for $ j = 1,2, \ldots, t$, $\SD(\sum_{i=1}^{j}Y_i, U) \leq (2\epsilon)^j$.

As $Y_1$ is $\epsilon$-close to uniform, this shows the base case.

Let $Y' = \sum_{i=1}^{j-1}Y_i$.
Suppose $\SD(Y', U ) \leq (2\epsilon)^{j-1}$.

Let $p' = (p'_0, p'_1, \ldots, p'_{M-1})$ be such that $p'_i = \Pr[Y' = i_b] = 1/M + \delta'_i$, for $i = 0,1,\ldots, M-1$, where $i_b$ is the binary form of $i$. 

We know that $\SD(Y', U ) = 1/2(\sum_{i=1}^{M-1} |\delta'_i|)   $ and $\sum_{i=0}^{M-1} \delta'_i = 0$.

Let $p = (p_0, p_1, \ldots, p_{M-1})$ be such that $p_i = \Pr[Y_j = i_b] =  1/M + \delta_i$ for $i = 0,1,\ldots, M-1$.  We know that $\SD(Y_j, U ) = 1/2(\sum_{i=1}^{M-1}| \delta_i|) $ and $\sum_{i=0}^{M-1} \delta_i = 0$.

So
\begin{equation}
\begin{split}
\Pr[Y' + Y_j = i_b]  | & =   \sum_{k=0}^{M-1} \Pr[Y' = k_b] \Pr[Y_j = (i-k)_b]\\
& =  \sum_{k=0}^{M-1} p'_k \cdot p_{i-k}  \\
&= 1/M + 2(\sum_{k=0}^{M-1} \delta_k)/M + \sum_{k=0}^{M-1} \delta_{k}\delta_{ i-k}\\
& = 1/M + \sum_{k=0}^{M-1} \delta_{k}\delta_{i-k}.
\end{split}
\end{equation}

Thus $|\Pr[Y'+ Y_j = i_b] - \Pr[U = i_b]| = |\sum_{k=0}^{M-1} \delta_{k}\delta_{i-k}|$.

As a result,
\begin{equation}
\begin{split}
\SD( Y'+Y_j, U ) &= 1/2\sum_{i=0}^{M-1} |\Pr[Y'\oplus Y_j = i_b] - \Pr[U = i_b]|  \\
& = 1/2 \sum_{i=0}^{M-1} |\sum_{k=1}^{M-1} \delta_k \delta_{i-k}| \\
&\leq 1/2(\sum_{k=0}^{M-1}\sum_{l=0}^{M-1} |\delta_k \delta_l|)\\
& = 1/2(\sum_{i=1}^{M-1} |\delta'_i|)(\sum_{i=1}^{M-1}| \delta_i|)\\
&\leq (2\epsilon)^j.
\end{split}
\end{equation}

\end{proof}

\begin{theorem}
\label{erreductionDExt}
If $\Ext:\{0,1\}^n \rightarrow \{0,1\}^m$ is a  $(k, \epsilon)$-extractor for $(n, k)$-bit-fixing sources, then for every $l\in \mathbb{N}$, the function $\Ext':\{0,1\}^{ln} \rightarrow \{0,1\}^{m}$, given by $\Ext'(x_1, \ldots, x_l) = \oplus_{i\in [l]} \Ext(x_i)$ is a  $(2lk, \epsilon^{ \Theta( l) })$-extractor for $(ln, 2lk)$-bit-fixing sources.

\end{theorem}

\begin{proof}

Let $X = (X^{(1)}, \ldots, X^{(l)})$ be an $(ln, 2lk)$-bit-fixing source.
Then for $  \delta = k/n$ fraction of $j\in [l]$, $X^{(j)}$ is an $(n, \delta n)$-bit-fixing source. Because if not, the total number of random bits is at most $ \delta l \cdot  n + (1-\delta) l \delta n < 2\delta ln = 2lk$. Also we know that $X^{(1)}, X^{(2)}, \ldots, X^{(l)}$ are independent because $X$ is a bit-fixing source. We regard each $\Ext(X^{(i)})$ as a random element (coefficients of the corresponding polynomial) in $\mathbb{F}_{2^{m}}$. By Lemma \ref{xorreduceerror}, $\Ext'(X)$ is $\epsilon^{\Theta(l)}$-close to uniform.

\end{proof}

By using the deterministic extractor in Theorem \ref{basicDExt}, combining with Theorem \ref{erreductionDExt}, adjusting the parameters, we get the following result.

\begin{theorem}
\label{smallerrDExt}
For any constant $a, c \in \mathbb{N}$ , there exists an explicit deterministic $(k  = \Theta(n/\log^a n), \epsilon = 2^{-\log^c n})$-extractor $\Ext:\{0,1\}^{n} \rightarrow \{0,1\}^m$ that can be computed by $\AC^0$ circuits of depth $ \Theta(\frac{\log m}{\log \log n} + a + c ) $, for any $(n, k)$-bit-fixing sources, where $m$ can be any $ \poly \log n$.
\end{theorem}

Finally we do output length optimization by applying the same technique as that in \cite{goldreich2015randomness}.  The technique is   given by Gabizon et al.\cite{GabizonRS04}.

\begin{theorem}

For  any constant $a, c \in \mathbb{N}$  and any constant $\gamma \in (0,1]$, there exists an explicit deterministic $(k  = \Theta(n/\log^a n), \epsilon = 2^{-\log^c n})$-extractor $\Ext:\{0,1\}^{n} \rightarrow \{0,1\}^{(1-\gamma)k}$ that can be computed by $\AC^0$ circuits of depth $\Theta(a + c +1)  $, for any $(n, k)$-bit-fixing sources. 
\end{theorem}

\begin{proof}[proof sketch]

The difference between our construction and \cite{goldreich2015randomness} Theorem 5.12 is that, for the three crucial components in the construction, we  use the deterministic extractor of Theorem \ref{smallerrDExt}, the seeded extractor of Theorem \ref{lenthm} and the averaging sampler in Lemma \ref{NC1sampler} instead. We briefly describe the construction as the follows.

\begin{itemize}

\item A deterministic $\epsilon_1$-error extractor $\Ext_1 :\{0,1\}^n \rightarrow \{0,1\}^{r+r_2}$ for $(n, \mu's)$-bit-fixing sources, by Theorem \ref{smallerrDExt};

\item A seeded $\epsilon_2$-error extractor $\Ext_2:\{0,1\}^n \times \{0,1\}^{r_2} \rightarrow \{0,1\}^m$ for $(n, \mu n-s)$-bit-fixing sources, by Theorem \ref{lenthm};

\item An $(\mu, \mu', \theta)$-averaging sampler $\Samp:\{0,1\}^r \rightarrow [n]^s$, by Lemma \ref{smallerersamp} .

\end{itemize}

We set $\mu = k/n, \mu' = \mu/2, s = k/2$ such that $\mu's = \Theta(n/\log^{2a} n)$ and $\mu n - s=  \Theta(n/\log^a n)$. Also we set $\epsilon_1 =  2^{-\log^{3c} n} $, $\epsilon_2 = \theta = 2^{-\log^{2c} n}$, $m = (1-\gamma)k$, $r_2 = (\log n)^{\Theta(a+c)}$ and $r = \Theta(\log^{2c} n)$.

Theorem 7.1 of \cite{GabizonRS04} constructs a deterministic extractor $\Ext:\{0,1\}^n \rightarrow \{0,1\}^m$ for $(n, \mu n)$-bit-fixing sources, where the error is $\epsilon = \epsilon_2+ 2^{r+3} \epsilon_1 + 3\theta$. So $\epsilon \leq 2^{-\log^c n}$. The construction is $\Ext(x) = \Ext_2(x_{[n] \backslash S(Z_1)}\circ 0^t, Z_2)$, where $Z_1$ is the first $r$ bits of $\Ext_1(X)$ and $Z_2$ is the last $r_2$ bits of  $\Ext_1(X)$.

%

Here we only need to compute the depth of the circuit.
We know that $r + r_2 = (\log n)^{\Theta(a+c)}$. The depth  of the final extractor is the sum of the depths of all three components. By Theorem \ref{smallerrDExt}, the depth for the deterministic extractor is $\Theta(a+c+1) $ . By theorem \ref{lenthm}, the depth for the seeded extractor is also ${\Theta(a+c+1)}$. By Lemma \ref{smallerersamp}, the depth for the sampler is 
$\Theta(c+1)$. So the overall depth is $\Theta(a+c+1)$.

\end{proof}

%% file: ExpanderExt.tex
\section{Randomness Condenser with Small Locality}\label{sec:localext}

In this section, we consider constructions of extractor families with small locality by first constructing a condenser family with small locality. Our condenser will be based on random walks on expander graphs and pseudorandom generators for space bounded computation. 

\subsection{Basic Construction}

We give a condense-then-extract procedure by first constructing a randomness condenser.

\begin{theorem}[Hitting Property of Random Walks \cite{jukna2011extremal} Theorem 23.6]
\label{RW}
Let $G = (V, E)$ be a $d$-regular graph with $\lambda(G) = \lambda$. Let $B \subseteq V$ such that $|B| = \beta |V|$. Let $(B,t) $ be the event that a random walk of length $t$ stays in $B$. Then $\Pr[(B,t)]\leq (\beta + \lambda)^t$.

\end{theorem}

\begin{lemma}[\cite{arora2009computational} implicit]
\label{constrExpanderG}
For every $n\in \mathbb{N}$ and for every $0<\alpha<1$,
there is an explicit construction of d-regular Graphs $G_n$ which have the following properties.
\begin{enumerate}
\item $\lambda \leq \alpha$.
\item $G_n$ has $n$ vertices.
\item $d$ is a constant.
\item There exists a $\poly(\log n)$-time algorithm that given a label of a vertex $v$ in  $G_n$ and an index $i\in d$, output the $i$th neighbor of $v$ in $G_n$. 
\end{enumerate}
\end{lemma}

\begin{lemma}[\cite{skorski2015shannon}]
\label{entropytrans}
Let $H_2(X) = \log ( 1/\coll(X))$, $ \coll(X) = \Pr_{X_1, X_2}[X_1 = X_2]$, where $X_1, X_2$ are independent random variables having the same  distribution as $X$.

For any random variable $X$,
$$ 2H_{\infty}(X) \geq H_2(X).$$
\end{lemma}

Recall that $w(\cdot)$ denotes the weight of the input string as we defined in Section 2.
\begin{lemma}
\label{weightrange}
Given any $(n, k)$-source $X$ and any string $x\in \{0,1\}^n$, with probability $1- 2^{-0.5k}$, $w(X \oplus x) \geq k/(c_1\log n)$ for any constant $c_1\geq 2$; with probability $1- 2^{-0.5k}$, $w(X \oplus x) \leq n - k/(c_1\log n)$ for any constant $c_1\geq 2$.
\end{lemma}

\begin{proof}
The number of strings which have $i$ digits different from $x$ is $n \choose i$. So the number of strings which have at most $l = k/(c_1\log n)$ digits different from $x$ is at most $\sum_{i=0}^{l} {n\choose i} \leq (\frac{en}{l})^l \leq 2^{0.5k}$ for any constant $c_1\geq 2$. So with probability at least $1- 2^{-0.5k}$, $w(X \oplus x) \geq l$.

Also as $\sum_{i= n-l}^{n}{n\choose i} = \sum_{i= 0}^{l}{n\choose i} $, with probability $1- 2^{-0.5k}$, $w(X \oplus x) \leq n - l$.
\end{proof}

\begin{lemma}
\label{simulation}

Consider a random vector $v\in \{0,1\}^n$ where $v_1,\ldots, v_n$ are independent random bits and $\forall i, \Pr[v_i = 1] = p = 1/\poly(n)$. For any $\epsilon = 1/\poly(n)$, there is an explicit  function $f:\{0,1\}^{l} \rightarrow \{0,1\}^n$ where $l = O(n\log n)$, such that 
$$\forall i\in [n], |\Pr[f(U)_i = 1] - p| \leq \epsilon$$
where $U$ is the uniform distribution of length $l$. 

There exists an algorithm $A$ which runs in $O(\log n)$ space and can compute $f(s)_i$, given input $s \in \{0,1\}^l$ and $i\in [n]$. 

\end{lemma}

\begin{proof}

We give the algorithm $A$ which runs in $O(\log n)$ space and can compute $f(s)_i$, given input $s \in \{0,1\}^l$ and $i\in [n]$. 

Assume the binary expression of $p$ is $0.b_1b_2b_3\ldots$. The algorithm $A$ is as follows. Intuitively, $A$ divides $s$ into $n$ blocks and uses the $i$th block to generate a bit which simulates $v_i$ roughly according to the probability $p$.

\begin{enumerate}

\item
Assume that $s$ has  $ n $ blocks. The $i$th block is $s_i$ where $|s_i| = t = c\log n$ for some constant $c$. Let $j=1$.
\item If $j = t + 1$, go to step 3.
If $s_{i,j} < b_j$, then set $f(s)_i = 1$ and stop; if $s_{i,j} > b_j$, set $f(s)_i = 0$ and stop; if $s_{i,j} = b_j$, set $j = j+1$ and go to step 2.  
\item
Set $f(s)_i = 1$ and stop.

\end{enumerate}

For every $i$, the probability 
$$\Pr[f(s)_i = 1 ] = \Pr[0.s_{i,1}\ldots s_{i,t} \leq 0.b_1b_2\ldots b_{t}]  = 0.b_1b_2\ldots b_{t}.$$ 
As a result,
$$\forall i\in [n], |\Pr[f(s)_{i} = 1] - \Pr[v_i = 1]|  \leq 0.00\ldots b_{t+1}b_{t+2}\ldots \leq 2^{-t} = 2^{-c\log n} .$$

For any $\epsilon = 1/\poly(n)$, if   $c$ is large enough, then  $2^{-t} = 2^{-c\log n} \leq \epsilon$.

The input length for $f$ is $l = n\times t = O(n\log n)$. 

As all the iterators and variables in $A$ only need $O(\log n)$ space, $A$ runs in space $O(\log n)$ . This proves our conclusion.

\end{proof}

\begin{construction}
\label{expanderCond}
For any $ k =  \Omega(\log^2 n)$, we construct an $(n,k, t = 10k, 0.1k,  \epsilon_c = 2^{-0.1k})$-condenser $\Cond:\{0,1\}^n \times \{0,1\}^d$ $\rightarrow \{0,1\}^t$ with $d = O(n\log n)$ and locality $ c = n/l, l = \frac{k}{2\log n}$.

\begin{enumerate}

\item Construct an expander graph $G  = (V, E)$ where $V = \{0,1 \}^{r_0=O(n\log n)}$ and $\lambda = 0.01$.

\item Use a uniform random string $U_1$ of length $r_0$ to select a vertex $v_1$ of $V$. 

\item Take a random walk on $G$ starting from $v_1$ to get $v_2, \ldots, v_t$ for $t = 10k$. 

\item For $i\in [t]$, get an $n$-bit string $v'_i = f(v_i)$ such that $\forall j\in [n], \Pr[v'_{i,j} = 1] - 1/l| \leq 1/n^2$, where $f:\{0,1\}^{r_0} \rightarrow \{0,1\}^n$ follows from Lemma \ref{simulation}, $r_0 = O(n\log n)$.

\item Let $ M = (v'_1, \ldots, v'_t)^T$.

\item Let $\Cond(x, u) =  Mx$. 

\end{enumerate}

\end{construction}

Let $V' = \{0,1\}^n$.
Let $B_i = \{ v\in V': w(v) = i\}, i\in [d]$.
Let $A_{x,j} = \{ v \in B_j: \langle v, x\rangle = 0  \}$. 

For any $x\in \{0,1\}^n$, let $V_x = \{v\in V: \langle f(v),x\rangle = 0\}$. Let $T = \{v\in V: w(f(v)) \in [0.8c, 1.2c]\}$.

\begin{lemma}
\label{weightconcentrate}
In Construction \ref{expanderCond}, with probability $1- 2\exp\{ -\Theta(c ) \} $, $w(v'_1) \in [0.8c, 1.2 c]$. That is, 
$$  \frac{|T|}{|V|} \geq 1- 2\exp\{ -\Theta(c ) \} .$$

\end{lemma}

\begin{proof}

According to our construction, $\forall i\in [n], \Pr[v'_{1,i}=1] \in [ 1/l - 1/n^2, 1/l+ 1/n^2]$ . As a result, $\mathbf{E}w(v'_1)  \in [n/l- 1/n, n/l+ 1/n ] = [ c - 1/n, c+ 1/n]$. According to Chernoff Bound, we know that $\Pr[w(v'_1) \in [0.8c, 1.2 c]] \geq \Pr[w(v'_1) \in [0.9\mathbf{E}w(v'_1), 1.1 \mathbf{E}w(v'_1)]] \geq 1- 2\exp\{ -\Theta(c ) \}$.

\end{proof}

\begin{lemma}
\label{expandermainlem}
In construction \ref{expanderCond},
we have the following conclusions.
\begin{enumerate}
\item For any $x\in \{0,1\}^n$ with $w(x) \in [l, n-l]$, for any integer $j \in [0.8c, 1.2c]$,  $\beta_{x, j} = |A_{x,j}|/|B_j| \leq   5/6$.
\item For any $x\in \{0,1\}^n$ with $w(x) \in [l, n-l]$, $|V_x|/|V| \leq 5/6 + 1/\poly(n)$.
\item $\Pr[MX_1 = MX_2] = p_0 \leq 2^{-0.5 k+1} + (5/6  + 1/\poly(n) + \lambda)^t$ where $X_1, X_2$ are independent random variables that both have the same distribution as $X$.

\item  $U_d\circ \Cond(X, U_d)$ is $p_0^{1/3}$-close to $U_d \circ W$. Here $U_d$ is a uniform distribution of length $d$. For every $u$, $W|_{U_d = u}$ has entropy $\frac{1}{3}\log \frac{1}{p_0} $.
\item $U_d\circ \Cond(X, U_d)$ is $\epsilon_c= 2^{-0.1k}$-close to $U_d \circ W$ where $\forall u \in \{0,1\}^d$, $W|_{U_d = u}$ has entropy $d + 0.1k$.

\end{enumerate} 

\end{lemma}
\begin{proof}

We know that
$$\beta_{x,j} = \frac{1}{2} (1 + \frac{\sum_{i=0}^{\min\{w(x),j\}} {w(x)\choose i}{n-w(x) \choose j-i} (-1)^i }{ {n \choose j}}).$$
Because $\langle v, x\rangle = 0$ happens if and only if $|\{i\in [n]: x_i = v_i = 1\}|$ is even.

Let $\Delta = \sum_{i=0}^{\min\{w(x),j\}} {w(x)\choose i}{n-w(x) \choose j-i} (-1)^i $.

Consider the series $\Delta_i = {w(x)\choose i}{n-w(x) \choose j-i}, i = 0, 1, \ldots, \min\{w(x), j\}$. Let $\Delta_i \geq \Delta_{i+1}$, we can get
$$  i \geq \frac{jw(x) - n + w(x) + j -1}{n+2} \in [\frac{0.8w(x)}{l} -3, \frac{1.2w(x)}{l} ] $$
Let $\Delta_{i-1} \leq \Delta_i$, we can get
$$ i \leq \frac{jw(x) + w(x) + j + 1}{n-2} \in [\frac{0.8w(x)}{l} , \frac{1.2w(x)}{l} + 3].$$

So series $\Delta_i, i = 0, 1, \ldots, \min\{w(x), j\}$ has its maximum for some $i\in [ \frac{0.8w(x)}{l} - 3 , \frac{1.2w(x)}{l} + 3]$.

To make it simpler, first we consider the situation that $0 < \frac{w(x)}{l} - 3$ and $ j > \frac{w(x)}{l} + 3 $.

Let $i' = \arg \max_i\{\Delta_i\}$ which is in $ [ \frac{w(x)}{l} - 3 , \frac{w(x)}{l} + 3]$. Consider $\Delta_{i'} / \Delta_{i'-1}$. Let $i' = \frac{\theta w(x)}{l} + \delta$ for some $\theta\in [0.8, 1.2], \delta \in [-3,3]$.

\begin{equation}
\begin{split}
\frac{\Delta_{i'}}{\Delta_{i'-1}} &= \frac{w(x) - i' + 1}{i'} \cdot \frac{j-i'+1}{n-w(x) - j +i'}\\
&= \frac{w(x) - \frac{\theta }{l}w(x) - \delta + 1}{\frac{\theta}{l}w(x) + \delta} \cdot \frac{ j - \frac{\theta}{l}w(x)-\delta + 1 }{ n-w(x) - j + \frac{\theta}{l}w(x) + \delta }\\
\end{split}
\end{equation}

The first term 
$$ \frac{w(x) - \frac{\theta }{l}w(x) - \delta + 1}{\frac{\theta}{l}w(x) + \delta} = l+O(1). $$
As  $j\in [0.8c , 1.2 c]$, we know
$$\frac{ n-w(x) - j + \frac{\theta}{l}w(x) + \delta }{ j - \frac{\theta}{l}w(x)-\delta + 1 } \geq \frac{ n-w(x) - 0.8c + \frac{\theta}{l}w(x) + 3 } { 0.8c - \frac{\theta}{l}w(x)-2 }\geq \frac{5l}{6}$$

So $\frac{\Delta_{i'}}{\Delta_{i'-1}} \leq 2$.
 
As ${n \choose c}\geq \Delta_{i'}+ \Delta_{i'-1}$, we know $\frac{\Delta_{i'}}{{n \choose c}} \leq 2/3$. Thus $\beta_x \leq 1/2(1 + \frac{\Delta_{i'}}{{n \choose c}} ) \leq 5/6$. 

If either $0$ or $j$ is in $[ \frac{w(x)}{l} - 3 , \frac{w(x)}{l} + 3]$, to prove our conclusion, we only need to check the situation that $i' = 0$ and $i' = j$.

If $i' = 0 $ or $ j$ then,
$$\beta_{x,j} \leq \frac{1}{2}(1 + \frac{{n-l \choose j}}{{n\choose j}}) \leq  \frac{1}{2}(1 + (1 - \frac{l}{n})^j) \leq  \frac{1}{2}(1 + (1 - \frac{l}{n})^{0.8n/l}) \leq 3/4. $$


For the second assertion, let's consider the expander graph $G = (V, E)$. Assume $v$ is a random node uniformly drawn from $V$. For any $i\in [n]$, the conditional random variable  $f(v)|_{w(f(v)) = i}$ is uniformly distributed on $B_i$. This is because $v$ is uniform, thus $v'_j, j\in [n]$ are independently identically distributed according to Lemma \ref{simulation}. So $ \Pr[f(v)|_{w(f(v)) = i} = v'], v'\in  B_i $ are all equal. 

According to the Union Bound,
\begin{equation}
\begin{split}
|V_x|/|V| &\leq (1-\Pr[w(f(v))\in [0.8c, 1.2c]]) + \sum_{i\in [0.8c, 1.2c]} \Pr[w(f(v)) = i] \frac{|A_{x, i}|}{|B_i|} \\
&\leq (1-\Pr[w(f(v))\in 0.8c, 1.2c]) +    \sum_{i\in [0.8c, 1.2c]} \Pr[w(f(v)) = i]\times 5/6 \\
&\leq  2\exp\{ -\Theta(c ) \} + 5/6.
\end{split}
\end{equation}
Our assertion follows as $c = n/l \geq 2 \log n$.

For the 3rd assertion, 
let's consider $\Pr[MX = 0]$ when $ l \leq w(X) \leq n-l$.

By Theorem \ref{RW}, for any $x$ such that $ l \leq w(x) \leq n-l$, $\Pr[Mx = \mathbf{0}] \leq (\frac{|V_x|}{|V|} + \lambda)^t $. Here $\frac{|V_x|}{|V|} \leq 5/6 + 1/\poly(n)$. 

Let $X_1, X_2$ be  independent random variables and have the same distribution as $X$.

$$ p_0 =   \Pr_{X_1, X_2}[MX_1 = MX_2] 
=   \sum_{x_2 \in \supp(X_2)} \Pr[X_2 = x_2] \cdot \Pr[MX_1 = Mx_2 ] $$

For any fixed $x_2\in \supp(X_2)$, let $X' = X_1 \oplus x_2$. So $\Pr[MX_1 = Mx_2] = \Pr[MX' =  \mathbf{0}]$. We know that $X'$ is also an $(n, k)$-source. 
As a result, we have the following.
\begin{equation}
\begin{split}
&Pr[M(X') = 0]   \\
\leq & \Pr[w(X') \notin [l, n-l]] + \Pr[w(X') \in [l, n-l]] \times \Pr[MX' = \mathbf{0}|_{w(X') \in [l, n-l]}] \\
\leq & \Pr[w(X') \notin [l, n-l]] +  \Pr[MX' = \mathbf{0}|_{w(X') \in [l, n-l]}] \\
\leq & \Pr[w(X') \notin [l, n-l]] + (\frac{|V_x|}{|V|} + \lambda)^t \\
\leq & \Pr[w(X') \notin [l, n-l]] + (5/6  + 1/\poly(n) + \lambda)^t\\
\leq &  2\times 2^{-0.5 k} + (5/6 + 1/\poly(n) + \lambda)^t.
\end{split}
\end{equation}

Thus, 
\begin{equation}
\begin{split}
p_0 = & \sum_{x_2 \in \supp(X_2)} \Pr[X_2 = x_2] \cdot  \Pr[M(X_1\oplus x_2) = \mathbf{0} ] \\
\leq &  2^{-0.5 k+1} + (5/6 + 1/\poly(n) + \lambda)^t.\\
\end{split}
\end{equation}

For the 4th conclusion, let's fix a $M_u \in \supp(M)$. We consider $ H_2(M_uX) $ as the following. 
\begin{equation}
\begin{split}
H_2(M_u X) &= -\log \Pr[M_uX_1 = M_uX_2]\\
& = -\log \sum_{x_1,x_2 \in \supp(X)} \Pr[X_1 = x_1] \Pr[X_2 = x_2] I_{M_u x_1 = M_u x_2}\\
\end{split}
\end{equation}
Here $I_e$ is the indicator function  such that $I_e = 1$ if and only if the event $e$ happens. Here for each $x_1, x_2$, $I_{M_u x_1 = M_u x_2}$ is a fixed value (either 0 or 1, not a random variable because $M_u$ is fixed).
 
Next let's consider $M$ to be a random variable generated by the seed $U_d$.

Let $Z_M =  \sum_{x_1,x_2 \in \supp(X)} \Pr[X_1 = x_1]\Pr[X_2 = x_2] I_{Mx_1 = Mx_2}$.
We know that 
$$\mathbf{E} Z_M =   \sum_{x_1,x_2 \in \supp(X)} \Pr[X_1 = x_1] \Pr[X_2 = x_2] \Pr[Mx_1 = Mx_2]     = p_0.$$ 

So according to Markov's Inequality, 
$$\Pr[Z_M \geq p_0^{2/3}] \leq p_0^{1/3}.$$ 
So with probability at least $1 -  p_0^{1/3}$ over $M$ (over $U_d$), $Z_M \leq p_0^{2/3}$.

Let's fix $M_u \in \supp(M)$ such that  $Z_{M_u} \leq p_0^{2/3}$.

Thus 
\begin{equation}
\begin{split}
H_{\infty}(M_uX) &\geq 1/2H_2(M_uX)\\
& = 1/2 (-\log (Z_{M_u})) \\
& \geq  -1/2\log(p_0^{2/3})\\
& = \frac{1}{3} \log \frac{1}{p_0}.
\end{split}
\end{equation}
 
This concludes that $U_d\circ \Cond(X, U_d) = U_d \circ MX$ is $p_0^{1/3}$-close to $U_d \circ W$ where for every $u$, $W|_{U_d = u}$ has entropy $1/3\log \frac{1}{p_0}$.

As we know  $p_0\leq 2^{-0.5k+1} + (5/6 + 1/\poly(n) + \lambda)^t$, $k = \Omega(\log^2 n)$. So if $t = 10k$, then $p_0^{1/3} \leq  2^{-0.1k}$.

According to conclusion 5, 
$U_d\circ \Cond(X, U_d)$ is $\epsilon_c$-close to $U_d \circ W$ where for every $u$, $W|_{U_d = u}$ has entropy at least $0.1k$ where $\epsilon_c = 2^{-0.1k}$. This proves the last assertion.
 
\end{proof}

\subsection{Seed Length Reduction}

The seed length can be shorter by applying the following PRG technique.
\begin{theorem}[Space Bounded PRG\cite{Nisan92}]
\label{spacePRG}
For any $s>0$, $ 0< n \leq 2^s$,
there exists an explicit PRG $g: \{0,1\}^r \rightarrow \{0,1\}^n$, such that for any algorithm $A$ using space $s$,
$$ |\Pr[A(g(U_r)) = 1] - \Pr[A(U_n) = 1]| \leq 2^{-s}.$$
Here $r  = O(s\log n)$, $U_r$ is uniform over $\{0,1\}^r$, $U_n$ is uniform over $\{0,1\}^n$.

\end{theorem}

\begin{lemma}
\label{PRGgraph}
Let $g:\{0,1\}^{r= O(\log^2 n)} \rightarrow \{0,1\}^{r_0 = O(n\log n)}$ be the PRG from Lemma \ref{spacePRG} with error parameter $\epsilon_g = 1/\poly(n)$.
Replace the 1-3 steps of Construction \ref{expanderCond} with the following 3 steps.
\begin{enumerate}
\setcounter{enumi}{0}
\item Construct an expander graph $\tilde{G}  = (\tilde{V}, \tilde{E})$ where $\tilde{V} = \{0,1\}^r$ and $ \lambda = 0.01 $.  
\item Use a uniform random string $U_1$ of length $r$ to select a vertex $\tilde{v}_1$ of $\tilde{V}$.
\item Take a random walk on $\tilde{G}$ starting from $\tilde{v}_1$ to get $\tilde{v}_2,\ldots, \tilde{v}_t$, for $t = 10k$ .
Let $v_i = g(\tilde{v}_i), i = 1,\ldots, t$.
\end{enumerate}
Let $\tilde{V}_x = \{ v\in \tilde{V}:  \langle f(g(v)),x \rangle = 0 \} $.
We have the following conclusions.

\begin{enumerate}
\item For any $x\in \{0,1\}^n$ such that $w(x) \in [l, n-l]$, 
$$ |\frac{|\tilde{V}_x|}{|\tilde{V}|} -  \frac{|V_x|}{|V|}| \leq \epsilon_g .$$
\item $p_0 = \Pr[MX_1 = MX_2] \leq 2^{-0.5 k+1} + (5/6  + 1/\poly(n) + \lambda)^t$ where $X_1, X_2$ are independent random variables that both have the same distribution as $X$.

\item  $U_d\circ \Cond(X, U_d)$ is $p_0^{1/3}$-close to $U_d \circ W$. Here $U_d$ is a uniform distribution of length $d$. For every $u$, $W|_{U_d = u}$ has entropy $\frac{1}{3}\log \frac{1}{p_0} $.
\item $U_d\circ \Cond(X, U_d)$ is $\epsilon_c= 2^{-0.1k}$-close to $U_d \circ W$ where $\forall u \in \{0,1\}^d$, $W|_{U_d = u}$ has entropy $d + 0.1k$.
\end{enumerate}

\end{lemma}

\begin{proof}

Consider the following algorithm $A$ which decides whether $v \in V_x$ on input $(v, x)$.

\begin{algorithm}[H]
\SetKwInOut{Input}{Input}
\SetAlgoLined
\Input{$v\in \{0,1\}^{r_0}$ and $x\in \{0,1\}^n$}
res = 0 \;
\For{$i= 1$ to $n$}
{
	compute $f(v)_i$ \;
	\If{$f(v)_i = 1$ }
	{
		res = res + $f(v)_i \cdot x_i$ \;
	}
}
\lIf{ $res = 0$}
{ Output 1
}
\lElse
{
Output 0
}
\caption{$A(v, x)$}
\end{algorithm}

We can see that algorithm $A$ runs in space $O(\log n)$ because $f(v)_i, i=1,\ldots, n$ can be computed sequentially by using $O(\log n)$ space according to Lemma \ref{simulation} and all the other variables need $O(\log n)$ space to record. Also $A(v, x) =  1$ if and only if $v\in V_x$. So $\Pr[A(U_{r_0}, x) = 1] = \frac{|V_x|}{|V|}$. 

Similarly, we can see $\Pr[ A(g(U_r), x) = 1] = \frac{|\tilde{V}_x|}{|\tilde{V}|}$.

According to the definition of our PRG $g$, we know that,
$$ |\Pr[ A(g(U_r), x) = 1] - \Pr[A(U_{r_0}, x) = 1] | \leq \epsilon_g. $$
So
$$ |\frac{|\tilde{V}_x|}{|\tilde{V}|} -  \frac{|V_x|}{|V|}| \leq \epsilon_g .$$

For the 2nd assertion, let's consider $\Pr[MX = 0]$ when $ l \leq w(X) \leq n-l$.

By Theorem \ref{RW}, for any $x$ such that $ l \leq w(x) \leq n-l$, $\Pr[Mx = \mathbf{0}] \leq (\frac{|\tilde{V}_x|}{|\tilde{V}|}  + \lambda)^t $.

Let $X_1, X_2$ be independent random variables and have the same distribution as $X$.

$$p_0
=   \sum_{x_2 \in \supp(X_2)} \Pr[X_2 = x_2] \cdot \Pr[MX_1 = Mx_2 ] $$

For any fixed $x_2\in \supp(X_2)$, let $X' = X_1 \oplus x_2$. So $\Pr[MX_1 = Mx_2] = \Pr[MX' =  \mathbf{0}]$. We know that $X'$ is also an $(n, k)$-source. 
As a result, we have the following.
\begin{equation}
\begin{split}
&\Pr[M(X') = 0]   \\
\leq & \Pr[w(X') \notin [l, n-l]] + \Pr[w(X') \in [l, n-l]] \times \Pr[MX' = \mathbf{0}|_{w(X') \in [l, n-l]}] \\
\leq & \Pr[w(X') \notin [l, n-l]] +  \Pr[MX' = \mathbf{0}|_{w(X') \in [l, n-l]}] \\
\leq & \Pr[w(X') \notin [l, n-l]] + (\frac{|\tilde{V}_x|}{|\tilde{V}|} + \lambda)^t \\
\leq & \Pr[w(X') \notin [l, n-l]] + (\frac{|V_x|}{|V|} + \epsilon_g + \lambda)^t \\
\leq & \Pr[w(X') \notin [l, n-l]] + (5/6  + 1/\poly(n) + \lambda)^t\\
\leq &  2\times 2^{-0.5 k} + (5/6 + 1/\poly(n) + \lambda)^t.
\end{split}
\end{equation}

Thus, 
\begin{equation}
\begin{split}
p_0 = & \sum_{x_2 \in \supp(X_2)} \Pr[X_2 = x_2] \cdot  \Pr[M(X_1\oplus x_2) = \mathbf{0} ] \\
\leq &  2^{-0.5 k+1} + (5/6 + 1/\poly(n) + \lambda)^t.\\
\end{split}
\end{equation}

Conclusion 3 and 4 follow the same proof as that of Lemma \ref{expandermainlem}. 

\end{proof}

\begin{theorem}
\label{condenserlemma1}
For any  $k = \Omega(\log^2 n)$, there exists an explicit construction of an $(n, k, 10k, 0.1k,2^{-0.1k})$-condenser with seed length $\Theta(k)$. 

\end{theorem}

\begin{proof}
According to Lemma \ref{PRGgraph}, it immediately follows that the function $\Cond$ in Construction \ref{expanderCond} is an $(n, k, t, 0.1k, \epsilon_c)$-condenser for $t = 10k$, $\epsilon_c = 2^{-0.1k}$. 

Now consider the seed length. We know that $|U_1| = \Theta(\log^2 n)$. For the random walks, the random bits needed have length $\Theta(t) = \Theta(k)$. So the seed length is $|U_1| +   \Theta(t) =\Theta(k)$.
\end{proof}

\subsection{Locality Control}

There is one problem left in our construction. We want the locality of our extractor to be small. However, in the current construction, we cannot guarantee that the locality is small, because the random walk may hit some vectors that have large weights. We bypass this barrier by setting these vectors to be $\mathbf{0}$.

We need the following Chernoff Bound for random walks on expander graphs.
\begin{theorem}[\cite{healy2008randomness} Theorem 1]
\label{ExpanderChernoff}
Let $G$ be
a regular graph with $N$ vertices where the second largest eigenvalue is $\lambda$.
For every $i\in [t]$, let $f_i: [N] \rightarrow [0,1]$ be any function. Consider a random walk $v_1, v_2, \ldots, v_t$ in $G$ from a uniform start-vertex $v_1$. Then for any $\epsilon > 0$,
$$ \Pr[|\sum_{i=1}^t f_i(v_i)- \sum_{i=1}^t \mathbf{E}f_i| \geq \epsilon t] \leq 2e^{-\frac{\epsilon^2(1-\lambda)t}{4}}.$$

\end{theorem}

Now we give our final construction.

\begin{construction}
\label{expanderCondfinal}
For any $ k =  \Omega(\log^2 n)$, we construct an $(n, k, t = 10k, 0.08k,  \epsilon_c)$-condenser $\Cond:\{0,1\}^n \times \{0,1\}^d$ $\rightarrow \{0,1\}^t$ with $d = \Theta(k)$ , locality $ c = n/l, l = \frac{k}{2\log n}$, $\epsilon_c = 2^{-k/500000}$.

Let $g:\{0,1\}^{r= O(\log^2 n)} \rightarrow \{0,1\}^{r_0 = O(n\log n)}$ be the PRG from Lemma \ref{spacePRG} with error parameter $\epsilon_g = 1/\poly(n)$.
\begin{enumerate}
 
\item Construct an expander graph $\tilde{G}  = (\tilde{V}, \tilde{E})$ where $\tilde{V} = \{0,1\}^r$ and $\lambda(\tilde{G}) = 0.01$ where $r = O(\log^2 n)$.  
\item Using a uniform random string $U_1$ of length $r$ to select a vertex $\tilde{v}_1$ of $\tilde{V}$.
\item Take a random walk on $\tilde{G}$ to get $\tilde{v}_2,\ldots, \tilde{v}_t$.
Let $v_i = g(\tilde{v}_i), i = 1,\ldots, t$.

\item For $i\in [t]$, get an $n$-bit string $v'_i = f(v_i)$ such that $\forall j\in [n], \Pr[v'_{i,j} = 1] - 1/l| \leq 1/n^2$, where $f:\{0,1\}^{r_0} \rightarrow \{0,1\}^n$ follows from Lemma \ref{simulation}, $r_0 = O(n\log n)$. 

\item Let $ M = ( v'_1, \ldots, v'_t)^T$.

\item Construct the matrix $M' = (\bar{v}_1,\ldots, \bar{v}_t)$ such that for $i\in [t]$, if  $w(v'_i) > 1.2c $,  $\bar{v}_i = \mathbf{0}$, otherwise $\bar{v}_i = v'_i$.

\item Let $\Cond(x, u) = M'x$. 

\end{enumerate}

\end{construction}

Let $ \tilde{T} = \{ v\in \tilde{V}: w(f(g(v))) \in [0.8c, 1.2c] \}$. 

\begin{lemma}
\label{localitylemma}
In Construction \ref{expanderCondfinal},
with probability $1-2e^{-k/500000}$, 
$$ |\{i: w(v'_i) \in [0.8c, 1.2 c] \}| \geq 0.998t .$$

\end{lemma}

\begin{proof}

Consider the following algorithm $A$. Given $v \in \{0,1\}^{r_0}$, $A$ tests whether $w(f(v)) \in [0.8c, 1.2 c]$. 

\begin{algorithm}[H]
\SetKwInOut{Input}{Input}
\SetAlgoLined
\Input{$v\in \{0,1\}^{r_0}$}
count = 0 \;
\For{$i= 1$ to $n$}
{
	compute $f(v)_i$ \;
	\If{$f(v)_i = 1$ }
	{
		count++\;
	}
}
\lIf{count is in $[0.8c, 1.2 c]$ }
{ Output 1
}
\lElse
{
Output 0
}
\caption{$A(v)$}
\end{algorithm}

It can be seen that $A$ runs in space $O(\log n)$. Because $f(v)_i, i = 1,\ldots, n$ can be computed sequentially using space $O(\log n)$ according to Lemma \ref{simulation}. Also all the iterators and variables used during the computation require only $O(\log n)$ space.

As a result, according to the definition of space bounded PRG, for any $\epsilon_g = 1/\poly(n)$,
$$ |\Pr[A(g(U_r))=1] - \Pr[A(U_{r_0}) = 1] | \leq \epsilon_g .$$

We know that $ \Pr[A(g(U_r))=1] =  \frac{|\tilde{T}|}{|\tilde{V}|} $ and $  \Pr[A(U_{r_0}) = 1]  =  \frac{|T|}{|V|}$.
Thus, $ |\frac{|\tilde{T}|}{|\tilde{V}|} - \frac{|T|}{|V|} | \leq \epsilon_g $.

According to Lemma \ref{weightconcentrate},
$\frac{|T|}{|V|} \geq  1- 2\exp\{ -\Theta(c ) \}$.

As a result, $\frac{|\tilde{T}|}{|\tilde{V}|} \geq  1- 2\exp\{ -\Theta(c ) \}-\epsilon_g \geq 1- 1/\poly(n)$ .

Thus for each $i$, $\Pr[w(f(g(\tilde{v})))\in [0.8c, 1.2c]] \geq  1- 1/\poly(n) $. Let $\mathbf{E}I_{\tilde{v}_i\in \tilde{T}} = u$. Then $u \geq 1- 1/\poly(n)$.

According to our construction, we can assume $I_{\tilde{v}_i \in \tilde{T}} = h(\tilde{v}_i), i = 1,\ldots, t $ for some function $h$.
By Theorem \ref{ExpanderChernoff},
$$ \Pr[|\sum_{i=1}^t I_{\tilde{v}_i\in \tilde{T}} - \sum_{i=1}^t u |\geq 0.001t ] \leq 2e^{-\frac{ (0.001)^2(1-\lambda)t}{4}} \leq 2e^{-t/5000000}$$
 
We know that $t = 10k$ and $|\{i: w(v'_i) \in [0.8c, 1.2 c] \}|  = \sum_{i=1}^t I_{\tilde{v}_i\in \tilde{T}}$. So with probability at least $1-2e^{-k/500000}$, 
$$   |\{i: w(v'_i) \in [0.8c, 1.2 c] \}|  \geq \sum_{i=1}^t u - 0.001t  \geq ( 1- 1/\poly(n))t - 0.001t \geq 0.998t .$$ 

\end{proof}

\begin{lemma}
\label{expanderfinallemma}
The function $\Cond:\{0,1\}^n \times \{0,1\}^d$ in 
Construction \ref{expanderCondfinal} is an $(n, k, t, 0.08k, \epsilon_c)$-condenser with seed length $\Theta(k)$. 

\end{lemma}

\begin{proof}

According to Lemma \ref{condenserlemma1}, we know that for $\epsilon = 2^{-0.1k}$, $U_d \circ MX$ is $\epsilon$-close to $U_d\circ W$ where for every $a\in \{0,1\}^d$, $H_{\infty}(W|_{U_d = a}) = 0.1k$. Let $M'X = h(U_d, MX)$. According to our construction, we know that $h$ is a deterministic function. More specifically, $h(u, y)$ will set the $i$th coordinate of $y$ to be $0$ for any $i$ such that $ \tilde{v_i} \notin \tilde{T} $. The function $h$ can check $\tilde{v_i} \notin \tilde{T} $ according to $u$ deterministically.

As a result, 
\begin{equation}
\begin{split}
&\SD(U_d\circ M'X , U_d\circ h(U_d, W))\\
= &\SD(U_d\circ  h(U_d, MX), U_d\circ h(U_d, W))\\
\leq &\SD(U_d\circ MX, U_d\circ W)\\
\leq &\epsilon.
\end{split}
\end{equation}

Now let's consider the entropy of $ U_d\circ h(U_d, W)$. let $\epsilon_0 = 2e^{-k/500000}$. By Lemma \ref{localitylemma}, for  $1-\epsilon_{0}$ fraction of $u\in \{0,1\}^d$, there are at most $0.002t$ bits in $W|_{U_d=u}$ that are set to be 0. 

As a result, for  $1-\epsilon_{0}$ fraction of $u\in \{0,1\}^d$, $u\circ h(u, W|_{U_d=u})$  has entropy $0.1k - 0.002t  \geq 0.08k$. As a result, $ U_d\circ h(U_d, W)$ is $\epsilon_0$-close to  $U_d\circ W'$ where for every $u\in\{0,1\}^d$, $W'|_{U_d=u}$ has entropy $0.08k$. So $ U_d\circ M'X$ is $\epsilon+\epsilon_0  \leq  2^{-k/500000}$-close to $U_d\circ W'$  where for every $u\in\{0,1\}^d$, $W'|_{U_d=u}$ has entropy $0.08k$.
\end{proof}

\begin{lemma}
\label{expanderlocal}
The locality of Construction \ref{expanderCondfinal} is $1.2c = \Theta(\frac{n}{k}\log n)$.
\end{lemma}

\begin{proof}

As for every $M'_i$, the number of $1$s in it is at most $1.2c$, the locality is $1.2c = 1.2n/l = \Theta(\frac{n}{k}\log n)$

\end{proof}

\begin{theorem}

For any $ k =  \Omega(\log^2 n)$, there exists an $(n, k, t = 10k, 0.08k,  \epsilon_c)$-condenser $\Cond:\{0,1\}^n \times \{0,1\}^d$ $\rightarrow \{0,1\}^t$ with $d = \Theta(k)$ ,  $\epsilon_c = 2^{-k/500000}$ and the locality is $\Theta(\frac{n}{k}\log n)$.

\end{theorem}
\begin{proof}
It follows from Lemma \ref{expanderfinallemma} and Lemma \ref{expanderlocal}.

\end{proof}

\begin{theorem}
\label{localExtfromTHA}
For any $k= \Omega(\log^2 n)$, for any constant $\gamma \in (0,1)$,
there exists a strong $(k , \epsilon)$-extractor $\Ext:\{0,1\}^n \times \{0,1\}^d \rightarrow \{ 0,1 \}^m$, where $\epsilon$ can be as small as $ 2^{-k^{\Omega(1)}}$, $d =  \Theta(k), m = (1-\gamma)k$ and the locality is $\frac{n}{k}\log^2 (1/\epsilon) (\log n) (\log^{ O(1)} k) $.

\end{theorem}

\begin{proof}[Proof Sketch]

We combine our $(n, k, m_c = 10k, 0.08k, \epsilon_c = 2^{-0.1k} )$-condenser $\Cond:$ $\{0,1\}^n \times \{0,1\}^{r}$ $\rightarrow \{0,1\}^{m_c}$ where $r = \Theta(k)$ from Lemma \ref{condenserlemma1} with the $(0.08k, \epsilon_0)$-extractor $\Ext_0: \{0,1\}^{m_c} \times \{0,1\}^{d_0 } \rightarrow \{0,1\}^{m_0}$ from Theorem \ref{poorlenththm1}   for  $\epsilon_c = 2^{-k^{\Omega(1)}}$. 

Let $\Ext(X, U) = \Ext_0(\Cond(X, U_1), U_2)$, where $U = U_1 \circ U_2$. We know that $U\circ\Ext(X,U) $ is $\epsilon = \epsilon_c + \epsilon_0 =  2^{-k^{\Omega(1)}}$-close to uniform distribution over $\{0,1\}^{\Theta(k)}$.

The locality  of $\Cond $ is $\Theta(\frac{n}{k}\log n )$. The locality of $\Ext_0$ is $ \log^2 (1/\epsilon) (\log^{ O(1)} m_c) $. So the overall locality is $\frac{n}{k}\log^2 (1/\epsilon) (\log n) (\log^{ O(1)} k) $. The seed length is $|U| = |U_1|+|U_2| = d_0 + \Theta(k) =  \Theta(k)$ by setting $\epsilon$ to be large enough, since $d_0=O(\frac{k \log (1/\epsilon_c)}{m_c^{1/10800} \log m_c })$.

Our theorem holds by applying the extraction in parallel technique in Lemma \ref{multiext} to increase the output length to $(1-\gamma)k$.
\end{proof}

Since we note that the extractor in  \cite{papakonstantinou2016true} (Supplementary Information Theorem 3) also has small locality, we can also combine our condenser with their extractor to get the following.
\begin{theorem}
\label{localExtfromRRB}
For any constant $\gamma \in (0,1)$, any $k= \Omega(\log^2 n)$,  $\epsilon = 2^{-o(k)}$,
there exists an explicit strong $(k , \epsilon)$-extractor $\Ext:\{0,1\}^n \times \{0,1\}^d \rightarrow \{ 0,1 \}^m$, $d =  \Theta(k), m = (1-\gamma)k$ and the locality is $\frac{n}{k}  (\log n) \log (k/\epsilon)$.

\end{theorem}

\begin{proof}

We combine our $(n, k, m_c = 10k, 0.08k, \epsilon_c = 2^{-0.1k} )$-condenser $\Cond:$ $\{0,1\}^n \times \{0,1\}^{r}$ $\rightarrow \{0,1\}^{m_c}$ where $r = \Theta(k)$ from Lemma \ref{condenserlemma1} with the $(0.08k, \epsilon_0)$-extractor $\Ext_0: \{0,1\}^{m_c} \times \{0,1\}^{d_0 } \rightarrow \{0,1\}^{m_0}$ from   \cite{papakonstantinou2016true} (Supplementary Information Theorem 3), where $  \epsilon_0 = O( \epsilon),  d= O(\log m_c \log (m_c/\epsilon_c)), m_0 = O(m_c)$. 

Let $\Ext(X, U) = \Ext_0(\Cond(X, U_1), U_2)$, where $U = U_1 \circ U_2$, with $U_1$ being uniform over $\{0,1\}^r$, $U_2$ uniform over $\{0,1\}^{d_0}$. We have that $U\circ\Ext(X,U) $ is $\epsilon = \epsilon_c + \epsilon_0 \leq \epsilon$-close to uniform distribution over $\{0,1\}^{\Theta(k)}$, by setting $ \epsilon_0 $ to be small enough.

The locality  of $\Cond $ is $\Theta(\frac{n}{k}\log n )$. The locality of $\Ext_0$ is $\log (m_c/\epsilon_0)$. So the overall locality is  $\frac{n}{k}  (\log n) \log (k/\epsilon)$. The seed length is $|U| = |U_1|+|U_2| = \Theta(k)+  d_0    =  \Theta(k)$.

Our theorem holds by applying the extraction in parallel technique in Lemma \ref{multiext} to increase the output length to $(1-\gamma)k$.
\end{proof}

%% file: Application.tex
\section{Applications}\label{sec:app}
In this section, we give some constructions of PRGs based on our AC0 extractor.

\subsection{PRG in AC0 Based on Random Local One-way Function}

Our first construction is based on random local one-way functions following the method of Applebaum \cite{applebaum2013pseudorandom}.

Let $\dist(\cdot)$ denotes the hamming distance between the two input strings (with equal length).

\begin{definition}[Hypergraphs \cite{applebaum2013pseudorandom}]
\label{hypergraph}
An $(n, m, d)$ hypergraph is a graph over $n$ vertices and $m$ hyperedges each of cardinality $d$. For each hyperedge $S = (i_0, i_1,\ldots, i_{d-1})$, the indices $i_0,i_1,\ldots, i_{d-1}$ are ordered. The hyperedges of $G$ are also ordered. Let $G$ be denoted as $([n], S_0, S_1, \ldots, S_{m-1})$ where for $i = 0, 1,\ldots, m-1$, $S_i$ is a hyperedge.

\end{definition}

\begin{remark}

Here we do not require the indices $i_0, i_1, \ldots, i_{d-1}$ to be distinct. This setting is the same as that in \cite{bogdanov2012security} and \cite{goldreich2011candidate} (Random Construction).

\end{remark}

\begin{definition}[Predicate]

A $d$-ary predicate $Q:\{0,1\}^d \rightarrow \{0,1\}$ is a function which partitions $\{0,1\}^d$ in to $V_0$ and $V_1$, where $V_a = \{w\in \{0,1\}^d | Q(w) = a\}$ for $a = 0,1$. 

Let $H_Q = (V_0\cup V_1, E)$ be a bipartite graph where $(u, v)\in V_0 \times V_1$ is an edge if $\dist(u,v) = 1$. Let $\mathcal{M}$ be all the possible matchings of $H_Q$. The size of the maximum matching of $H_Q$ is 
$$ \Match(Q) = \max_{M \in \mathcal{M}} \Pr_{v}[\exists u, (u,v)\in M \mbox{ or } (v, u) \in M ] = \max_{M \in \mathcal{M} } 2|M|/2^d,$$
where $v$ is uniformly distributed in $V_0\cup V_1$.
\end{definition}
%
%
%
%

\begin{definition}[Collection of Functions]

For $s = s(n), m = m(n)$, a collection of functions $F:\{0,1\}^{s} \times \{0,1\}^n \rightarrow \{0,1\}^m$ takes an input $(k, x)$ and outputs $F(k, x)$. Here $k$ is a public index and $x$ can be viewed as the input for the $k$th function in the collection. We also denote $F(k,x)$ as $F_{k}(x)$ where $F_k$ is the $k$th function in the collection.

\end{definition}
\begin{remark}
For simplicity, we usually consider $n$ as an exponential of $2$.

\end{remark}

In the following paragraph, an efficient adversary is defined to be a probabilistic polynomial time Turing Machine. Also the term efficient means in probabilistic polynomial time.

\begin{definition}[Approximate One-way Function for Collection of Functions]

For $\delta = \delta(n) \in (0,1)$ and $\epsilon = \epsilon(n) \in (0,1)$, a collection of functions $F:\{0,1\}^{s}\times \{0,1\}^n \rightarrow \{0,1\}^m$ is an $(\epsilon,\delta)$-approximate one-way function if for every efficient adversary $A$ which outputs a list of $\poly(n)$ candidates and for sufficiently large $n$'s, we have that
$$ \Pr_{k, x, y = F_k(x)}[\exists z \in A(k, y), z' \in F^{-1}_k(y), \dist(z, z')/n\leq \delta] < \epsilon, $$
where $k$ and $x$ are independent and uniform. Specially, when $\delta = 0$, we say the collection $F$ is $\epsilon$-one-way.
\end{definition}

\begin{definition}[Goldreich's Random Local Function \cite{goldreich2011candidate}]
\label{rlf}

Given a predicate $Q: \{0,1\}^d \rightarrow \{0,1\}$ and an $(n,m,d)$ hypergraph $G = ([n], S_0, \ldots, S_{m-1} )$,
the function $f_{G,Q}: \{0,1\}^n \rightarrow \{0,1\}^m$ is defined as follows: for input $x$, the $i$th output bit of $f_{G, Q}(x)$ is $f_{G, Q}(x)_i = Q(x_{S_i})$. 

For $m = m(n)$, the function collection $F_{Q,n,m}:\{0,1\}^{s} \times \{0,1\}^{n} \rightarrow \{0,1\}^{m}$ is defined via the mapping $(G,x) \rightarrow f_{G, Q}(x)$. 

\end{definition}

\begin{lemma}
\label{RLFinAC0}
For every $ d = O(\log n)$, every $ m = \poly(n) $ and every predicate $Q:\{0,1\}^d \rightarrow \{0,1\}$, the random local function $F_{Q,n, m}$, following Definition \ref{rlf}, is in $\AC^0$.

\end{lemma}

\begin{proof}

For every $i\in [m]$, we claim that the $i$th output bit of $F_{Q,n,m}(G, x)$ can be computed in $\AC^0$. The reason is as follows. We know that $F_{Q, n, m}(G, x)_i = Q(x_{S_{i}})$. So it is determined by $d$ bits of $x$ and $S_i$ which corresponds to $d \log n$ bits of $G$. 
Thus for $S_i = (j_0, j_1, \ldots, j_{d-1})$, $\forall l\in [d]$, the $l$th input for $Q$ is 
$$ x_{j_l} = \bigvee_{k=0}^{n} (I_{j_l = k} \wedge x_k) = \bigwedge_{k=0}^{n} (I_{j_l \neq k} \vee x_k ) .$$
As $|j_l| = \log n$, $I_{j_l = k}$ and $I_{j_l \neq k}$ can be computed in $\AC^0$ by Lemma \ref{xorpolylog}.  So every input bit in $ x_{S_i}$ can be computed in $\AC^0$.
As $d = O(\log n)$, we know that $F_{Q, n, m}(G, x)_i = Q(x_{S_i})$ can be computed in $\AC^0$. Thus $F_{Q,n,m}(G, x)$ can be computed in $\AC^0$.

\end{proof}

\begin{definition}

Two distribution ensembles $Y = \{Y_n\}$ and $Z = \{Z_n\}$ are $\epsilon$-indistinguishable if for every efficient adversary $A$,
$$|\Pr[A(1^n, Y_n) = 1] - \Pr[A(1^n, Z_n) = 1]| \leq \epsilon(n).$$ 

Here the subscript of a random variable indicates its length.

\end{definition}

\begin{definition}[PRG for a Collection of Functions]

Let $m = m(n)$. A collection of functions $F:\{0,1\}^s \times \{0,1\}^n \rightarrow \{0,1\}^m$ is an $\epsilon$-PRG, if $(K, F_K(U_n))$ is $\epsilon$-indistinguishable from the uniform distribution. Here $K$ is uniform over $\{0,1\}^s$, $U_n$ is uniform over $\{0,1\}^n$.

A collection of functions $F:\{0,1\}^s \times \{0,1\}^n \rightarrow \{0,1\}^m$ is $\epsilon$-unpredictable generator (UG) if for every efficient adversary $A$ and every sequence of indices $\{i_n\}_{n\in \mathbb{N}}$ where $i_n \in [m(n)]$, we have that
$$ \Pr_{k\leftarrow U_s, x\leftarrow U_n}[A(k, F_k(x)_{[0,\ldots, i_n-1]}) = F_k(x)_{i_n} ] \leq \epsilon(n)  $$
for sufficiently large $n$'s. Here $F$ is $\epsilon$-last-bit unpredictable generator (LUG) if $i_n = m(n)-1$.

\end{definition}

\begin{remark}

Let $t = t(r)$. A function $G:\{0,1\}^r \rightarrow \{0,1\}^t$ is a classic $\epsilon$-PRG, if $(K, F_K(U_n))$ is $\epsilon$-indistinguishable from the uniform distribution. Here $K$ is uniform over $\{0,1\}^r$, $U_n$ is uniform over $\{0,1\}^n$.

The definition of PRG for a collection of functions implies the classic definition of PRG. Following our definition, if there exists an explicit $\epsilon$-PRG $F(\cdot, \cdot) $ for a collection of functions, we know  $(U_s, F_{U_s}(U_n))$ is $\epsilon$-indistinguishable from uniform distributions. Let $G:\{0,1\}^{r = s+n} \rightarrow \{0,1\}^{t = s+m}$ be such that $\forall k\in \{0,1\}^s, \forall x\in \{0,1\}^n, G(k\circ x) = k\circ F(k, x)$. We know that $G(U_r)$ is indistinguishable from uniform distributions. So $G$ is a classic $\epsilon$-PRG.

\end{remark}

\begin{definition}

An $\epsilon$-LPRG is an $\epsilon$-PRG whose output length is linear of its input length (including the index length, $m > (1+\delta)(n+s)$  for some constant $\delta$).

An $\epsilon$-PPRG is an $\epsilon$-PRG whose output length is a polynomial of its input length (including the index length, $m > (n+s)^{(1+\delta)}$ for some constant $\delta$).

\end{definition}

\begin{lemma}
\label{PRGenlength}
For every $c \in \mathbb{N}^+$, an $\epsilon$-PRG $G:\{0,1\}^r \rightarrow \{0,1\}^t$ in $\AC^0$ can be transformed to an $(c \epsilon)$-PRG $G':\{0,1\}^r \rightarrow \{0,1\}^{t(t/r)^c}$.

Here $G'(\cdot) = G^{(c)}(\cdot)$, where  $G^{(i+1)}(\cdot) = G^{(i)}(\cdot), \forall i \in \mathbb{N}^+$ and $G^{(1)}(\cdot) = G(\cdot)$.

If $c$ is a constant, then $G'$ is in $\AC^0$.

\end{lemma}

\begin{proof}

We use inductions. Assume the the output length for $G^{(i)}$ is $t^{(i)}$.

For the basic step, as $G$ is an $\epsilon$-PRG, $G^{(1)} = G$ is an $\epsilon$-PRG.

For the induction step, assume for $i$, $G^{(i)}$ is an $(i\epsilon)$-PRG. Suppose there exists an efficient adversary $A$ such that
$$ |\Pr[A(G^{(i+1)}(U_r)) = 1] - \Pr[A(U_{t^{(i+1)}}) = 1]| > (i+1)\epsilon. $$  
We know that
\begin{equation}
\begin{split}
& |\Pr[A(G^{(i+1)}(U_r)) = 1] - \Pr[A(U_{t^{(i+1)}}) = 1]| \\
 \leq & |\Pr[A(G^{(i+1)}(U_r)) = 1] - \Pr[A(G(U_{t^{(i)}})) = 1]| +  |\Pr[A(G(U_{t^{(i)}})) = 1] - \Pr[A(U_{t^{(i+1)}}) = 1]|
\end{split}
\end{equation}
As $|\Pr[A(G(U_{t^{(i)}})) = 1] - \Pr[A(U_{t^{(i+1)}}) = 1]| \leq \epsilon$,
$$|\Pr[A(G^{(i+1)}(U_r)) = 1] - \Pr[A(G(U_{t^{(i)}})) = 1]| > i\epsilon$$
contradicting the the induction assumption.
So $G^{(i+1)}$ is an $(i+1)\epsilon$-PRG.
\end{proof}

\begin{theorem}
\label{PRGthm}
For any $d$-ary predicate $Q$,
if the random local function $F_{Q,n,m}$ is $\delta$-one-way for some constant $\delta \in (0,1)$, then we have the following results.

\begin{enumerate}

\item For some constant $c = c(d) > 1$, if $m > cn$ , then there exists a $\epsilon$-LPRG in $\AC^0$ with $\epsilon$ being negligible.

\item For any constant $c > 1$, if $m > n^{c}$, then there exists a $\epsilon$-PPRG in $\AC^0$ with $\epsilon$ being negligible.

\end{enumerate}

\end{theorem}

Before we prove Theorem \ref{PRGthm}, we first use it to obtain our main theorem in this subsection.
\begin{theorem}
\label{PRGthmfinal}
For any $d$-ary predicate $Q$,
if the random local function $F_{Q,n,m}$ is $\delta$-one-way for some constant $\delta \in (0,1)$, then we have the following results.

\begin{enumerate}

\item For some constant $c > 1$, if $m > cn$ , then for any constant $ a >1  $, there exists a $\epsilon$-LPRG $G: \{0,1\}^r \rightarrow \{0,1\}^t$ in $\AC^0$, where $t \geq ar$ and $\epsilon$ is negligible.

\item For any constant $c > 1$, if $m > n^{c}$, then for any constant $a>1 $ there exists a $\epsilon$-PPRG $G: \{0,1\}^r \rightarrow \{0,1\}^t$ in $\AC^0$, where $t \geq r^a$ and $\epsilon$ is negligible.

\end{enumerate}

\end{theorem}

\begin{proof}
For the first assertion, let the LPRG in Theorem \ref{PRGthm} be $G_0:\{0,1\}^{r_0} \rightarrow \{0,1\}^{t_0}$ with $t_0 > c_0 r_0$ for some constant $c_0 > 1$. We apply the construction in Lemma \ref{PRGenlength} to obtain $G^{(c_1)}$ such that $ c_0^{c_1} \geq a $. So $c_1$ is a constant. By Lemma \ref{PRGenlength} we know that $G^{(c_1)}$ is a $c_1 \epsilon$-PRG in $\AC^0$. This proves the first assertion.
 
By the same reason, the second assertion also holds.
\end{proof}

\begin{construction}
\label{ParallelRLF}

Let $F_{Q,n,m}:\{0,1\}^{s} \times \{0,1\}^{n} \rightarrow \{0,1\}^{m}$ be the random local function following Definition \ref{rlf}. We construct $F':\{0,1\}^{s} \times \{0,1\}^{n'} \rightarrow \{0,1\}^{m'}$ where $n' = tn, m' = tm, t = n$.

\begin{enumerate}

\item Draw $G$ uniformly from $\{0,1\}^s$.

\item Draw $x^{(1)}, x^{(2)}, \ldots, x^{(t)}$ independently uniformly from $\{0,1\}^n$. Let $x = (x^{(1)}, x^{(2)}, \ldots, x^{(t)})$.

\item Output $F'(G, x) = \bigcirc_{i=1}^{t} G(x^{(i)}) $.

\end{enumerate}

\end{construction}

\begin{lemma}

In Construction \ref{ParallelRLF}, for every constant $d \in \mathbb{N}^+$, every predicate $Q:\{0,1\}^d \rightarrow \{0,1\}$, every $m  = \poly(n)$ and every $\epsilon = 1/\poly(n)$, if $F_{Q, n, m}$ is $(\frac{1}{2}+\epsilon)$-last-bit unpredictable then $F'$ is $(\frac{1}{2} + \epsilon(1+ 1/n))$-unpredictable.

\end{lemma}

\begin{proof}

Suppose there exists a next-bit predictor $P$ and a sequence of indices $\{i_n\}$ such that 
$$ \Pr_{x \leftarrow U_n, G \leftarrow U_s, y = F'(G, x)}[P(G, y_{0,\ldots, i_n-1}) = y_{i_n}] \geq \frac{1}{2} + \epsilon(n) (1+1/n)$$
for sufficiently large $n$'s. 

Now we construct a last-bit predicator $P'$ which can predicate the last bit of $F_{Q, n, m}$ with success probability $1/2+\epsilon$.

By Remark 3.2 of \cite{applebaum2013pseudorandom}, $P'$ can find an index $j\in [m']$ by running a randomized algorithm $M$ in polynomial time, such that, with probability $1- 2^{-\Theta(n)}$ over the random bits used in $M$, 
$$\Pr_{G, x, y = F'(G, x)}[P(G, y_{[0,\ldots,j-1]}) = y_j] > \frac{1}{2} + \epsilon(n) + \frac{\epsilon(n)}{2n}.$$
Recall that in  Remark 3.2 of \cite{applebaum2013pseudorandom}, $M'$ tries every index and pick the best one.

According to Construction \ref{ParallelRLF}, assume $j = an + b$ for some $a, b\in \mathbb{N}, b< n $.

Given $(G, y_{[0,\ldots, m-2]})$, $P'$ generates $x^{(1)}, x^{(2)}, \ldots, x^{(a-1)}$ independently uniformly over $\{0,1\}^n$. Also $P'$ constructs a hypergraph $G'$ by swapping $S_b$ and $ S_{m-1} $ of $G$. Next, $P'$ computes $y' = \bigcirc_{i=1}^{a}G'(x^{(i)}) \circ y_{[0,\ldots, b-2]}$. Finally $P'$ outputs $P(G', y')$. As $G$ is uniform, $G'$ is also uniform. Also as $x^{(1)}, \ldots, x^{(a)}$ are uniform, $(G', y')$ has the same distribution as $(G, y_{[0,\ldots, j-1]})$. So
$$ \Pr[P'(G', y') = y_{m-1}] \geq \frac{1}{2} +  \epsilon(n) + \frac{\epsilon(n)}{2n} - 2^{\Theta(n)} > \frac{1}{2} + \epsilon(n).$$
This contradicts that  $F_{Q, n, m}$ is $(\frac{1}{2}+\epsilon)$-last-bit unpredictable.
\end{proof}

\begin{theorem}[\cite{applebaum2013pseudorandom}, Section 5]
\label{OWFtoUG}
For every constant $d \in \mathbb{N}$, predicate $Q : \{0,1\}^d \rightarrow
\{0,1\}$, and constant $\epsilon \in (0,\Match(Q)/2)$, there exists a constant $c > 0$ such that for every polynomial $m > cn$ the following holds. If the collection $F_{Q,n,m}$ is $\epsilon/5$-one-way then it is a $(1-\Match(Q)/2+\delta)$-last-bit UG where
$\delta = \epsilon(1-o(1))$. Thus it is also a $(1-\Match(Q)/2+\epsilon)$-UG.

\end{theorem}

\begin{remark}

Our definition of random local function has only one difference with the definition of \cite{applebaum2013pseudorandom}. That is, for each hyperedge we do not require the incoming vertices to be distinct. This difference does not affect the correctness of Theorem \ref{OWFtoUG}.


\end{remark}

\begin{construction}[Modified from \cite{applebaum2013pseudorandom} Construction 6.8]
\label{ConstUGtoPRG}
Let $F:\{0,1\}^{s(n)} \times \{0,1\}^n \rightarrow \{0,1\}^{m(n)}$ be a UG and $\Ext:\{0,1\}^{n_1} \times \{0,1\}^{d_1} \rightarrow \{0,1\}^{m_1} $ be a strong $(k = \alpha n_1, \epsilon_1)$-extractor following Theorem \ref{lenthm} where $n_1 = n$, $\alpha$ is some constant, $\epsilon = 1/2^{\Theta(\log^a n)}$ for some large enough constant $a\in \mathbb{N}^+$, $d_1 = (\log a)^{\Theta(a)}$, $m = 0.9k$. 

We construct the following UG $H:\{0,1\}^{sn}\times \{0,1\}^{n^2 + d_1 n} \rightarrow \{0,1\}^{mn}$.

\begin{enumerate}

\item Index: Generate $G_0, G_1, \ldots, G_{n-1}$ independently uniformly over $\{0,1\}^{s}$. Generate extractor seeds $u_0, u_1, \ldots, u_{m-1}$ independently uniformly over $\{0,1\}^{d_1}$. Denote $G = (G_0, G_1, \ldots, G_{n-1})$ and $u = (u_0, u_1,\ldots, u_{m-1})$.

\item Input: Generate $x^{(0)}, x^{(1)}, \ldots, x^{(n-1)}$ independently uniformly over $\{0,1\}^{n}$.

Denote $x = (x^{(0)}, x^{(1)}, \ldots, x^{(n-1)})$.

\item Output: Compute the $n\times m$ matrix $Y$ whose $i$th row is $G_{i}(x^{(i)})$. Let $Y_i$ denote the $i$th column of $Y$. Output $H(G, u, x) = \Ext(Y_0, u_0)\circ \Ext(Y_1, u_1) \circ \cdots \circ \Ext(Y_{m-1}, u_{m-1})$. 
\end{enumerate}

\end{construction}

\begin{remark}

There are 2 differences between our construction and Construction 6.8 of \cite{applebaum2013pseudorandom}. First, we use our AC0-extractor to do extraction. As our extractor is strong, its seed can also be regarded as part of the public key (index). Second, our construction is for any $m$, while their construction only considers $m$ as a linear function of $n$.

\end{remark}


\begin{lemma}
\label{UGtoPRG}
For any constant $\epsilon \in [0, 1/2)$,
if $F$ is $(\frac{1}{2} + \epsilon)$-unpredictable, then the mapping $H$ is a PRG with negligible error.

\end{lemma}

\begin{proof}[Proof Sketch]

The proof is almost the same as that of Lemma 6.9 of \cite{applebaum2013pseudorandom}. 

By the same argument of Lemma 6.9 of \cite{applebaum2013pseudorandom}, we know that for every sequence of efficiently computable index family $\{i_n\}$ and every efficient adversary $A$, there exists a random variable $W\in \{0,1\}^n$ jointly distributed with $G $ and $Y$ such that
\begin{itemize}
\item the min-entropy of $W$, given any fixed $G$ and the first $i_n$ columns of $Y$, is at least $n(1-2\epsilon - o(1))$.

\item $A$ cannot distinguish between $(G, Y_{[0,\ldots, i_n]})$ and $(G, [Y_{[0,\ldots,i_n-1]} W])$ with more than negligible advantage even when $A$ is given an oracle which samples the  distribution $(G, Y, W)$. Here $[Y_{[0,\ldots,i_n-1]} W]$ is a matrix such that the first $i_{n}$ columns are $Y_{[0,\ldots,i_n-1]}$ and the last column is $W$.

\end{itemize}

By the definition of strong extractors, for every family $\{i_n\}$, the distribution 
$$(G, u , Y_{[0, \ldots, i_n-1]}, \Ext(Y_{i_n}, u_{i_n}))$$
is indistinguishable from $(G, u , Y_{[1, \ldots, i_n-1]}, U_{m_1})$. Otherwise, suppose there is an adversary $B$ that can distinguish the two distributions. We construct another adversary $A$ as the follows. First $A$ generates a uniform $u$ as seeds for the extractors and invokes $B$ on $(G, u, y, \Ext(v, u))$ where $G$ is generated from uniform, $y$ is drawn from $Y_{[0,\ldots, i_n-1]}$. If $v$ is drawn from $Y_{i_n}$ then $B$ gets a sample from $(G, u, Y_{[0,\ldots, i_n-1]}, \Ext(Y_{i_n}, u))$. If $v$ is drawn from $ W $, then $B$ gets a sample from $(G, u, Y_{[0,\ldots, i_n-1]}, \Ext(W, u))$ which is $\epsilon_0$-close to $(G, u,  Y_{[0,\ldots, i_n-1]}, U_{m_1})$ by the definition of strong extractors, where $\epsilon_0$ is negligible according to our settings in Construction \ref{ConstUGtoPRG} and $U_{m_1}$ is the uniform distribution of length $m_1$.  So $A$ can distinguish $(G, Y_{[0,\ldots, i_n]})$ and $(G, [Y_{[0,\ldots,i_n-1]} W])$, having the same distinguishing advantage as $B$ does (up to a negligible loss). This is a contradiction.

As a result, for every family $\{i_n\}$, the distributions 
$$ (G, u, H(G, u, x)_{[0,\ldots, i_n]}) \mbox{ and } (G, u, H(G, u, x)_{[0,i_{n-1}]}\circ U_{m_1})   $$
are indistinguishable. So $H$ is a $(1/2 + \negl(n))$-UG. By Fact 6.1 (Yao's theorem) of \cite{applebaum2013pseudorandom}, $H$ is a PRG.
\end{proof}

\begin{proof}[Proof of Theorem \ref{PRGthm}]

We combine Construction \ref{ParallelRLF} and Construction \ref{ConstUGtoPRG} together by using the UG of Construction \ref{ParallelRLF} in Construction \ref{ConstUGtoPRG}. By Theorem \ref{OWFtoUG} and Lemma \ref{UGtoPRG}, we know that our construction gives a PRG (with negligible error). Assume the PRG is $H:\{0,1\}^{s_H} \times \{0,1\}^{n_H} \rightarrow \{0,1\}^{m_H}$.

Next we mainly focus on the stretch. The output length of $H$ is $m_H = \Theta(ntm)$, the input length (including the index length) is $s_H + n_H = sn + d_1n + n^2 t$. Here we know that $s = m\log n$, $t = n$. 

Assume $m > cn $ for some constant $c>1$. We know that $\frac{ m_H}{s_H + n_H} = c' > 1$, for some constant $c'$. 

For the polynomial stretch case,  assume $m > n^{c}$ for some constant $c > 1$. We know that $m_H \geq (s_H + n_H)^{c'}$ for some constant $c' > 1$.

For both cases, the construction is in $\AC^0$. The reason is as follows. By  Lemma \ref{RLFinAC0}, the random local function is in $\AC^0$. In Construction \ref{ParallelRLF} and Construction \ref{ConstUGtoPRG}, we compute $O(nt)$ random local functions (some of them share the same index) in parallel. Also our extractor is in $\AC^0$. So the overall construction is in $\AC^0$.

This proves the theorem.

\end{proof}

\subsection{PRG in AC0 for Space Bounded Computation}

In this subsection, we give an $\AC^0$ version of the PRG in \cite{NisanZ96}.

\begin{theorem}

For every constant $c\in \mathbb{N}$ and every $m = m(s) = \poly(s)$, there is an explicit PRG $g: \{0,1\}^{r = O(s)} \rightarrow \{0,1\}^{m}$ in $\AC^0$, such that for any randomized algorithm $A$ using space $s$,
$$ | \Pr[A(g(U_r)) = 1] - \Pr[A(U_m) = 1] | = \epsilon \leq  2^{-\Theta(\log^c s)},$$
where $U_r$ is the uniform distribution of length $r$,  $U_m$ is the uniform distribution of length $m$.

\end{theorem}

\begin{proof}[Proof Sketch]

We modify the construction of \cite{NisanZ96} by replacing their extractor with the extractor from Theorem \ref{lenthm} for some constant entropy rate and with error parameter $\epsilon' = 2^{-\Theta(\log^c s)}$. In the PRG construction of \cite{NisanZ96}, it only requires an extractor for constant entropy rate.  As our extractor meets their requirement, the proof in \cite{NisanZ96} still holds under this modification.

For the security parameter $\epsilon$,
according to \cite{NisanZ96}, $\epsilon = \poly(s) (\epsilon' + 2^{-s})$. As a result, $\epsilon = 2^{-\Theta(\log^c s)}$.

\end{proof}

%% file: appendix.tex
\appendix

\section{Proof of Lemma \ref{amp3}}\label{appen}

\begin{lemma}[Lemma \ref{amp3} restated, implicit in \cite{iw:bppequalsp}]
\label{amp3restate}
For any  $\gamma \in (0, 1/30)$,
if there is a boolean function $f:\{0,1\}^l\rightarrow \{0,1\}$ that is $1/3$-hard for circuit size $g = 2^{\gamma l}$, then there is a boolean function $f':\{0,1\}^{l' = \Theta(l)} \rightarrow \{0,1\}$ that is $(1/2-\epsilon)$-hard for circuit size $g' =  \Theta( g^{1/4} \epsilon^2 l^{-2})$  
where $\epsilon \geq (500l)^{1/3} g^{-1/12} $.
$$f'(a,s,v_1,w) = \langle s, f(a|_{S_1} \oplus v_1) \circ f(a|_{S_2} \oplus v_2) \circ \cdots f(a|_{S_l} \oplus v_l) \rangle$$

Here $(S_1,\ldots, S_l)$ is an $(|a|,l, \gamma l/4,l)$-design where $|a| = \lfloor \frac{40l}{\gamma}\rfloor$. The vectors $v_1,\ldots,v_l$ are obtained by a random walk on an expander graph, starting at $v_1$ and walking according to $w$ where $|v_1| = l, |w| = \Theta(l)$. The length of $s$ is $l$. So $ l' = |a| + |s| + |v_1|+|w| = \Theta(l)$.

\end{lemma}

In order to clearly compute the circuit size, we need the following theorem.
\begin{theorem}[Circuit Size of Majority Function \cite{wegener1987complexity} Page 76, Theorem 4.1]
\label{majsize}

The circuit size of the majority function on $n$ input bits is $O(n)$.

\end{theorem}

We need the following version of Goldreich Levin Theorem.

\begin{theorem}[Goldreich-Levin Algorithm \cite{GoldreichL89}, Circuit Version]
\label{GLC}
For any $\epsilon > 0$, for any function $f:\{0,1\}^{l} \rightarrow \{0,1\}^t$,
if there is a circuit $C: \{0,1\}^{l+t} \rightarrow \{0,1\}$ of size $g$ such that we have 
$$ \Pr_{x,r}[C(x, r) = \langle f(x), r\rangle] \geq 1/2 + \epsilon,$$
where $r$ is uniformly random distributed over $\{0,1\}^t$, $x$ is uniformly distributed over $\{0,1\}^l$,
then there is a circuit $C':\{0,1\}^{l} \rightarrow \{0,1\}^t$ of size $O(g t^2/\epsilon^2)$ such that
$$ \Pr_x[C'(x) = f(x)]  \geq \epsilon^3/(500t).$$

\end{theorem}

\begin{proof}

As 
$$ \Pr_{x,r}[C(x, r) = \langle f(x), r\rangle] \geq 1/2 + \epsilon,$$
for at least $\epsilon/2$ fraction of all $x \in \{0,1\}^l$,
$$ \Pr_r[C(x, r) = \langle f(x), r\rangle] \geq 1/2 + \epsilon/2.$$

Let $T = \{x: \Pr_r[C(x, r) = \langle f(x), r\rangle] \geq 1/2 + \epsilon/2\}$.
We have $|T|/2^{l} \geq \epsilon/2$.

For every $x$ in $T$, consider the following Goldreich-Levin algorithm $GL$.

\begin{algorithm}[H]
\SetKwInOut{Input}{Input}
\SetAlgoLined

\Input{$x \in \{0,1\}^{l}$}

Let $L_x = \emptyset$

Generate uniform random strings $r_{0}, r_{ 1}, \ldots, r_{k-1}$ over $\{0,1\}^t$, where $k = \lceil \log (100t/\epsilon^2 + 1)\rceil$.

For every  $S\subseteq [k], S\neq \emptyset$, let $r_S = \sum_{j\in S} r_j$.

Let $ R =  \{ r_S\in \{0,1\}^t: \emptyset \neq S\subseteq [k]\} $

\For { $b_0, b_1, \ldots, b_{k-1} \in \{0,1\}$}
{
	For every  $S\subseteq [k], S\neq \emptyset$, let $b_S = \sum_{j\in S}b_j$
	
	Let $ B = \{b_S \in \{0,1\}: \emptyset \neq S\subseteq [k] \}$
	
	\For{$i= 0$ to $t-1$}
	{
		$y_i = \maj_{\emptyset \neq S\subseteq [k]}\{C(x, r_S \oplus e_i)\oplus b_S \}$
	}
 	Add $y$ to $L_x$
	
}

\Return $L_x$

\caption{$GL(x)$}
\end{algorithm}

We claim that $\forall x\in T$, with probability at least $0.99$ over the random variables used in $GL$, $x\in L_x$. The reason is as follows.

In the algorithm $GL$, we try all the possibilities of $b_0,\ldots, b_{k-1}$. Consider one special choice of them, saying $b_j = \langle f(x), r_j \rangle, \forall j \in [k]$. As a result, $b_S = \langle f(x), r_S \rangle, \forall S \subseteq [k], S\neq \emptyset$. Now we fix an  $x\in T$ and an $i \in [t]$. If $C(x, r_S \oplus e_i) = \langle f(x), r_S\oplus e_i\rangle$, then $C(x,  r_S \oplus e_i) \oplus b_S =  f(x)_i$.
We know that, as $x\in T$,
$$\Pr_{r_S}[C(x, r_S\oplus e_i) = \langle f(x), r_S \oplus e_i\rangle] = 1/2 + \epsilon'/2$$
for $\epsilon' \in [\epsilon, 1]$.
So 
$$ \Pr_{ r_S}[C(x, r_S\oplus e_i) \oplus b_S = f(x)_i] = 1/2+ \epsilon'/2. $$
According to the construction of $r_S$ in our algorithm, $r_S$'s, $\forall S \subseteq [k], S \neq \emptyset$ are pairwise independent.
Let $I_S$ denote the indicator such that $I_S = 1$ if $C(x,  r_S\oplus e_i)\oplus b_S = f(x)_i$ and $I_S = 0$ otherwise. Let $I = \sum_{\emptyset \neq S\subseteq [k]} I_S$.
Thus $\mathbf{E} I_S  = 1/2+ \epsilon'/2$. So $\mathbf{E}I = (1/2 + \epsilon'/2)(2^k-1)$. 
Also, the variance of $I_S$ is $ \mathbf{Var}(I_S) = \mathbf{E} I^2_S - (\mathbf{E} I_S)^2 = \mathbf{E} I_S - (\mathbf{E} I_S)^2 \leq 1/4 $. 
So $\mathbf{Var}(I) = \sum_{\emptyset \neq S\subseteq [k]} \mathbf{Var}(I_S) \leq \sum_{\emptyset \neq S\subseteq [k]} \frac{1}{4} = (2^k-1)/4$.
So according to the Chebyshev's inequality,
$$\Pr[I \leq 1/2(2^k-1)] \leq \Pr[|I - \mathbf{E} I | \geq \frac{\epsilon'}{2} (2^k - 1) ] \leq \frac{\mathbf{Var}(I)}{( \frac{\epsilon'}{2} (2^k - 1))^2} = \frac{1}{\epsilon'^2(2^k-1)} \leq 1/(100t),$$
as $2^k - 1 \geq 100t/\epsilon^2 $ and $\epsilon' \geq \epsilon$.
Thus $\Pr[y_i \neq f(x)_i] \leq \Pr[I \leq 1/2(2^k-1)]\leq 1/(100t)$.
According to the union bound, $\Pr[y = f(x)] \geq 1- t/(100t)= 0.99$. 

We know that as we have guessed all the possible values of $b_0, b_1, \ldots, b_{k-1}$, one of them should be correct. So for every $x\in T$, with probability at least $0.99$ over $r_0, r_1,\ldots, r_{k-1}$, $x$ is in $L_x$.  

Now we modify this algorithm to construct the circuit $C'$. Notice that we only need to prove that there exists a circuit $C'$. So we are going to fix the random variables and the choice of $b_0, b_1, \ldots,b_{k-1}$ in the algorithm.

We first fix those random variables. Assume we use the same random variables $r_0, r_1,\ldots, r_{k-1}$ to get $L_x$ for every $x\in T$. For every $x\in T$, let $I_x$ denote the event that $x\in L_x$. We know that $\mathbf{E} I_x \geq 0.99$. So $\mathbf{E} \sum_{x\in T} I_x \geq 0.99 |T|$. So there exists $r'_0,r'_1,\ldots,  r'_{k-1}$ such that 
$$  (\sum_{x\in T} I_x)|_{ r_i = r'_i, \forall i \in [k]} \geq 0.99|T|.$$

Denote the event $r_i = r'_i, \forall i \in [k]$ as $\varphi$.

After we fix these random variables, we fix the choice of $b_0, b_1, \ldots, b_{k-1}$. We know that for at least $0.99|T|$ number of $x\in T$, $x \in L_x|_{\varphi}$.
For every $x$, let $ y_{x, b_0,\ldots, b_{k-1}} $ denote the element in $L_x|_{\varphi}$ corresponding to $b_0, b_1,\ldots, b_{k-1}$.
As the size of $L_x|_{\varphi} $ is $2^k \leq 200t/\epsilon^2 +2$, there exists $b'_0, b'_1,\ldots, b'_{k-1} \in \{0,1\}$ such that for at least $\frac{0.99|T|}{2^l 2^k} \geq \epsilon^3/(500t)$ fraction of $x \in \{0,1\}^l$, $ y_{x, b'_0, b'_1, \ldots, b'_{k-1}} = f(x) $. 

By fixing the random variables and $b_0, b_1,\ldots, b_{k-1}$, according to our algorithm, we have a circuit $C'$, such that for every $i \in [t]$,
$$ C'(x)_i =  \maj_{\emptyset \neq S\subseteq [k]}\{ C(x, r'_S \oplus e_i)\oplus b'_S \}$$
where $\forall S\subseteq [k], S\neq \emptyset$, $r'_S = \sum_{j\in S} r'_j$, $b'_S = \sum_{j\in S} b'_j$.

By Theorem \ref{majsize}, we know the circuit size for computing majority function over $2^k - 1$ input bits is $O(2^k)$.

As all the random variables and $b_0,\ldots, b_{k-1}$ are fixed, for every $i \in [l]$ and every $  S \subseteq [k], S \neq \emptyset$, $ r'_S \oplus e_i$ and $b'_S$ are fixed (given bits, not need to be computed by the circuit).

So we can see that for every $i\in [t]$, the circuit size of $C'(\cdot)_i$ is  $O( g\cdot 2^k  + 2^k) = O(gt/\epsilon^2 + t/\epsilon^2) = O(gt/\epsilon^2)$. So the circuit size of $C'$ is $O(gt^2/\epsilon^2 )$.

\end{proof}

\begin{lemma}[\cite{iw:bppequalsp} Section 5.3]
\label{amp3step1}
For any  $\gamma \in (0, 1/30)$,
if there is a boolean function $f:\{0,1\}^l\rightarrow \{0,1\}$ that is $1/3$-hard for circuit size $g = 2^{\gamma l}$, then there is a boolean function $\tilde{f}:\{0,1\}^{\tilde{l} = \Theta(l)} \rightarrow \{0,1\}^l$ that is $g^{-1/4}$-hard for circuit size $ g^{1/4}$.

$$\tilde{f}(a,v_1,w) = f(a|_{S_1} \oplus v_1) \circ f(a|_{S_2} \oplus v_2) \circ \cdots f(a|_{S_l} \oplus v_l)$$

Here $(S_1,\ldots, S_l)$ is an $(|a|,l, \gamma l/4,l)$-design where $|a| = \lfloor \frac{40l}{\gamma}\rfloor$. The vectors $v_1,\ldots,v_l$ are obtained by a random walk on an expander graph, starting at $v_1$ and walking according to $w$ where $|v_1| = l, |w| = \Theta(l)$. So $ \tilde{l} = |a|  + |v_1|+|w| = \Theta(l)$.

\end{lemma}

Now we prove Lemma \ref{amp3restate}.

\begin{proof}
By Lemma \ref{amp3step1}, for any  $\gamma \in (0, 1/30)$,
if there is a boolean function $f:\{0,1\}^l\rightarrow \{0,1\}$ that is $1/3$-hard for circuit size $g = 2^{\gamma l}$, then there is a boolean function $\tilde{f}:\{0,1\}^{\tilde{l} = \Theta(l)} \rightarrow \{0,1\}^l$ that is $\tilde{\epsilon} = g^{-1/4}$-hard for circuit size $\tilde{g} = g^{1/4}$.

$$\tilde{f}(a,v_1,w) = f(a|_{S_1} \oplus v_1) \circ f(a|_{S_2} \oplus v_2) \circ \cdots f(a|_{S_l} \oplus v_l)$$

Now consider $f':\{0,1\}^{\tilde{l} + l} \rightarrow \{0,1\}$ which is as follows.
$$f'(a,s,v_1,w) = \langle s, f(a|_{S_1} \oplus v_1) \circ f(a|_{S_2} \oplus v_2) \circ \cdots f(a|_{S_l} \oplus v_l) \rangle$$
where $|s| = l$.


By Theorem \ref{GLC}, $f'$ is $(1/2-\epsilon)$-hard for circuits of size $g' = \Theta(\tilde{g}\epsilon^2 l^{-2} ) = \Theta( g^{1/4} \epsilon^2 l^{-2})$  
where $ \epsilon$ can be such that $\epsilon^3/(500l) \geq \tilde{\epsilon}$. That is $\epsilon \geq (500l)^{1/3} g^{-1/12}$.

\end{proof}

%
%
%
%